\documentclass{aa}  
\usepackage{graphicx}
\usepackage{txfonts}
\newcommand{\Mv}{\mbox{$M_{V}\,$}}
\newcommand{\Mk}{\mbox{$M_{K}\,$}}
\newcommand{\Mj}{\mbox{$M_{J}\,$}}

\newcommand{\Mi}{\mbox{$M_{I}\,$}}
\newcommand{\Wvi}{\mbox{$W_{VI}\,$}}
\newcommand{\Wjk}{\mbox{$W_{JK}\,$}}

\newcommand{\FeH}{\mbox{[Fe/H]}\,}

\newcommand{\kms}{\mbox{$\mbox{km\,s}^{-1}$}\,}

\newcommand{\logP}{\mbox{$\log P$}\ }

\begin{document}

   \title{The effect of metallicity on Cepheid Period-Luminosity
relations from a Baade-Wesselink analysis of Cepheids in the Milky
Way and Magellanic Clouds\thanks{Based on data obtained with ESO-LP-190.D-0237, and 
programmes 097.D-0150 and 097.D-0151}}

\titlerunning{The effect of metallicity on the Cepheid PL relations}
\authorrunning{W.~Gieren et al.}

   %\subtitle{A better title is needed}

\author{
W. Gieren\inst{1,2},
J. Storm\inst{3},
P. Konorski\inst{4},
M. G\'orski\inst{1,2},
B. Pilecki\inst{5},
I. Thompson\inst{6},
G. Pietrzy\'nski\inst{5,1},
D. Graczyk\inst{1,2,5},
T.G. Barnes\inst{7},
P. Fouqu\'e\inst{8,9},
N. Nardetto\inst{10},
A. Gallenne\inst{11},
P. Karczmarek\inst{4},
K. Suchomska\inst{4},
P. Wielg\'orski\inst{5},
M. Taormina\inst{5},
B. Zgirski\inst{5}
}

   \institute{
	     Universidad de Concepci\'on, Departamento
             Astronom\'ia, Casilla 160-C, Concepci\'on, Chile
             \email{wgieren@astro-udec.cl}
         \and
	     Millenium Institute of Astrophysics, Santiago, Chile
	 \and
             Leibniz-Institut f\"ur Astrophysik Potsdam (AIP),
             An der Sternwarte 16,
             D-14482 Potsdam, Germany,
             %\email{jstorm@aip.de}
%%             \thanks{The university of heaven temporarily does not
%%                     accept e-mails}
	\and
           Obserwatorium Astronomiczne, Uniwersytet Warszawski, Al.
           Ujazdowskie 4, 00-478, Warsaw, Poland
         \and
            Centrum Astronomiczne im. Mikołaja Kopernika (CAMK), PAN, 
            Bartycka 18, 00-716 Warsaw, Poland
	\and
	    Carnegie Observatories, 813 Santa Barbara Street, Pasadena,
            CA 911101-1292, U.S.A.
	 \and
	    University of Texas at Austin, McDonald Observatory, 1
            University Station, C1402, Austin, TX 78712-0259, U.S.A.
	 \and
            IRAP, Universit\'e de Toulouse, CNRS, 
            14 av. E. Belin, F-31400 Toulouse, France
	 \and
	    CFHT Corporation, 65-1238 Mamalahoa Hwy, Kamuela, 
            HI 96743, U.S.A.
	 \and
	    Université C\^{o}te d'Azur, 
	    Observatoire de la C\^{o}te d'Azur, CNRS, 
	    Laboratoire Lagrange, France
	 \and
	    European Southern Observatory, Alonso de C\'ordova 3107, 
            Casilla 19001, Santiago, Chile
          }

%   \date{Received September 15, 1996; accepted March 16, 1997}
    \date{Received; accepted}

\abstract{The extragalactic distance scale builds on the
Cepheid period-luminosity (PL) relation. Decades of work have not
yet convincingly established the sensitivity of the PL relation to
metallicity. This currently prevents a determination of the Hubble
constant accurate to 1\% from the classical Cepheid-SN Ia method. } 
{In this paper, we want to carry out a strictly differential comparison
of the absolute PL relations obeyed by classical Cepheids in the Milky
Way (MW), LMC and SMC galaxies. Taking advantage of the substantial metallicity
difference among the Cepheid populations in these three galaxies,
we want to establish a possible systematic trend of the PL relation
absolute zero point as a function of metallicity, and determine the
size of such an effect in optical and near-infrared photometric bands.}
{We are using the IRSB Baade-Wesselink type method as calibrated by Storm et
al. to determine individual distances to the Cepheids in our samples in
MW, LMC and SMC. For our analysis, we use a greatly enhanced sample of
Cepheids in the SMC (31 stars) as compared to the small sample (5 stars)
available in our previous work. We use
the distances to determine absolute Cepheid PL relations in optical
and near-infrared bands in each of the three galaxies.} {Our
distance analysis of 31 SMC Cepheids with periods from 4-69 days yields
tight PL relations in all studied bands, with slopes consistent with the 
corresponding LMC and MW relations.
Adopting the very accurately determined LMC slopes for the optical 
and near-infrared bands, we
determine the zero point offsets between the corresponding absolute PL
relations in the 3 galaxies. We find that in all bands the metal-poor SMC
Cepheids are intrinsically fainter than their more metal-rich counterparts
in the LMC and MW.  In the $K$ band 
the metallicity effect is $-0.23\pm0.06$~mag/dex while
in the $V,(V-I)$ Wesenheit index it is slightly stronger, 
$-0.34\pm0.06$~mag/dex.  We find some evidence that the PL relation
zero point-metallicity relation might be nonlinear, becoming steeper for lower
metallicities.} 
{Using sizeable Cepheid samples in the MW,
LMC and SMC with very accurate photometric and radial velocity data we
establish the metallicity sensitivity of the Cepheid PL relations in the
optical and near-infrared regimes. We find a significant effect in all
bands in the sense that the more metal-poor SMC Cepheids are intrinsically
fainter than their LMC and Galactic counterparts.  We find suggestive evidence
that the metallicity sensitivity of the PL relation might be nonlinear,
being small in the range between solar and LMC Cepheid metallicity,
and becoming steeper towards the lower-metallicity regime.}

   \keywords{Stars: variables: Cepheids --
	Stars: distances --
	Stars: fundamental parameters --
	Magellanic Clouds --
	Galaxies: distances and redshifts
        }

   \maketitle
%
%-------------------------------------------------------------------

%\section{Introduction}
\section{Introduction} One of the most important current challenges in
astrophysics is the quest for a 1\% determination of the local value of
the Hubble constant $H_0$. The traditional route \citep[e.g.][and
references therein]{Freedman01, Sandage06} to determine the value of
the local Hubble constant is to use reddening insensitive photometric
observations of classical Cepheid variables in nearby type Ia supernovae
(SN Ia) host galaxies.  The distances to these galaxies can then be
determined from the Cepheid period-luminosity (PL) relation, calibrating
in this way the SN Ia peak luminosities.  The most recent application
of this method has led to an accuracy of $H_0$ of 2.4\% \citep{Riess16,
Riess18}.

It is obviously of fundamental importance in this process to provide a
very accurate absolute calibration of the Cepheid PL relation. This can
be achieved in different ways, using Cepheids in our own Galaxy having
accurate parallax measurements, or using extragalactic Cepheids for which
the distances of their host galaxies have been very accurately determined
with some Cepheid-independent method. A critical aspect of the calibration
of the PL relation is a precise determination of its possible dependence
on the metallicity of the Cepheid variables. Without an accurate knowledge
of this "metallicity effect" a distance measurement to a galaxy accurate
to 1\% with a Cepheid PL relation is clearly not possible. 
In the past,
a lot of work has been done to determine the metallicity dependence of
the Cepheid PL relation with a variety of methods. Early work includes
the studies of \cite{Gould94}, \cite{Kennicutt98} and \cite{Sakai04}.
More recent studies on the metallicity effect are those of
\cite{Shappee11}, \cite{Mager13}, and \cite{Fausnaugh15}, without this list
being exhaustive. A very detailed compilation of metallicity effect 
determinations prior to 2008 can be found in Table 1 of 
\cite{Romaniello08}. 
These studies seemed to indicate
that the metallicity effect in the near-infrared $JHK$ bands is small and
perhaps even vanishing while there is a significant effect in optical
and mid-infrared photometric bands \citep[e.g.][]{FM11}.  While most
studies yielded a negative sign of the metallicity effect in optical
bands meaning that more metal-rich Cepheids are intrinsically brighter
than their more metal-poor counterparts of the same pulsation period, 
the work of \cite{Romaniello08} has yielded
the opposite sign for the effect in the V band, meaning that metal-poor Cepheids
are intrinsically brighter than their more metal-rich counterparts of the same
pulsation period. 
Theoretical studies \citep[e.g][]{Caputo00, Bono08}
seemed to support Romaniello's results, but the uncertainty on these determinations
of the metallicity effect from pulsation theory still seems to be rather substantial.
In the most recent work on the subject, \cite{Wielgorski17},
using the extremely well-established Cepheid PL relations in the
Magellanic Clouds in optical and near-infrared bands combined with
the accurate distance determinations to the LMC \citep{Pietrzynski13}
and SMC \citep{Graczyk14} from late-type eclipsing binaries, found a
metallicity effect compatible with zero in all bands. As a conclusion,
there is still not a consensus about the true effect of metallicity
on Cepheid absolute magnitudes in different spectral regions.  It is
especially important to obtain a truly accurate determination of the
metallicity effect in the near-infrared bands since these are used, due
to their much lower sensitivity to extinction, in the space-based work
on $H_0$ with the Hubble Space Telescope during the last two decades,
and the near future with the James Webb Space Telescope.

We present here new and improved measurements of
the metallicity effect by direct distance determinations to sizeable
samples of Cepheids in the Milky Way, LMC and SMC galaxies, using the
Infrared Surface Brightness (IRSB) Technique originally introduced
by \cite{FG97}.  This technique was applied to Galactic Cepheids for
the first time by \cite{Gieren97,Gieren98} demonstrating the great
improvement in the accuracy of the Cepheid distances as compared to
the earlier version of the technique which had used the $(V-R)$ colour
index as a surface brightness indicator \citep{Gieren93}. The IRSB
technique was later improved by \cite{Gieren05}, and by \cite{Storm11a}
and applied to extra galactic Cepheids \citep{Storm04}. In \cite{Storm04}
we demonstrated that the IRSB technique itself is capable of yielding
distances to Cepheids which are independent of their metallicities.
In \cite{Storm11b}, we analyzed samples of Cepheids in the Milky Way
(MW), LMC and a few in the SMC and found no significant metallicity
effect in the $K$-band, and a marginally significant effect of $-0.23 \pm
0.10$~mag/dex in the optical $V,(V-I)$ Wesenheit index.  This result
is valid for the metallicity range between solar and $-0.35$ dex,
which is the mean metallicity of classical Cepheids in the LMC
\citep[e.g.][]{Luck98}. However, since the Cepheid populations in the outer
part of massive spiral galaxies tend to typically have metallicities
comparable to those in the SMC \citep[e.g.][]{Bresolin09}, it is very
important to extend the determination of the metallicity effect down
to the $-0.73$~dex metallicity of SMC Cepheids. A perfect opportunity
to do so is a distance determination to a sample of SMC Cepheids using
the very same IRSB technique we had used before on MW and LMC Cepheids,
and compare the absolute PL relations defined by these distances with
the ones we had previously obtained for the Cepheid samples in the more
metal-rich MW and LMC galaxies.  We note here that this approach is
purely differential and does not depend directly on the true value of the
Cepheid absolute magnitude scale. It is thus robust even if the Cepheid
absolute magnitude scale might change a bit as more accurate geometric
parallaxes to Cepheids become available \citep[e.g.][]{Clementini17,
Riess18}.  We show in the following sections
that we now obtain small but significant metallicity effects in both,
near-infrared and optical bands, in the sense that the metal-poorest
sample of Cepheids (the SMC sample) exhibits PL relations shifted to
fainter absolute magnitudes as compared to MW and LMC.

\section{The Data}
\label{sec.data}

\begin{figure*}
\centering
\includegraphics[width=18cm]{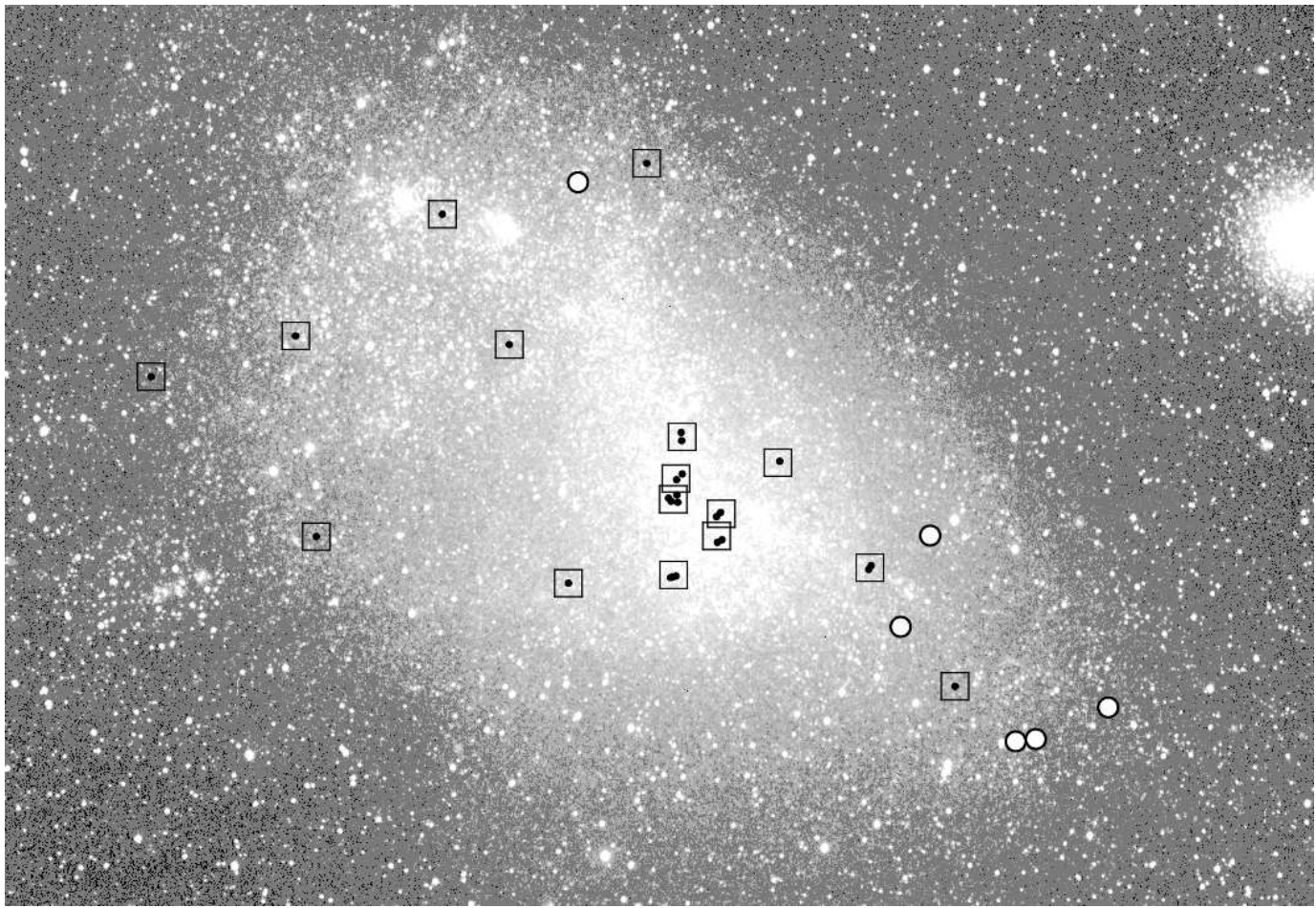}
\caption{The location of the Cepheids in the SMC have been marked with
black dots for the new stars and filled circles for the six stars
studied previously. The boxes are indicative of the field size of the
SOFI near-IR imager. The underlying image was obtained by ASAS and originates
from \cite{Udalski08}
\label{fig.SMC_finder}}
\end{figure*}

\subsection{The sample}
\label{sec.sample}
Based on the OGLE survey \citep{Udalski08} we have selected a sample of 26
Cepheids distributed over the face of the SMC which appeared to be little
affected by crowding and spanning the range of pulsation periods from four
to seventy days. In addition to these stars we also include the five stars
HV~822, HV~1328, HV~1333, HV~1335, and HV~1345 analyzed by \cite{Storm04}
and the star HV~837 which was analysed by \cite{Groenewegen13}. To make
efficient use of the near-IR imager we looked for fields which contained
several Cepheids. We succeeded in finding seven fields containing more
than one Cepheid. In Tab.\ref{tab.Kbandfields} the individual fields
have been listed and in Fig.\ref{fig.SMC_finder} we show the location of
the individual stars as well as the actual fields of the SOFI near-IR
imager. It can be seen that about half of the stars are in the central
part of the SMC while the other half samples the outer parts.

While all of the SMC Cepheids in our sample had previous $V$- and $I$-band
light curves from the OGLE Project, the application of the near-IR
surface brightness technique requires full and precise radial curves and
$K$-band light curves. ESO granted us time through a Large Program which
we have supplemented with additional observations from other facilities.
These new data are described and discussed in the following subsections.

\begin{table}
\caption{\label{tab.Kbandfields} The Cepheids in the sample grouped
according to field. The first group is all the fields containing a
single Cepheid. The next fields show the grouping of the remaining
stars. The table gives the OGLE-SMC-CEP identifier, the coordinates 
(J2000.0), the number of nights ($N_{\mbox{\scriptsize n}}$) 
the Cepheid has been observed, 
the number of those nights, $N_{\mbox{\scriptsize cal}}$,
 which also had standard star observations. Finally the
standard deviation of the photometric zero points from those nights
and the resulting estimated zero point uncertainty are given.}
\scriptsize
\begin{tabular}{c c c c c c c}
\hline\hline
ID & RA & DEC & $N_{\mbox{\scriptsize n}}$ &
$N_{\mbox{\scriptsize cal}}$ & std.dev. & $\sigma(ZP)$\\
\hline
0320 &  00:39:33.82 &  -73:44:54.3 & 25 & 12 &  0.015 &  0.006\\
1977 &  00:52:56.82 &  -71:55:03.3 & 26 & 14 &  0.011 &  0.004\\
2533 &  00:56:20.90 &  -73:23:13.9 & 21 & 10 &  0.018 &  0.008\\
2905 &  00:58:55.06 &  -72:33:07.5 & 31 & 14 &  0.015 &  0.005\\
3311 &  01:01:49.49 &  -72:05:45.4 & 25 & 14 &  0.010 &  0.004\\
4017 &  01:08:11.60 &  -72:31:18.3 & 24 & 13 &  0.012 &  0.005\\
3927 &  01:07:17.54 &  -73:13:26.1 & 24 & 12 &  0.020 &  0.008\\
4444 &  01:14:28.07 &  -72:39:53.6 & 17 &  9 &  0.011 &  0.005\\
0958 &  00:47:10.43 &  -72:57:37.9 & 24 & 14 &  0.014 &  0.005\\
\hline 
%00:43:15.50 & -73:20:00.0 & GR1 & 22 & 13 & 0.013 \\
0518 &  00:43:12.35 &  -73:19:31.8 & 22 & 13 &  0.013 &  0.005\\
0524 &  00:43:18.77 &  -73:20:19.8 & 22 & 13 &  0.013 &  0.005\\
\hline
%00:51:25.00 & -72:52:30.0 & GR6 & 28 & 17 & 0.014 \\
1686 &  00:51:26.08 &  -72:53:18.4 & 28 & 17 &  0.014 &  0.004\\
1693 &  00:51:27.36 &  -72:51:35.1 & 28 & 17 &  0.014 &  0.004\\
%\hline
%00:47:15.50 & -72:57:55.0 & GR2 & 24 & 14 & 0.014 \\
%00:47:10.43 & -72:57:37.9 & 0958 & 24 & 14 & 0.014 \\
%00:47:21.65 & -72:58:13 & 0981 # & 24 & 14 & 0.014 \\
\hline
%00:49:42.60 & -73:08:40.0 & GR3 & 27 & 17 & 0.020 \\
%00:49:31.73 & -73:07:56.9 & 1343 # & 27 & 17 & 0.020 \\
1385 &  00:49:44.60 &  -73:08:23.1 & 27 & 17 &  0.020 &  0.006\\
1410 &  00:49:55.28 &  -73:09:16.2 & 27 & 17 &  0.020 &  0.006\\
\hline
%00:49:55.00 & -73:13:20.0 & GR4 & 24 & 14 & 0.021 \\
1365 &  00:49:40.95 &  -73:14:07.0 & 24 & 14 &  0.021 &  0.008\\
1403 &  00:49:52.88 &  -73:14:41.0 & 24 & 14 &  0.021 &  0.008\\
%00:50:09.54 & -73:11:53.3 & 1454  # & 24 & 14 & 0.021 \\
\hline
%00:51:41.00 & -73:01:15.0 & GR7 & 28 & 15 & 0.020 \\
1680 &  00:51:24.40 &  -73:00:17.9 & 28 & 15 &  0.020 &  0.007\\
1723 &  00:51:39.32 &  -73:01:30.6 & 28 & 15 &  0.020 &  0.007\\
%00:51:58.69 & -73:02:24.1 & 1785 & 28 & 15 & 0.020 \\
\hline
%00:51:47.00 & -73:21:30.0 & GR5 & 26 & 17 & 0.015 \\
1729 &  00:51:40.67 &  -73:21:44.3 & 26 & 17 &  0.015 &  0.005\\
1750 &  00:51:49.15 &  -73:21:55.5 & 26 & 17 &  0.015 &  0.005\\
1765 &  00:51:54.98 &  -73:22:04.0 & 26 & 17 &  0.015 &  0.005\\
%\hline
%00:51:25.00 & -72:52:30.0 & GR6 & 28 & 17 & 0.014 \\
%00:51:26.08 & -72:53:18.4 & 1686 & 28 & 17 & 0.014 \\
%00:51:27.36 & -72:51:35.1 & 1693 & 28 & 17 & 0.014 \\
\hline
%00:51:48.00 & -73:05:37.0 & GR8 & 33 & 19 & 0.010 \\
1712 &  00:51:36.22 &  -73:06:15.1 & 33 & 19 &  0.010 &  0.003\\
1717 &  00:51:38.26 &  -73:04:43.4 & 33 & 19 &  0.010 &  0.003\\
1761 &  00:51:53.58 &  -73:06:01.6 & 33 & 19 &  0.010 &  0.003\\
1797 &  00:52:00.35 &  -73:05:22.3 & 33 & 19 &  0.010 &  0.003\\
\hline
 & & & & & mag & mag \\
\hline
\end{tabular}
\end{table}

\subsection{The $K$-band photometric data}
\label{sec.photometry}
The $K$-band data presented in this paper were collected with
the ESO NTT telescope at the La Silla observatory in Chile, equipped
with the SOFI near-IR camera \citep{Moorwood98}. 
With the ,,{\tt LARGE\_FIELD\_IMAGING}'' setup
that we used, the field of view was 4.9x4.9 arcmin, with a scale of
0.288~arcsec/pixel. A total of 754 data points were collected 
including the data for three additional stars (OGLE-SMC-CEP-1413, 
-1696, and -1812) for which we do not have radial velocity data. 
All the data is given in Tab.\ref{tab.Kdata}.

The observations were performed between Oct.~10, 2012 and Sep.~23, 2017.
26 selected Cepheids were grouped in 16 fields, containing from
one to four Cepheids in each field.  During this period each field was
observed between 17 and 33 times.  If the conditions were photometric, a set of
different UKIRT \citep{Hawarden01} standard stars spanning a wide
range of colors was observed. It allowed us to calibrate our measurements
to the standard system for between 9 and 19 nights for each star
depending on the field.

In order to account for frequent sky level variations, the observations
were performed with a jittering technique.  The number of jittering
positions, integration time and consecutive observations were chosen
each night for each field separately to provide the best quality of
images. The integration time varied from 3s to 8s in the $K$-band, with
one to four consecutive observations, and 21-25 jittering positions.

The reduction process of all images followed the approach
described in \cite{Pietrzynski02G}.  After basic calibration
routines (dark correction, bad pixel correction), sky subtraction
was performed with a two-step process implying masking stars with the
XDIMSUM IRAF package.  Subsequently the single images were flat-fielded
and stacked into the final images.

After the reduction process we performed PSF photometry using DAOPHOT
\citep{Stetson87} and ALLSTAR \citep{Stetson94} routines, 
following the procedure described in 
\cite{Pietrzynski02GU}.  The list of star positions was created
for each file, based on the best quality images in our sample. This
star position list was consequently used to obtain photometry for each
field for all nights. In all images around 30 candidate stars for the
PSF model were selected, and the PSF model was obtained in an iterative
way with subtracting neighbouring stars.  The aperture corrections were
obtained with aperture photometry of 30 previously selected isolated
stars, with subtracted neighbouring objects.

To obtain the light curve, the coordinates of stars on separate
images were transformed and crossmatched using
DAOMATCH and DAOMASTER \citep{Stetson94}.

Cepheid differential brightness was calculated by comparison with
the selected sample of 26 stars in each field.  The random photometric 
uncertainty on the zero points, $\sigma(ZP)$, is reported in 
Tab.\ref{tab.Kbandfields} for each field.

For about half of the light curve points at different phases, the
calibrated brightness on the UKIRT standard system was available. We used
this to shift the light curve to obtain the light curve on the UKIRT standard
system.  The Cepheid calibrated brightness dispersion varies for each
object from 0.01 and 0.02 mag. The corresponding calibration accuracy of
the light curve for each Cepheid is below 0.008 mag.

The SOFI data are originally time stamped with the Modified Julian Date 
(MJD) which we have here converted to Julian Date (JD) by 
adding 2400000.5~days. 

Our data are on the UKIRT system \citep{Hawarden01} but for the IRSB analysis 
we convert the data to the SAAO system \citep{Carter90} to be consistent
with the $(V-K)$ surface-brightness relation used in the analysis.
We do this using the transformations given by \cite{Carpenter01}.

\begin{table}
\caption{\label{tab.Kdata}The $K$-band data in the UKIRT system as
observed with SOFI. The photometry is given together with the
Julian Date (JD), and the estimated measurement error. The full table 
is available in the electronic version of the paper and from the CDS.}
\begin{tabular}{c c c c}
\hline\hline
Identifier & JD & $K$ & $\sigma$ \\
\hline
OGLE-SMC-CEP-0320 & 2456211.57648 & 12.505 & 0.002\\
OGLE-SMC-CEP-0320 & 2456211.78426 & 12.514 & 0.004\\
OGLE-SMC-CEP-0320 & 2456212.55166 & 12.509 & 0.004\\
OGLE-SMC-CEP-0320 & 2456213.67309 & 12.507 & 0.002\\
OGLE-SMC-CEP-0320 & 2456214.78703 & 12.517 & 0.002\\
OGLE-SMC-CEP-0320 & 2456233.51368 & 12.532 & 0.003\\
OGLE-SMC-CEP-0320 & 2456234.53185 & 12.551 & 0.003\\
OGLE-SMC-CEP-0320 & 2456235.51983 & 12.577 & 0.004\\
OGLE-SMC-CEP-0320 & 2456247.67128 & 12.515 & 0.004\\
OGLE-SMC-CEP-0320 & 2456529.76325 & 12.536 & 0.002\\
...\\
\hline
 & days & mag & mag \\
\hline
\end{tabular}
\end{table}

%({\em check what is the
%final source, the web or from Maria-Rosa, Marek happens to have slightly
%different and worse data}) Actually my data agree with the web based
% data as far as I can tell (20180212).

We have supplemented these $K$-band data with the data from the VMC survey
\citep{Cioni11,Ripepi16} wherever available, except for a few bright
stars which show excessive scatter (OGLE-SMC-CEP-1797, -1977, -3311,
-3927), likely caused by saturation issues for the VMC data.  In this
way all of the stars have well sampled light curves in the $K$-band with
between 20 and 40 data points.  The VMC data is on the VISTA system so
we convert it first to the 2MASS system as described in \cite{Ripepi16}
and then to the SAAO system as described in \cite{Carpenter01}. We then
applied small zero point shifts to the VMC magnitudes for each star
to ensure the best agreement between the two data sets.  One star
(OGLE-SMC-CEP-1365) showed a significant offset of $+0.18$~mag.
For the remaining stars the mean offset is only $-0.01$~mag with
a standard deviation of 0.01~mag which gives us confidence that the
photometric zero point is very well established.  Our $K$-band 
photometric data is tabulated in Tab.\ref{tab.Kdata}
and light curves for stars can be found in the figures
\ref{fig.cep0320-data}-\ref{fig.cep4444-data} in appendix
\ref{app.lightcurves}.  Where appropriate the VMC data with the proper
zero point offsets applied have been overplotted in those figures.

The VMC survey \citep{Ripepi16} also provides mean $J$-band magnitudes 
based on a few ($\approx 6$) phase points which have been fitted with 
a template light curve. These mean magnitudes have been adopted here 
for deriving the PL-relation in the $J$-band as well as for the 
Wesenheits index in $(J-K)$. \cite{Ripepi16} estimate that for 93\% of
the stars the error on the thus derived $J$-band mean magnitude is below
0.02~mag.

\subsection{Optical light curves}
$V$ and $I$-band light curves were obtained from the OGLE-III and
OGLE-IV surveys \citep{Udalski08, Soszynski08, Udalski15, Soszynski15}. 
We adopted in all cases the photometric zero point defined by the
OGLE-III survey and shifted the OGLE-IV data onto the same system.
Plots of the $V$-band light curves can be found in appendix
\ref{app.lightcurves}.

\subsection{The Harvard variables from earlier studies}

\begin{table}
\caption{\label{tab.crossid}The cross-identification of the Harvard
variables from earlier studies and the OGLE data base.}
\begin{tabular}{c c}
\hline\hline
OGLE identifier & HV identifier \\
\hline
OGLE-SMC-CEP-0431 & HV822 \\
OGLE-SMC-CEP-2470 & HV837 \\
OGLE-SMC-CEP-0152 & HV1328 \\
OGLE-SMC-CEP-0230 & HV1333 \\
OGLE-SMC-CEP-0246 & HV1335 \\
OGLE-SMC-CEP-0368 & HV1345 \\
\hline
\end{tabular}
\end{table}

For the five stars which have been studied earlier in \cite{Storm04}
we have used the data from \cite{Storm04data} and \cite{Welch87},
and for HV837 we have used the data from \cite{Udalski08} ($V$-band)
and \cite{LaneyStobie86} ($K$-band). The cross-identification with the
OGLE catalogue is given in Tab.\ref{tab.crossid}.  For all these stars
the data have been supplemented with the VMC \citep{Ripepi16} and OGLE
\citep{Udalski15, Soszynski15} data.  The star HV1345 showed a phase
shift corresponding to 0.3~days between the old data and the newer data,
we have shifted the the old data accordingly. For the $V$-band data we
have kept the old photometric zero points as they were in good agreement
for three of the stars, but for the two stars HV1333 and HV1345 we shift
the OGLE-IV data by $-0.04$ and $-0.05$mag respectively. Where necessary
we have shifted the old $K$-band data to the VMC system appropriately
shifted to the SAAO system as described in the previous section. In
the case of HV~1345 the shift was $+0.11$~mag while for the remaining
stars the shifts were on average $-0.01$~mag with a standard deviation
of 0.03~mag.  Finally we have shifted these $K$-band magnitudes by the
mean VMC offset of $-0.01$~mag to be consistent with the adopted SOFI
photometric zero point.

\subsection{Reddening}
\label{sec.reddening}
For the reddening law we proceed exactly as in \cite{Storm11a} but
note that assuming a universal reddening law for the three galaxies
is not strictly correct \citep[e.g.][]{Alonso-Garcia17}. 
We adopt the reddening law by \cite{Cardelli89}
and use a ratio of total to selective absorption of $R_V = 3.23$,
$R_I=1.96$, $R_J=0.292R_V$, and $R_K=0.119R_V$ respectively following
the discussion in \cite{Fouque07}.  The reddening of our stars is based
on the photometric $E(V-I)$ reddening maps derived by \cite{Haschke11}
based on OGLE photometry \citep{Udalski08,Soszynski08} of red clump
stars. 
These reddenings are not based on intrinsic Cepheid colors so
they are independent of the metallicities of the Cepheids which is
relevant in the present study as we are looking for metallicity 
effects on the PL relations.
We determine
$E(B-V)$ from $E(B-V)=E(V-I)/(R_V - R_I)=E(V-I)/1.27$. To verify that we
are on the same reddening system as adopted in our previous work, we have
compared the reddening values for the 26 LMC Cepheids in \cite{Storm11b}
which are within the area of the Haschke et al. reddening map. The two
reddening systems show excellent agreement with an average difference
of $0.006\pm0.007$mag with a standard deviation of 0.035mag.
\cite{Inno16} using multi-band photometry ranging from the optical
($VI$) over the near-IR ($JHK_s$) to the mid-IR Wise $w1$ band for a
large sample of LMC Cepheids have determined individual reddenings for
these Cepheids based on the multi-band Period-Luminosity relations. They
also find that their values are in reasonable agreement with the maps from
\cite{Haschke11} even if they are derived in an entirely independent way.
\cite{Inno16} report an even better agreement with the reddenings for
eclipsing binary stars as reported by \cite{Pietrzynski13} based on the
work of \cite{Graczyk12}.  They used the equivalent width of the NaI D1
line in spectra of eclipsing binary stars to determine the reddening
directly, as well as employing a calibration of the spectroscopically
determined effective temperatures with the photometric $(V-I)$ and $(V-K)$
colors to infer the reddening.  This work has been further developed by
\cite{Graczyk14, Graczyk18} and we have adopted the reddening values from
latter work and determined the unreddened $(V-I)_0$ color for red
clump stars around these eclipsing binaries.  This color could then be
used to determined the reddening for red clump stars in the immediate
neighbourhood of our Cepheids. We found a shift with respect to the
\cite{Haschke11} values of $+0.05\pm0.01$mag and $+0.04\pm0.01$mag
respectively for the SMC and LMC samples.

Similarly recent work by \cite{Turner16} employing space reddenings 
suggests that the reddenings adopted for the galactic sample by 
\cite{Storm11b}, which we use here, should be transformed by the 
linear relation
\begin{equation}
E(B-V)_{\mbox{\scriptsize Turner}} = 0.020_{\pm0.006} +
1.067_{\pm0.019}E(B-V)_{\mbox{\scriptsize Storm11}}
\end{equation}
The galactic sample has a mean reddening of about $E(B-V)=0.5$mag which
would lead to an overall increase in the reddening of about $+0.05$mag
so very similar to the shifts suggested for the Magellanic Cloud samples.
It thus seems as if the reddenings for all three samples should be shifted
by about the same amount and consequently the relative luminosities, which
is our main concern in the present work, would not change by much. We
thus prefer to continue the use of the original reddening scales adopted
in \cite{Storm11b} but include the uncertainty on the reddening estimate
in the final error estimate. Considering the systematic offsets between
recent works we estimate the uncertainty to be
$\sigma_{\mbox{\scriptsize sys}}(E(B-V)) = 0.05$~mag.

The adopted reddening values from \cite{Haschke11} are tabulated in
Tab.\ref{tab.absmag}. We can see that the reddenings are in general very
small ($\approx 0.05$mag) so the exact choice of reddening law is not
critical for the SMC stars themselves.

\subsection{The radial velocity data}
\label{sec.rvdata}
During five years between October 2012 and October 2017 we have
obtained a total of 714 radial velocity observations of our sample of
26 SMC Cepheids, using three different high-resolution spectrographs:
HARPS \citep{Mayor03} at the 3.6-m telescope at ESO-La Silla; MIKE
\citep{Bernstein03} attached to the 6.5-m Clay telescope at Las Campanas
Observatory; and UVES \citep{Dekker00} mounted at the ESO-VLT at the
Paranal site of the European Southern Observatory. Great care was taken
to schedule the observations in a way so as to assure an optimum phase
coverage of the radial velocity curves of the Cepheids with the help
of the observation planning software written by one of us (BP).  65\%
of the data were obtained with HARPS, 29\% with MIKE, and 6\% with UVES.
We adjusted the integration times in a way as to achieve S/N ratios in the
range 3-8 for the Cepheids which is high enough to measure very accurate
radial velocities, according to our previous extensive experience with
the spectrographs used in this study. For the shortest-period and faintest
Cepheids we set a limit of 1800 seconds in the integration times, to keep
them below 1\% of their periods.  The HARPS spectra were reduced using
the on-site pipeline; MIKE data were reduced with software developed by
Dan Kelson \citep{Kelson03}, and the UVES spectra were reduced with a
publicly available ESO pipeline \citep{Freudling13}.

The radial velocities were measured with the RaveSpan software
\citep{Pilecki17}.
This code uses the Broadening Function technique originally introduced
by \cite{Rucinski92, Rucinski99} and a set of synthetic spectra from
\cite{Coelho05} as radial velocity reference templates. For consistency,
all spectra were processed in a similar manner, being properly
continuum-normalized and analyzed in the same wavelength range within
4125-6800{\AA} which contains numerous metallic lines. The individual
velocities were typically accurate to 250~m/s.

\begin{table*}
\caption{\label{tab.rvdata}The radial velocity measurements for the
stars are tabulated together with the HJD of the measurement, the
estimated error and the instrument used. The full table is available
in the electronic version of the paper and from the CDS.}
\begin{tabular}{c c c c r}
\hline\hline
Identifier & HJD & Radial velocity & $\sigma$ & Instrument \\
\hline
OGLE-SMC-CEP-0320 & 2456211.81673 & 160.26 & 0.03 & HARPS \\
OGLE-SMC-CEP-0320 & 2456213.72001 & 164.27 & 0.06 & HARPS \\
OGLE-SMC-CEP-0320 & 2456214.79017 & 166.60 & 0.07 & HARPS \\
OGLE-SMC-CEP-0320 & 2456240.53387 & 164.64 & 0.20 & HARPS \\
OGLE-SMC-CEP-0320 & 2456241.52979 & 161.74 & 0.03 & HARPS \\
OGLE-SMC-CEP-0320 & 2456529.84546 & 160.37 & 0.04 & HARPS \\
OGLE-SMC-CEP-0320 & 2456579.62039 & 170.02 & 0.05 & HARPS \\
OGLE-SMC-CEP-0320 & 2456605.55640 & 159.00 & 0.03 & HARPS \\
OGLE-SMC-CEP-0320 & 2456879.84336 & 169.07 & 0.03 & HARPS \\
OGLE-SMC-CEP-0320 & 2456908.72710 & 159.38 & 0.04 & HARPS \\
...\\
\hline
 & days & km/s & km/s & \\
\hline
\end{tabular}
\end{table*}

In the Figures \ref{fig.cep0320-data}-\ref{fig.cep4444-data} in appendix
\ref{app.lightcurves} we show the radial velocity curves for Cepheids.
The full set of individual radial velocity data 
is given in Tab.\ref{tab.rvdata}.

A significant number of the stars (seven) turned out to show velocity
variations indicative of orbital motion.  The amplitudes are too large to
be caused by amplitude variations as described by e.g.  \cite{Anderson16}
and the smooth variation of the long term trends also suggest orbital
motion.  As we do not have enough data to determine a proper binary
orbit solution we have instead attempted to simply shift the data from
different epochs to take out the putative orbital motion. We appeal
to continuity arguments and try to fix the offset for a given epoch
by requiring data points at the same phase but for different epochs
give approximately the same radial velocity. We thus assume that the
variation of the orbital motion is small for a certain period of time,
typically hundreds of days, and thus that the orbital period is much
longer than the pulsational period. In appendix \ref{app.orbitalrv}
we present the detailed analysis for the individual stars. The stars in
question are OGLE-SMC-CEP-1680, -1686, -1693, -1729, -1977, -2905 and
HV837. In the case of OGLE-SMC-CEP-1686 the radial velocity amplitude
is quite low and it was difficult to find consistent velocity offsets
so we prefer to leave the data unchanged for this star. In the case of
OGLE-SMC-CEP-1693 the offsets seems to be large but we did not manage
to find sensible offsets and this star had to be disregarded in the
further analysis.  We nevertheless include the observed radial velocities
in Tab.\ref{tab.rvdata}.

For the stars which were analyzed previously by \cite{Storm04} we have
used the radial velocities from \cite{Storm04data} and for HV837,
analyzed by \cite{Groenewegen13}, we have used the radial velocity
measurements from \cite{Imbert89}.

\section{Analysis}
\label{sec.analysis}

\begin{figure}
\centering
\includegraphics[width=9cm]{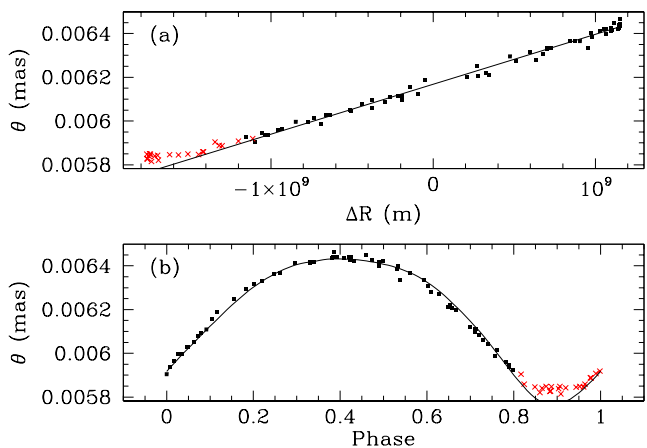}
\caption{An example of the actual fit of the angular diameters from
photometry and the radius variation from the radial velocity curve (panel a) 
for the 5.6~day period star OGLE-SMC-CEP-1765. 
In panel (b) the corresponding match of the angular 
diameters curve from the photometry (points) and from the radial velocity 
curve is plotted. The points marked with red crosses have not been
considered in the fit.
\label{fig.cep1765_fit}}
\end{figure}

\begin{figure}
\centering
\includegraphics[width=9cm]{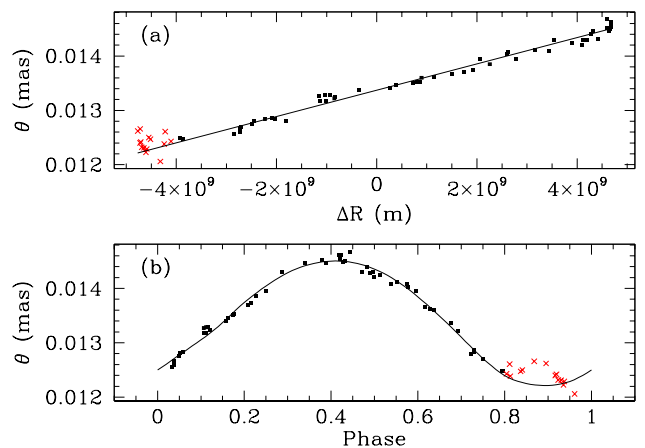}
\caption{An example of the actual fit of the angular diameters from
photometry and the radius variation from the radial velocity curve (panel a) 
for the 16~day period star OGLE-SMC-CEP-1385. 
In panel (b) the corresponding match of the angular 
diameters curve from the photometry (points) and from the radial velocity 
curve is plotted. The points marked with red crosses have not been
considered in the fit.
\label{fig.cep1385_fit}}
\end{figure}

We apply the near-IR surface-brightness method (IRSB) to the data
following exactly the same procedures and calibrations as adopted and
described by \cite{Storm11a}. In this way we ensure that we can perform 
a purely differential comparison with the results for the Milky Way and 
LMC samples from that study. 

We note that \cite{Merand15} have recently
developed a new implementation of the method in a code named SPIPS, which can 
utilize more observational data (more photometric bands, interferometric
data) and has a more complex data processing. As we do not have
additional data for the SMC Cepheids and want to make a purely
differential comparison with the LMC and Milky Way samples we proceed
with the IRSB method as calibrated in \cite{Storm11a}.

The IRSB method is a Baade-Wesselink type of technique which utilizes the
stellar radial pulsation to determine the distance and radius of the
star. This is achieved by geometrically matching the angular diameter
variation of the star with the absolute radius variation:

\begin{equation}
\label{eq.theta}
\theta(\phi) = 2R(\phi) / d  = 2 (R_0 + \Delta R(\phi)) / d
\end{equation}
\noindent
where $\theta$ is the angular diameter, $\phi$ is the pulsation phase, 
$d$ is the distance and $R$ is the stellar radius, $R_0$ the average
radius, and $\Delta R(\phi)$ the radius difference with respect to the
average radius.

The angular diameter is determined from the surface-brightness, $F_V$, through
the relation:

\begin{equation}
\label{eq.Fv}
F_V(\phi) = 4.2207 - 0.1 V_0(\phi) - 0.5 \log \theta(\phi)
\end{equation}
\noindent
where $V_0$ is the dereddened $V$-band magnitude at a given phase. 

As shown first by \cite{Welch94} $F_V$ is a nearly linear function of the
$(V-K)_0$ color index.  This is the near-IR version of the Barnes-Evans
method \cite{BarnesEvans76}.  \cite{Welch94} also pointed out the
significantly reduced scatter in the relation compared to purely optical
color indices while maintaining the low sensitivity to reddening errors.
In fact an error of 0.06mag in $E(B-V)$ causes only an error of 0.03mag
in the derived distance.
The calibration has later been refined by \cite{FG97},
\cite{Kervella04irsb} and others based on more modern interferometrically
determined stellar radii and in the case of \cite{Kervella04irsb} also
using interferometrically determined angular diameters of Cepheids.  
Several groups \citep[e.g.][]{Boyajian13,Challouf14,Graczyk17}
have presented much expanded empirical calibrations for
non-variable stars which potentially will improve the method. In the 
present work we need to stay consistent with \cite{Storm11a, Storm11b}
and we thus adopt the calibration of the surface-brightness 
as a function of the $(V-K)_0$ color index from \cite{Kervella04irsb}:
\begin{equation}
\label{eq.FvVK}
F_V = -0.1336 (V-K)_0 + 3.9530
\end{equation}

The $(V-K)_0$ surface-brightness relation has been shown to be practically
metallicity independent \citep{Storm04, Thompson01}. \cite{Thompson01}
find only a 1\% effect going from \FeH=0.0 to $-2.0$.

The other observable in Eq.\ref{eq.theta} is the radial velocity,
$V_r(\phi)$, at a given phase. This can be used to determine the radius
variation by integrating the pulsational velocity:
\begin{equation}
\label{eq.dR}
\Delta R(\phi)  = \int -p[V_r(\phi)-V_\gamma]d\phi
\end{equation}

To compute the pulsational velocity it is necessary to subtract the
systemic velocity, $V_\gamma$, and multiply by the so called projection
factor, $p$, which takes into account that the observed radial velocity
refers to the integrated light from the observed hemisphere of the star.
$p$ is mostly a geometric effect but is also affected by limb darkening
and velocity gradients in the dynamical atmosphere of the pulsating
star, see e.g. \cite{Nardetto17} and reference therein. 
\cite{Storm11a} calibrated the $p$-factor applicable for the IRSB method 
and we adopt here the exact same relation:
\begin{equation}
\label{eq.pfactor}
p = 1.55 - 0.186 \log (P)
\end{equation}

Other recent empirical efforts to determine the projection factor for
Cepheids, and its dependence on pulsation period, agree with our
findings, (e.g. \cite{Pilecki18}), have produced a milder
dependence (e.g.  \cite{Gallenne17}), or even values consistent with a
zero period dependence of the $p$-factor \citep{Kervella17}. However,
these studies show a large scatter among the projection factors of
individual Cepheids, hinting at large systematic uncertainties on the
individual determinations \citep{Kervella17}, and/or a possible
intrinsic dispersion of the $p$ factors of Cepheids of similar periods.
As shown by the hydrodynamical Cepheid models by \cite{Nardetto11} the
$p$-factor shows no significant dependence on metallicity. This 
means that the differential metallicity effect which we determine here
is largely independent on the exact choice of the $p$-factor relation.

In Fig.\ref{fig.cep1765_fit}-\ref{fig.cep1385_fit} the fits for two
typical stars with different pulsation periods are shown. In the case
of the star OGLE-SMC-CEP-1385 (Fig.\ref{fig.cep1385_fit}) with a period
of about 16~days a typical bump in the photometric angular diameter
curve can be seen. As in the previous work we disregard the phase
interval from 0.8 to 1.0 when performing the fits to avoid this region
where shock waves are known to be present in the stellar atmospheres
thus possibly affecting the surface brightness-color relation.

\section{Metallicities}
\label{sec.metallicities}

As was the case for \cite{Storm11b} we do not have individual
metallicities for our LMC and SMC Cepheids. We proceed as in that 
analysis and we will adopt mean metallicities for each of the 
three samples of Milky Way, LMC and SMC Cepheids. In this way we treat
the three samples in the same way.

As we are looking for differential effects it is particularly important
that the values are on the same system. \cite{Romaniello08} made a
detailed study of Cepheids in all three galaxies. The Milky Way sample
consists of 32 stars, the LMC sample of 22 Cepheids and the SMC sample
of 14 Cepheids. The mean metallicities are $0.00\pm0.02$, $-0.34\pm0.03$,
$-0.75\pm0.02$ with dispersions of 0.12, 0.15, and 0.08~dex respectively.
For 25 of our Milky Way stars \cite{Romaniello08} reports metallicities
and they show an average value of $-0.01\pm0.02$~dex. 

For the Milky Way sample \cite{Groenewegen13} has compiled a
list of individual metallicities based on the measurements by
\cite{LuckLambert11, Luck11, FryCarney, Andrievsky03} and \cite{Romaniello08}.  
We found metallicities for 64 of our MW Cepheids in this
list and after applying the offsets between samples as determined in
that paper, we find an average value of $\FeH=+0.07\pm0.01$.
This value is slightly different from that found by Romaniello et al. 
but comparable
to the offsets between works found by \cite{Romaniello08} and
\cite{Groenewegen13}. If we look for the 25 stars also present in the 
study of \cite{Romaniello08} we find here a mean value of 
$+0.07\pm0.03$, in good agreement with the value from the full sample, 
so this smaller sample is still very representative of our full MW sample.

Recently \cite{Lemasle17} have measured metallicities for four SMC Cepheids as
well as for six Cepheids in the LMC young blue cluster NGC1866. These
values are in good agreement with the values from \cite{Romaniello08}
with a slight offset of $+0.03$~dex leading to mean abundances of
the combined samples of $-0.73\pm0.02$ and $-0.33\pm0.03$~dex respectively. 
\cite{Molinaro12} also measured metallicities for three Cepheids in
NGC1866 and found a mean value of $-0.40\pm0.04$ in good agreement 
with the value of $-0.36\pm0.03$ from \cite{Lemasle17}.
\cite{Lemasle13} has similarly determined abundances for Milky Way 
Cepheids and for 12 stars in common with our sample the average 
metallicity is $-0.07\pm0.04$~dex. The metallicity range is in this case
slightly shorter than was the case when we employ the \cite{Romaniello08}
metallicities.  \cite{Genovali14} has measured metallicities for ten 
of our stars and they on the other hand find a mean metallicity of 
$+0.05\pm0.03$~dex.  So these four sources for the metallicity for 
Milky Way Cepheids range between $-0.07$, and $+0.07$. 
We prefer to stay with the \cite{Romaniello08} value of $0.00$, as it is 
obtained in a self consistent way with the LMC and SMC metallicities.
We do add in quadrature a systematic error contribution on the
metallicities of $0.05$ to reflect a possible uncertainty in the 
metallicity scale. To summarize we adopt for the MW, LMC and SMC samples the
values $+0.00\pm0.05$, $-0.34\pm0.06$, and $-0.75\pm0.05$dex.

\section{Results}
\label{sec.results}

\begin{figure}
\centering
\includegraphics[width=9cm]{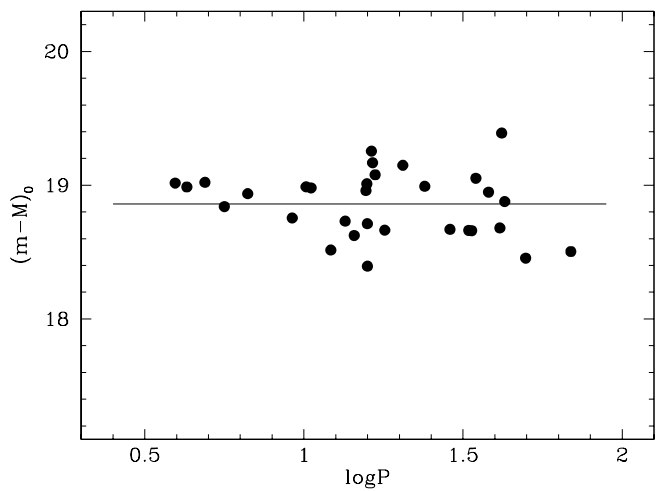}
\caption{The derived distance modulus as a function of $\log P$ for all the SMC
stars. The horizontal line indicates the average value.
\label{fig.logPmM}}
\end{figure}

\begin{table*}
\caption{\label{tab.absmag}For each star we give the identifier and
the logarithm of the period in days,
$\log (P)$. The resulting distances and their associated
formal fitting errors are given in col. (3) and (4). In col (5) and (6)
the distance modulus and formal error is given. Col. (7) to (12) lists
the resulting absolute magnitudes, col (13) gives the adopted reddening
based on \cite{Haschke11}. Col. (14) gives the phase shift adopted in
the IRSB fitting.}
\scriptsize
\begin{tabular}{c c c c c c c c c c c c c r}
\hline\hline
(1) & (2) & (3) & (4) & (5) & (6) & (7) &(8) & (9) & (10) & (11) & (12)
& (13) & (14) \\
ID & $\log (P)$ & $d$ & $\sigma_{\mbox{FIT,d}}$ &
$(m-M)_0$ & $\sigma_{\mbox{FIT,(m-M)}}$ &
\Mv & \Mi & \Mj & \Mk & \Wvi & \Wjk & $E(B-V)$ & $\Delta \phi$ \\
\hline
   OGLE-SMC-CEP-1761 &   0.595415 &  63.6 &  1.2 & 19.02 & 0.04 & $-2.79$ & $-3.49$ & $-4.03$ & $-4.36$ & $-4.56$ & $-4.58$ & $ 0.024$ & $ 0.005$ \\
   OGLE-SMC-CEP-1729 &   0.631889 &  62.7 &  0.6 & 18.99 & 0.02 & $-2.71$ & $-3.46$ & $-4.10$ & $-4.37$ & $-4.61$ & $-4.55$ & $ 0.024$ & $ 0.025$ \\
   OGLE-SMC-CEP-1680 &   0.689176 &  63.7 &  1.2 & 19.02 & 0.04 & $-3.14$ & $-3.83$ & $-4.33$ & $-4.71$ & $-4.91$ & $-4.98$ & $ 0.031$ & $ 0.000$ \\
   OGLE-SMC-CEP-1765 &   0.750044 &  58.6 &  0.6 & 18.84 & 0.02 & $-2.87$ & $-3.62$ & $-4.15$ & $-4.55$ & $-4.78$ & $-4.83$ & $ 0.024$ & $ 0.010$ \\
   OGLE-SMC-CEP-1717 &   0.823504 &  61.3 &  0.8 & 18.94 & 0.03 & $-3.13$ & $-3.92$ & $-4.48$ & $-4.90$ & $-5.14$ & $-5.19$ & $ 0.024$ & $ 0.005$ \\
   OGLE-SMC-CEP-1410 &   0.963223 &  56.4 &  0.9 & 18.75 & 0.04 & $-3.71$ & $-4.48$ & $-4.97$ & $-5.44$ & $-5.67$ & $-5.76$ & $ 0.055$ & $ 0.005$ \\
   OGLE-SMC-CEP-1712 &   1.006814 &  62.7 &  0.5 & 18.99 & 0.02 & $-3.51$ & $-4.43$ & $-5.07$ & $-5.53$ & $-5.85$ & $-5.84$ & $ 0.024$ & $ 0.020$ \\
   OGLE-SMC-CEP-0524 &   1.022321 &  62.5 &  0.8 & 18.98 & 0.03 & $-3.74$ & $-4.54$ & $-5.08$ & $-5.53$ & $-5.78$ & $-5.83$ & $ 0.047$ & $ 0.020$ \\
   OGLE-SMC-CEP-2533 &   1.084483 &  50.5 &  1.0 & 18.52 & 0.04 & $-3.97$ & $-4.66$ & $-5.16$ & $-5.51$ & $-5.73$ & $-5.75$ & $ 0.024$ & $ 0.025$ \\
    HV1345 &   1.129670 &  55.7 &  1.4 & 18.73 & 0.05 & $-4.08$ & $-4.80$ & $-5.32$ & $-5.79$ & $-5.90$ & $-6.10$ & $ 0.031$ & $-0.015$ \\
    HV1335 &   1.157800 &  53.1 &  0.8 & 18.62 & 0.03 & $-3.94$ & $-4.71$ & $-5.22$ & $-5.67$ & $-5.89$ & $-5.98$ & $ 0.024$ & $-0.020$ \\
   OGLE-SMC-CEP-1365 &   1.194727 &  61.9 &  1.8 & 18.96 & 0.06 & $-4.56$ & $-5.34$ & $-6.05$ & $-6.33$ & $-6.55$ & $-6.51$ & $ 0.055$ & $ 0.035$ \\
   OGLE-SMC-CEP-0518 &   1.197888 &  63.4 &  0.8 & 19.01 & 0.03 & $-4.00$ & $-4.95$ & $-5.60$ & $-6.08$ & $-6.42$ & $-6.41$ & $ 0.055$ & $ 0.010$ \\
   OGLE-SMC-CEP-1385 &   1.199259 &  55.3 &  1.0 & 18.71 & 0.04 & $-4.12$ & $-4.96$ & $-5.57$ & $-6.03$ & $-6.25$ & $-6.34$ & $ 0.055$ & $ 0.020$ \\
    HV1328 &   1.199692 &  47.7 &  1.1 & 18.39 & 0.05 & $-4.32$ & $-5.02$ & $-5.45$ & $-5.83$ & $-6.11$ & $-6.09$ & $ 0.016$ & $ 0.015$ \\
    HV1333 &   1.212084 &  71.0 &  1.4 & 19.25 & 0.04 & $-4.65$ & $-5.46$ & $-6.02$ & $-6.50$ & $-6.71$ & $-6.82$ & $ 0.024$ & $-0.020$ \\
   OGLE-SMC-CEP-1723 &   1.215794 &  68.2 &  0.8 & 19.17 & 0.03 & $-4.25$ & $-5.22$ & $-5.89$ & $-6.36$ & $-6.70$ & $-6.68$ & $ 0.024$ & $ 0.010$ \\
     HV822 &   1.223807 &  65.4 &  1.9 & 19.08 & 0.06 & $-4.69$ & $-5.50$ & $-6.01$ & $-6.48$ & $-6.76$ & $-6.80$ & $ 0.039$ & $-0.020$ \\
   OGLE-SMC-CEP-0320 &   1.253893 &  54.1 &  0.4 & 18.66 & 0.02 & $-4.59$ & $-5.28$ & $-5.75$ & $-6.10$ & $-6.34$ & $-6.35$ & $ 0.031$ & $ 0.015$ \\
   OGLE-SMC-CEP-1750 &   1.310784 &  67.6 &  1.2 & 19.15 & 0.04 & $-4.62$ & $-5.56$ & $-6.18$ & $-6.71$ & $-7.01$ & $-7.08$ & $ 0.024$ & $-0.005$ \\
   OGLE-SMC-CEP-0958 &   1.379643 &  62.9 &  0.5 & 18.99 & 0.02 & $-4.81$ & $-5.67$ & $-6.29$ & $-6.82$ & $-7.01$ & $-7.17$ & $ 0.031$ & $ 0.005$ \\
   OGLE-SMC-CEP-1403 &   1.459128 &  54.2 &  0.4 & 18.67 & 0.02 & $-4.49$ & $-5.44$ & $-6.10$ & $-6.66$ & $-6.91$ & $-7.05$ & $ 0.055$ & $ 0.005$ \\
   OGLE-SMC-CEP-3927 &   1.517905 &  54.0 &  0.6 & 18.66 & 0.03 & $-5.08$ & $-5.85$ & $-6.50$ & $-7.00$ & $-7.03$ & $-7.34$ & $ 0.055$ & $ 0.020$ \\
   OGLE-SMC-CEP-4017 &   1.527325 &  54.0 &  1.5 & 18.66 & 0.06 & $-5.04$ & $-5.96$ & $-6.41$ & $-7.02$ & $-7.38$ & $-7.45$ & $ 0.039$ & $ 0.040$ \\
   OGLE-SMC-CEP-1686 &   1.540177 &  64.6 &  1.2 & 19.05 & 0.04 & $-5.70$ & $-6.53$ & $-7.02$ & $-7.53$ & $-7.80$ & $-7.88$ & $ 0.063$ & $ 0.030$ \\
   OGLE-SMC-CEP-2905 &   1.580231 &  61.6 &  0.8 & 18.95 & 0.03 & $-5.21$ & $-6.22$ & $-6.75$ & $-7.37$ & $-7.77$ & $-7.80$ & $ 0.079$ & $ 0.035$ \\
   OGLE-SMC-CEP-1797 &   1.615447 &  54.4 &  0.7 & 18.68 & 0.03 & $-4.97$ & $-6.00$ & $-6.55$ & $-7.27$ & $-7.58$ & $-7.76$ & $ 0.024$ & $ 0.025$ \\
   OGLE-SMC-CEP-4444 &   1.621557 &  75.5 &  1.4 & 19.39 & 0.04 & $-6.52$ & $-7.33$ & $-7.90$ & $-8.30$ & $-8.57$ & $-8.57$ & $ 0.031$ & $ 0.095$ \\
     HV837 &   1.630904 &  59.6 &  0.9 & 18.88 & 0.03 & $-5.72$ & $-6.64$ & $-7.11$ & $-7.71$ & $-8.07$ & $-8.12$ & $ 0.024$ & $ 0.035$ \\
   OGLE-SMC-CEP-3311 &   1.696679 &  49.1 &  0.8 & 18.45 & 0.04 & $-5.31$ & $-6.34$ & $-6.56$ & $-7.54$ & $-7.93$ & $-8.22$ & $ 0.024$ & $-0.045$ \\
   OGLE-SMC-CEP-1977 &   1.838831 &  50.2 &  1.0 & 18.50 & 0.04 & $-5.69$ & $-6.72$ & $-7.16$ & $-7.91$ & $-8.32$ & $-8.42$ & $ 0.031$ & $-0.010$ \\
\hline
 & P in days & kpc & kpc & mag & mag & mag & mag & mag & mag & mag & mag & mag & \\
\hline
\end{tabular}
\end{table*}

\begin{figure}
\centering
\includegraphics[width=9cm]{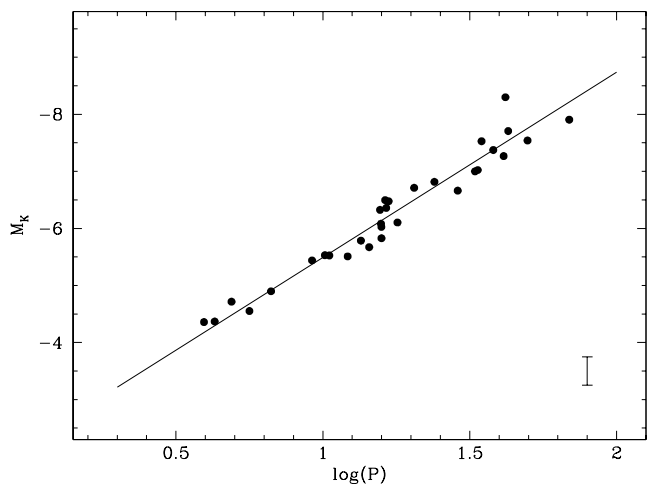}
\caption{The period-luminosity relation in the $K$-band for the SMC
Cepheids. The line represents the LMC slope as determined by
\cite{Macri15} shifted to match the SMC. 
A typical errorbar is shown in the lower right.
\label{fig.logPMk}}
\end{figure}

\begin{figure}
\centering
\includegraphics[width=9cm]{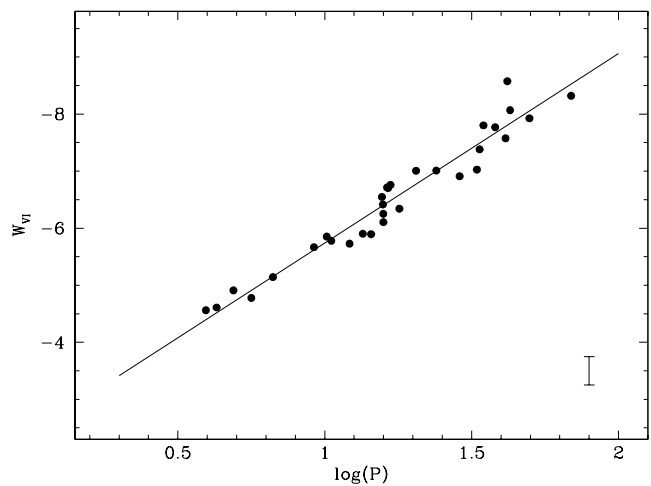}
\caption{The period-luminosity relation in the $V,(V-I)$ Wesenheits
index for the SMC Cepheids. The line represents the slope as determined 
by \cite{Storm11a} for their combined sample, shifted to match the SMC.
A typical errorbar is shown in the lower right.
\label{fig.logPWvi}}
\end{figure}

\begin{figure}
\centering
\includegraphics[width=9cm]{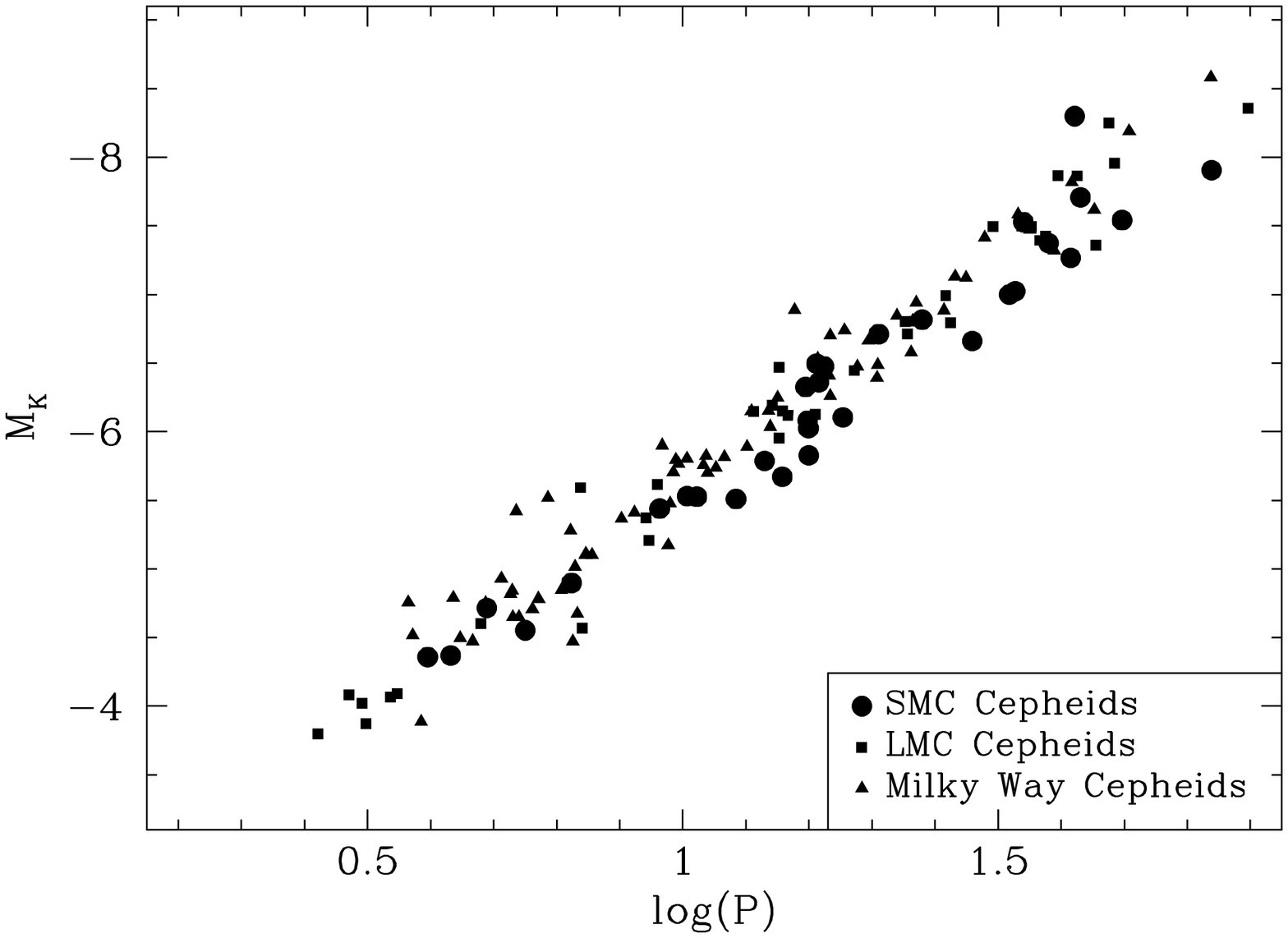}
\caption{The $K$-band Period-Lumninosity relation for the SMC stars
(black) overplotted on the Milky Way (blue triangles) and LMC (red
squares) samples.
\label{fig.logPMkall}}
\end{figure}

\begin{figure}
\centering
\includegraphics[width=9cm]{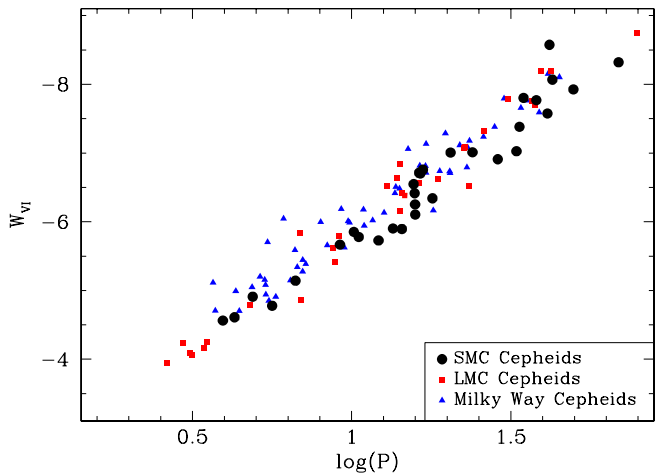}
\caption{The Wesenheit $V,(V-I)$ index Period-Luminosity relation for 
the SMC stars
(black) overplotted on the Milky Way (blue triangles) and LMC (red
squares) samples.
\label{fig.logPWviall}}
\end{figure}

\begin{figure*}
\centering
\includegraphics[width=16cm]{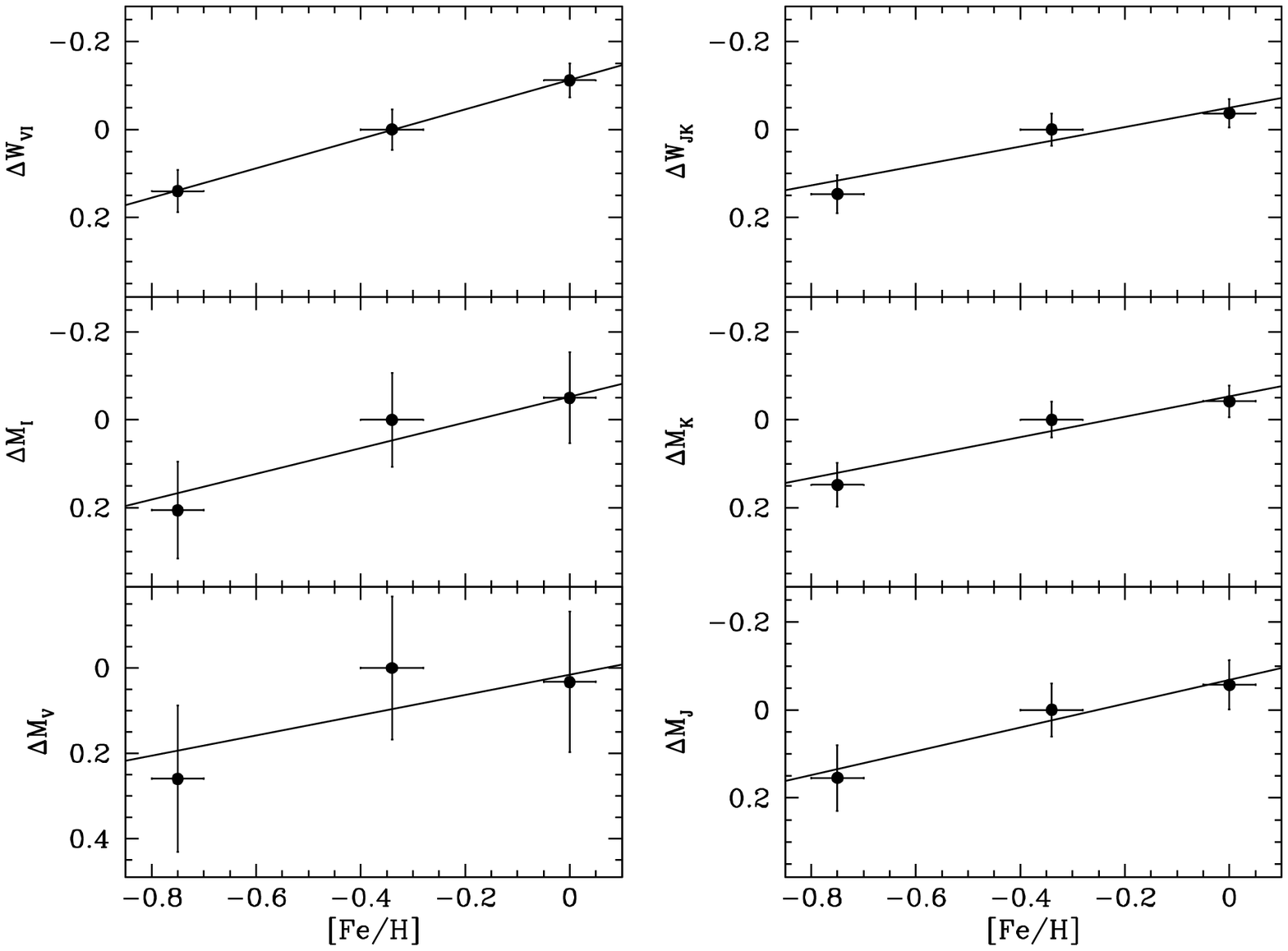}
\caption{The luminosity zero-point offsets in the different
passbands as a function of the metallicity of each of the
samples (SMC, LMC, and MW) with the associated error bars from
Tab.\ref{tab.PLrelallfix}.  The metallicities and their error bars
are taken from Sec.\ref{sec.metallicities} and the lines represent the
statistically most plausible model.  
\label{fig.FeHdMall}}
\end{figure*}

In Tab.\ref{tab.absmag} the derived distance moduli and associated
fitting errors are given for all the stars in the sample. We emphasize
that these are fitting errors only. \cite{Barnes05} compared these error
estimates with error estimates from a full bayesian analysis of the data
and found that, on average, these errors are underestimated by about a
factor of 3.4. In particular the fitting errors can sometimes be very
small as the data line up almost perfectly on a straight line. Such
small errors can significantly skew weighted fits so we prefer here to
use unweighted fits for our PL relations.  The table is sorted according
to pulsation period and it includes the resulting absolute magnitudes
and Wesenheit indices, adopted reddenings as well as the usually very
small phase shifts we needed to apply to obtain a best match between
angular diameter and linear displacement curves.

We have plotted the distance moduli of the stars versus $\log P$ in
Fig.\ref{fig.logPmM}. Linear regression gives a marginally significant
slope of $-0.15\pm0.13$ still compatible with no period dependence. 
We have computed the average of the distance
moduli to the Cepheids in the sample and find $(m-M)_0 = 18.86\pm0.04$~mag.
Discarding the seven possible binaries in our sample, the mean distance modulus
changes by just 0.01 mag which is not significant and supports our procedure
to correct the observed radial velocities of these stars for binary motion (see
Appendix \ref{app.orbitalrv}). We can also compute difference in
distance modulus between the LMC ($(m-M)_0=18.46\pm0.04$mag,
\cite{Storm11b}) and the SMC, and we find $\Delta (m-M)_0 =
0.40\pm0.06$mag.

On the basis of the data in Tab.\ref{tab.absmag} we have determined the
period-luminosity relation by linear regression for each band. These
relations in the form $M = \alpha (\log P -1.0) + \beta$ are tabulated in
Tab.\ref{tab.PLrel}. In Fig.\ref{fig.logPMk} and Fig.\ref{fig.logPWvi}
we plot the data in the $K$-band and the reddening insensitive 
Wesenheit $V,(V-I)$ index, $\Wvi = V - R_{Wvi}(V-I) - (m-M)_0$, 
as adopted in \cite{Storm11a}, where $R_{Wvi}=R_V/(R_V - R_I)=2.54$ with
the total-to-selective absorption ratios from Sec.\ref{sec.reddening}. 
We similarly compute the near-IR Wesenheits index 
$\Wjk = K - R_{Wjk}(J-K) - (m-M)_0$ where $R_{Wjk}=R_K/(R_J - R_K)=0.69$. 
The relations are very well defined and from Tab.\ref{tab.PLrel} we see
that the standard deviation around the best fit is about 0.24mag for the
reddening independent relations ($K$, \Wvi, \Wjk), in very good agreement
with the values found by \cite{Storm11a} for the LMC and MW samples.

As we intend to compare the slopes and zero points for the three
different samples, we have also determined the relations in the slightly
different form of $M = \alpha (\log P -1.18) + \beta$ where $\log P =
1.18$ is close to the mid-point of the period range under investigation.
In this way the zero point errors are minimized and they are the 
least correlated with errors in the derived slopes.
These relations can be found in Tab.\ref{tab.PLrelall} together with the
relations which we have redetermined for the Milky Way and LMC samples
from \cite{Storm11a}. From the table it is clear that the slopes of
the PL-relations in the three different galaxies are in excellent
agreement which justifies the assumption of a common,
metallity-independent slope of the PL relation is each band. 
In Fig.\ref{fig.logPMkall} and Fig.\ref{fig.logPWviall}
we have overplotted the data from the three samples in the $K$-band
and in the Wesenheit $V,(V-I)$ index. It can be seen that the slopes of
the different samples appear very similar but also that there are small
shifts of the zero points.

To determine the zero point offsets between the three samples we adopt a
slope for the given band and fit the three samples individually. We thus
get three different zero point values which we tabulate as $\beta$ in
Tab.\ref{tab.PLrelallfix}. We adopt the LMC zero point as the reference
and subtract it from the SMC and MW zero points respectively to derive
$\Delta M$. In the $V$, and $I$-bands, we have adopted the reference
slopes from the OGLE samples determined by \cite{Soszynski15} of
$-2.690\pm0.018$, and $-2.911\pm0.014$. For the Wesenheit $V,(V-I)$
index we have adopted $-3.32\pm0.08$ from \cite{Storm11a}. In the $J$ and
$K$ band we have adopted the values from \cite{Macri15} of
$-3.156\pm0.004$ and $-3.247\pm0.004$ respectively which also agrees
very well with the relations from \cite{Persson04}. For \Wjk we adopted
a value of $-3.36\pm0.1$ based on the Milky Way sample from 
Tab.\ref{tab.PLrelall}. 

Our adopted PL relation slope values do agree mostly well with
other modern determinations e.g. \cite{Subramanian15, Macri15,
Bhardwaj16, Inno16, Ripepi17}.
Small non-linearities of the LMC PL relations in different
optical bands at around 10~d have been reported by \cite{Tammann03}, 
\cite{Kanbur04}, \cite{Bhardwaj16} and references therein.
There might be a small non-linearity also in PL relations in
near-infrared bands \citep{Bhardwaj16}. 
These nonlinearities are so small however that
they do not affect our conclusions in any significant way. We
are also aware of the fact that the Magellanic Cloud PL relations in
the literature, like the ones based on the OGLE samples, are based
on Cepheid samples which contain a much larger number of short-period
Cepheids than Cepheids with periods longer than 10 d. However, cutting
out the short-period Cepheids in the PL relation solutions does not
change significantly the slopes. An important consideration is also that
Cepheid samples detected in distant galaxies (beyond several Mpc) which
are important in the context of the determination of the Hubble constant,
always consist of long-period Cepheids because the short-period variables
are too faint to be detected.

We plot in Fig.\ref{fig.FeHdMall} the zero point offsets from
Tab.\ref{tab.PLrelallfix} against the adopted metallicity values from
Sec.\ref{sec.metallicities}. To obtain realistic error estimates we
have used the Python package {\tt emcee} \citep{emcee13}, to perform
Markov chain Monte Carlo simulations based on a linear model between
metallicity and the magnitude offset including the estimated errors
in both parameters. Following the discussion in
Sec.\ref{sec.reddening} we have added in quadrature to the statistical
errors the contribution from the uncertainty in the
zero points of the reddening scales, $\sigma_{\mbox{\scriptsize
sys}}(M_\lambda)=R_\lambda \times \sigma_{\mbox{\scriptsize sys}}(E(B-V))$ 
where $R_\lambda$ is the ratio of total to
selective absorption as given in Sec.\ref{sec.reddening} and 
$\sigma_{\mbox{\scriptsize sys}}(E(B-V))=0.05$~mag.
The lines overplotted in the figures are the
resulting relations which are tabulated with their associated errors
in Tab.\ref{tab.FeHdM}. In \Wvi we find a linear relation with a slope
of $-0.34\pm0.06$mag/dex. In the $K$-band we also find a significant
variation from SMC to MW metallicity with a slope of $-0.23\pm0.06$mag/dex
in the sense that metal-poor stars are fainter than metal-rich stars
for a given pulsation period. From Fig.\ref{fig.FeHdMall} it can also
be seen that the relation might not be entirely linear in all bands but
might be steeper for lower metallicities.

\begin{table}
\caption{ \label{tab.PLrel}The PL-relation in the SMC for the various 
bands in the form $M = \alpha (\logP - 1.0) + \beta$.}
%\scriptsize
\begin{tabular}{r r r r r r}
\hline\hline
Band & $\alpha$ & $\sigma(\alpha)$ & $\beta$ & $\sigma(\beta)$ & std.dev. \\
\hline
   \Mk & $-3.179$ &  0.141 & $-5.509$ &  0.056 &  0.24 \\
   \Mj & $-2.856$ &  0.169 & $-5.098$ &  0.067 &  0.29 \\
  \Wjk & $-3.401$ &  0.133 & $-5.791$ &  0.053 &  0.23 \\
   \Mv & $-2.705$ &  0.177 & $-3.751$ &  0.070 &  0.30 \\
   \Mi & $-2.934$ &  0.156 & $-4.536$ &  0.062 &  0.27 \\
  \Wvi & $-3.287$ &  0.148 & $-5.746$ &  0.059 &  0.25 \\
\hline
 & & & mag & mag & mag \\
\hline
\end{tabular}
\end{table}

\begin{table}
\caption{ \label{tab.PLrelall}The PL-relations for the three different
samples and for each band the form $M = \alpha (\logP - 1.18) + \beta$.}
\scriptsize
\begin{tabular}{r r r r r r r}
\hline\hline
Band & Galaxy & $\alpha$ & $\sigma(\alpha)$ & $\beta$ & $\sigma(\beta)$ & std.dev. \\
\hline
\Mk & SMC & $-3.179$ &  0.141 & $-6.081$ &  0.046 &  0.24 \\
\Mk & LMC & $-3.282$ &  0.087 & $-6.225$ &  0.036 &  0.21 \\
\Mk & MW & $-3.258$ &  0.092 & $-6.268$ &  0.030 &  0.23 \\
\Mj & SMC & $-2.856$ &  0.169 & $-5.612$ &  0.055 &  0.29 \\
\Mj & LMC & $-3.220$ &  0.090 & $-5.750$ &  0.037 &  0.21 \\
\Mj & MW & $-3.114$ &  0.092 & $-5.802$ &  0.030 &  0.23 \\
\Wjk & SMC & $-3.401$ &  0.133 & $-6.404$ &  0.043 &  0.23 \\
\Wjk & LMC & $-3.324$ &  0.089 & $-6.552$ &  0.036 &  0.21 \\
\Wjk & MW & $-3.357$ &  0.097 & $-6.589$ &  0.032 &  0.24 \\
\Mv & SMC & $-2.705$ &  0.177 & $-4.238$ &  0.057 &  0.30 \\
\Mv & LMC & $-2.775$ &  0.111 & $-4.499$ &  0.045 &  0.26 \\
\Mv & MW & $-2.615$ &  0.100 & $-4.457$ &  0.033 &  0.25 \\
\Mi & SMC & $-2.934$ &  0.156 & $-5.064$ &  0.051 &  0.27 \\
\Mi & LMC & $-3.021$ &  0.101 & $-5.280$ &  0.041 &  0.21 \\
\Mi & MW & $-2.664$ &  0.098 & $-5.293$ &  0.031 &  0.21 \\
\Wvi & SMC & $-3.287$ &  0.148 & $-6.338$ &  0.048 &  0.25 \\
\Wvi & LMC & $-3.411$ &  0.112 & $-6.484$ &  0.046 &  0.24 \\
\Wvi & MW & $-3.084$ &  0.117 & $-6.562$ &  0.037 &  0.25 \\
\hline
 & & & & mag & mag & mag \\
\hline
\end{tabular}
\end{table}

\begin{table}
\caption{ \label{tab.PLrelallfix}The PL-relation with fixed slopes
for the various bands in the form $M = \alpha (\logP - 1.18) + \beta$.}
\scriptsize
\begin{tabular}{r r r r r r r r}
\hline\hline
Band & Galaxy & $\alpha$ & $\sigma(\alpha)$ & $\beta$ & $\sigma(\beta)$ & std.dev. &
$\Delta M$\\
\hline
\Mk & SMC & $-3.247$ &  0.141 & $-6.077$ &  0.046 &  0.24 & $ 0.148$\\
\Mk & LMC & $-3.247$ &  0.088 & $-6.225$ &  0.036 &  0.21 & $ 0.000$\\
\Mk & MW & $-3.247$ &  0.092 & $-6.267$ &  0.030 &  0.23 & $-0.042$\\
\Mj & SMC & $-3.156$ &  0.177 & $-5.595$ &  0.058 &  0.31 & $ 0.155$\\
\Mj & LMC & $-3.156$ &  0.091 & $-5.750$ &  0.037 &  0.22 & $ 0.000$\\
\Mj & MW & $-3.156$ &  0.092 & $-5.807$ &  0.030 &  0.23 & $-0.057$\\
\Wjk & SMC & $-3.360$ &  0.133 & $-6.406$ &  0.043 &  0.23 & $ 0.147$\\
\Wjk & LMC & $-3.360$ &  0.089 & $-6.552$ &  0.036 &  0.21 & $ 0.000$\\
\Wjk & MW & $-3.360$ &  0.097 & $-6.589$ &  0.032 &  0.24 & $-0.037$\\
\Mv & SMC & $-2.690$ &  0.177 & $-4.239$ &  0.057 &  0.30 & $ 0.260$\\
\Mv & LMC & $-2.690$ &  0.112 & $-4.498$ &  0.045 &  0.27 & $ 0.000$\\
\Mv & MW & $-2.690$ &  0.100 & $-4.465$ &  0.033 &  0.25 & $ 0.033$\\
\Mi & SMC & $-2.911$ &  0.156 & $-5.065$ &  0.051 &  0.27 & $ 0.206$\\
\Mi & LMC & $-2.911$ &  0.103 & $-5.270$ &  0.042 &  0.22 & $ 0.000$\\
\Mi & MW & $-2.911$ &  0.104 & $-5.320$ &  0.033 &  0.22 & $-0.050$\\
\Wvi & SMC & $-3.320$ &  0.148 & $-6.336$ &  0.048 &  0.26 & $ 0.140$\\
\Wvi & LMC & $-3.320$ &  0.114 & $-6.476$ &  0.046 &  0.24 & $ 0.000$\\
\Wvi & MW & $-3.320$ &  0.121 & $-6.588$ &  0.038 &  0.26 & $-0.112$\\
\hline
 & & & & mag & mag & mag & mag \\
\hline
\end{tabular}
\end{table}

\begin{table}
\caption{\label{tab.FeHdM}The $\Delta M$ versus \FeH relation for 
the various bands in the form $\Delta M = \gamma \FeH + \psi$. The
fitting errors $\sigma(\gamma)$ and $\sigma(\psi)$ based on Markov Chain
Monte Carlo analysis is given as well.} 
\begin{tabular}{r r r r r}
\hline\hline
Band & $\gamma$ & $\sigma(\gamma)$ & $\psi$ & $\sigma(\psi)$ \\
\hline
   \Mj & $-0.270$ &  0.108 & $-0.068$ &  0.044 \\
   \Mk & $-0.232$ &  0.064 & $-0.054$ &  0.024 \\
  \Wjk & $-0.221$ &  0.053 & $-0.049$ &  0.019 \\
   \Mv & $-0.238$ &  0.186 & $ 0.016$ &  0.111 \\
   \Mi & $-0.293$ &  0.150 & $-0.053$ &  0.076 \\
  \Wvi & $-0.335$ &  0.059 & $-0.113$ &  0.023 \\
\hline
 & mag/dex & mag/dex & mag & mag \\
\hline
\end{tabular}
\end{table}

\section{Discussion}
\label{sec.discussion}

%\subsection{Systematic effects} 
\cite{Storm11a} calibrated the $p$-factor relation for the IRSB method
to give distances independent of pulsation period and in agreement with
the \cite{Benedict07} HST parallaxes to nine galactic Cepheids. The LMC
distance modulus which results from the mean of the individual distances
to the sample of LMC Cepheids is $(m-M)_0 = 18.46\pm0.04$ (statistical
only) which is in excellent agreement with the very accurate modulus of
$(m-M)_0 = 18.493\pm0.047$ (statistical and systematic) determined by
\cite{Pietrzynski13} from late-type eclipsing binaries. The zero point
of the method thus seems to be very well established. We stress that
in the present work we are looking only for differential effects with
metallicity so any shift of the absolute zero point will have little,
if any, impact on the conclusions. 

As already discussed in Sec.\ref{sec.analysis} the effect of metallicity
on the method itself is also very small. Metallicity could potentially
affect the $p$-factor relation as well as the surface 
brightness-color relation. \cite{Nardetto11} showed that the $p$-factor is
largely independent of the metallicity and \cite{Thompson01} showed that
the effect on the surface brightness $(V-K)$ relation is very small
as well.  Similarly \cite{Storm04} showed that the method itself is
robust to metallicity variations.  Furthermore, any changes in either
the adopted $p$-factor relation or the surface brightness-color relation
would affect all three samples in equal measure and not significantly
change the result.

Apart from the method itself, we also have to consider systematic
differences between our samples, in particular that the reddening
and metallicities are on the same scale.  As already explained in
Sec.\ref{sec.analysis} the method is very robust to errors in the
reddening (an error of 0.05~mag leads to an error in the distance modulus
of 0.025~mag). 
Reddening errors however carry over directly in the luminosities so
reddening insensitive indices perform much better as distance
indicators. We have included the estimated systematic uncertainties in the
reddenings in the fits of the metallicity effect and it can be seen in
Tab.\ref{tab.FeHdM} that the reddening insensitive indices are indeed
very well constrained.

The metallicity scales as described in Sec.\ref{sec.metallicities} also
appear to be in very good agreement and we found the possible systematic
difference to be less than 0.03~dex which has been included in the
error propagation.

We are thus confident that the derived absolute magnitudes are all on
the same system and that the offsets in the zero points as determined in
Sec.\ref{sec.results} are real and significant. We also stress here that the
metallicity effects found in Sec.\ref{sec.results} do not in any way
rely on an ensemble distance to either the LMC nor the SMC but relies on
the individual distance estimates for each star. We are thus entirely
free of influence from any depth effects in the clouds. The main remaining
limitation is that we still rely on mean metallicity values for each sample.

The new data presented here allows us for the first time to determine the
PL relations directly for SMC Cepheids.  We now have SMC and LMC samples
of comparable size and spanning the full range of periods for classical
fundamental mode Cepheids.  The slopes are in excellent agreement with
the slopes for the LMC and MW samples and the large sample allows us to
constrain much better the metal poor magnitude zero point as was the
case in \cite{Storm11b}.  In fact the estimated uncertainties on the
metallicity effect is now pushed down to $0.06$~mag/dex, or almost half
of the value we obtained previously. The effect is now somewhat stronger
than reported there, and it is much more significant due to the reduced
uncertainty.  We note that the relations do not appear perfectly linear
and that there might be a stronger effect in the metal poor regime. The
data is still consistent with a negligible effect for metallicities in
the range from $-0.35$ to $0.0$. However the metal poor data point puts
a very strong constraint on the slope which is hardly possible with the
significantly smaller metal range covered by LMC and MW Cepheids.

Due to our significant sample size and the distribution of the
Cepheids across the face of the SMC we believe that we can determine
a reasonable distance to the SMC itself in spite of the fact that
the SMC exhibits significant depth effects (see e.g. \cite{Ripepi17},
\cite{Muraveva18}). In this way we can also determine the difference
in distance modulus between the two clouds. In Sec.\ref{sec.results}
we found a difference of $\Delta (m-M)_0 = 0.40$~mag.  This compares very
well with the value of 0.44 $\pm0.10$~mag determined by \cite{Cioni00}
from the Tip of the Red Giant Branch method in $IJK$ bands, the value
of $0.458\pm0.068$~mag determined by \cite{Graczyk14} from a combination
of indicators (Cepheids, RR~Lyr, Red clump, and eclipsing binaries),
and the value of $0.39\pm0.05$~mag derived by
\cite{Szewczyk09} from near-infrared photometry of RR Lyrae variables.
Of course the much elongated structure of the SMC along the line of
sight, with a range of 10 kpc from RR Lyrae stars \citep{Muraveva18}
and an even larger range from Cepheids \citep{Ripepi17} complicates the
determination of a sensible mean distance to the SMC, but at least we
can conclude that our IRSB-based distances to the Cepheids in our LMC
and SMC samples compare very well with the results obtained from other
distance indicators applied in a purely differential way.  We note here
that the observed dispersion around the SMC PL relation in the $K$-band
of 0.24~mag corresponds to a depth effect of 6~kpc, but this dispersion
is identical to what we observe for the LMC and MW samples suggesting
that depth effects are not significantly affecting our results.

Most of the previous observational determinations of the metallicity
effect on the Cepheid PL relation used the "inner/outer field method" -
magnitudes of Cepheid samples in a field close to the center of a spiral
galaxy were compared to the magnitudes of their Cepheid counterparts
in a field located at a much larger galactocentric distance. In all
such studies the inner-field Cepheids were found to be brighter than
the more metal-poor outer field Cepheids leading, together with an
adopted metallicity gradient in the disk of the galaxy, to a negative
sign of the metallicity effect, as in our present study.  There are
however two fundamental problems with this approach; firstly, there
are a number of calibrations of H II region oxygen abundances in the
literature yielding quite different results, so the size of the derived
metallicity effect depends on the adopted oxygen abundance calibration
(e.g. \citep{Bresolin09}). Secondly, Cepheids in the inner fields are
more strongly affected by crowding and blending problems, so at least
part of the systematically brighter magnitudes of inner field Cepheids
may be caused by close companion stars which are not resolved in the
photometry. An example is the (excellent!) work of \cite{Shappee11}
who find a metallicity effect of $-0.80$~mag/dex in optical $V$ and $I$
bands in M101 using HST/ACS images, which seems to be unreasonably large
and is probably significantly biased by crowding affecting the Cepheids
in their inner field in M 101. Deriving the metallicity effect in nearby
galaxies like the Magellanic Clouds where crowding is not a problem in
the photometry seems therefore to be a safer route to determine the true
size and sign of the Cepheid metallicity effect.

It is particularly interesting to compare the metallicity effect
we obtain in this study with the recent determination reported by
\cite{Wielgorski17} who used a completely different approach, and
basically obtained a zero effect in all bands.  Their determination
critically depends on the distance difference between LMC and SMC which
they assume to be $0.472\pm0.026$~mag, as obtained from similar late-type
eclipsing binary systems in both galaxies. While the LMC distance obtained
from this method is extremely well established \citep{Pietrzynski13},
the SMC distance reported in \cite{Graczyk14} is based on only five
systems. The average distance modulus obtained from this small number of
systems might not represent the SMC mean distance very well considering
the large spread of the SMC in the line of sight we discussed before. A
change of 0.07~mag in the mean SMC distance, from the 18.97~mag value obtained
by \cite{Graczyk14} to 18.90~mag, might thus be consistent with the
current uncertainty on the eclipsing binary  distance to the SMC. If the
conjecture is correct, it would bring
the metallicity effect determination in \cite{Wielgorski17} to
about $-0.2$~mag/dex in all bands, in excellent agreement with the value
derived in this paper. We therefore suspect that the apparent discrepancy
between the zero metallicity effect found by \cite{Wielgorski17}, and the
$-0.2$~mag/dex effect found in this paper, is due to an overestimated
SMC distance in \cite{Graczyk14} from their small number of systems
available for analysis.

\section{Conclusions}
\label{sec.conclusions}

We have obtained new and very accurate radial velocity and $K$-band light
curves of 26 SMC Cepheids, expanding our previous sample of five stars
in \cite{Storm04, Storm11b} to 31 Cepheids covering the full Cepheid
period range from 4 to 69 days. We complemented our new $K$-band light
curves with data from the VMC Survey. Using these data together with
the excellent $V$-band light curves of the variables from the OGLE
Project, we applied the IRSB Technique as calibrated by \cite{Storm11a}
and calculated the distances of the individual SMC Cepheids, and their
absolute magnitudes in near-infrared and optical bands. These magnitudes
define tight period-luminosity relations in the $V$, $I$, $J$, and $K$
bands as well as in the optical and near-infrared Wesenheit indices, with
dispersions practically identical to the relations we have previously
obtained for Cepheid samples in the Milky Way and LMC in \cite{Storm11b}
with the same technique.

We find very good agreement between the slopes of these PL relations
and the fiducial PL relations in the LMC obtained by \cite{Soszynski15}
in the optical, and by \cite{Macri15} in the near-infrared bands,
supporting the universality of the slopes of Cepheid PL relations in these
wavelength regimes. Our SMC Cepheid distances yield a mean SMC distance
of $18.86\pm0.04$~mag which compares very well with recent determinations
from other distance indicators. From the Cepheid samples analyzed with
the IRSB Technique in the LMC and SMC we obtain a distance difference
between the Clouds of $0.40$~mag, which again compares very well to other
recent estimates from different standard candles. The distance modulus
of $18.46\pm0.04$~mag we obtain for the LMC from our Cepheid sample in
this galaxy is in excellent agreement with the near-geometrical value
of $18.497$~mag established by \cite{Pietrzynski13} from late-type
eclipsing binaries.

We find that the absolute PL relations defined by the SMC Cepheids are
significantly displaced to fainter magnitudes, as compared to their MW
and LMC counterparts.  This is true for all near-infrared and optical
bands studied in this paper, and argues for a metallicity effect in all
bands in the sense that the more metal-poor Cepheids are intrinsically
fainter than their more metal-rich counterparts with similar pulsation
periods.  The metallicity effect we obtain is $-0.23\pm0.06$~mag/dex
in the near-infrared $K$ band, and slightly larger in the $J$, $I$
and the optical Wesenheit bands.  The uncertainties have been reduced by
almost a factor of two with respect to our previous work and the effect
is now very (3$\sigma$) significant.  Our data suggests that the change
of the PL relation zero points with metallicity might not be entirely
linear in the different studied bands, but might become {\em steeper}
for lower metallicities. We stress that our IRSB analyses of the Cepheids
in MW, LMC and SMC samples have been carried out following identical procedures 
leading to a strictly {\em differential} analysis between the absolute
magnitudes of the Cepheids in the three galaxies, making our results and
conclusions independent of eventual systematic errors on the distances
due to imperfections in the technique. We have also shown that there is
no systematic offset between the reddening scales adopted for the SMC
Cepheids in the present work, and for the LMC Cepheids in our 
previous work, which combined with the reddening insensitivity of the
method itself means that reddening effects on the final relations for
the $K$-band and Wesenheit indices are negligble.

We argue that the $K$-band Cepheid PL relation continues to be the best
tool to determine the distances to late-type galaxies. However, the mild
but significant metallicity effect determined in this paper should be 
taken into account, which obviously requires an estimate of the average 
metallicities of the Cepheid samples used in such determinations.

\begin{acknowledgements}
W.G. and G.P. gratefully acknowledge financial support for this work from the BASAL
Centro de Astrofisica y Tecnologias Afines (CATA) AFB-170002. W.G., D.G. and M.G.
also gratefully acknowledge financial support from the Millenium Institute of Astrophysics (MAS)
of the Iniciativa Cientifica Milenio del Ministerio de Economia, Fomento y Turismo
de Chile, Project IC120009. This research has also been supported by funding from
the European Union's 2020 research and innovation program (grant agreement No. 695099).

We are greatly indebted to the staffs at the European Southern Observatory La Silla and
Paranal sites, as well as the staff at Las Campanas Observatory, for their excellent
support during the many visitor mode runs which allowed us to collect the data
for this study. We are pleased to thank ESO and the Chilean TAC for large amounts of 
observing time allotted to this program.
\end{acknowledgements}

\bibliographystyle{aa} % style aa.bst
\bibliography{SMC_IRSB} % your references Yourfile.bib

\begin{appendix}
\section{The light and radial velocity curves}
\label{app.lightcurves}

\begin{figure}
\centering
\includegraphics[width=9cm]{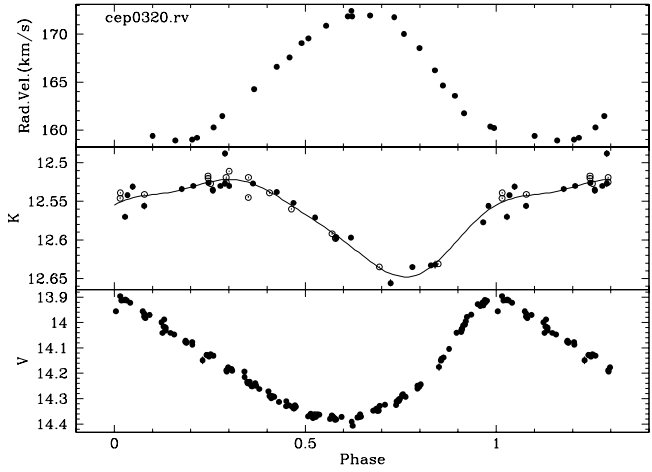}
\caption{The light and radial velocity curves for the star OGLE-SMC-CEP0320.
\label{fig.cep0320-data}}
\end{figure}

\begin{figure}
\centering
\includegraphics[width=9cm]{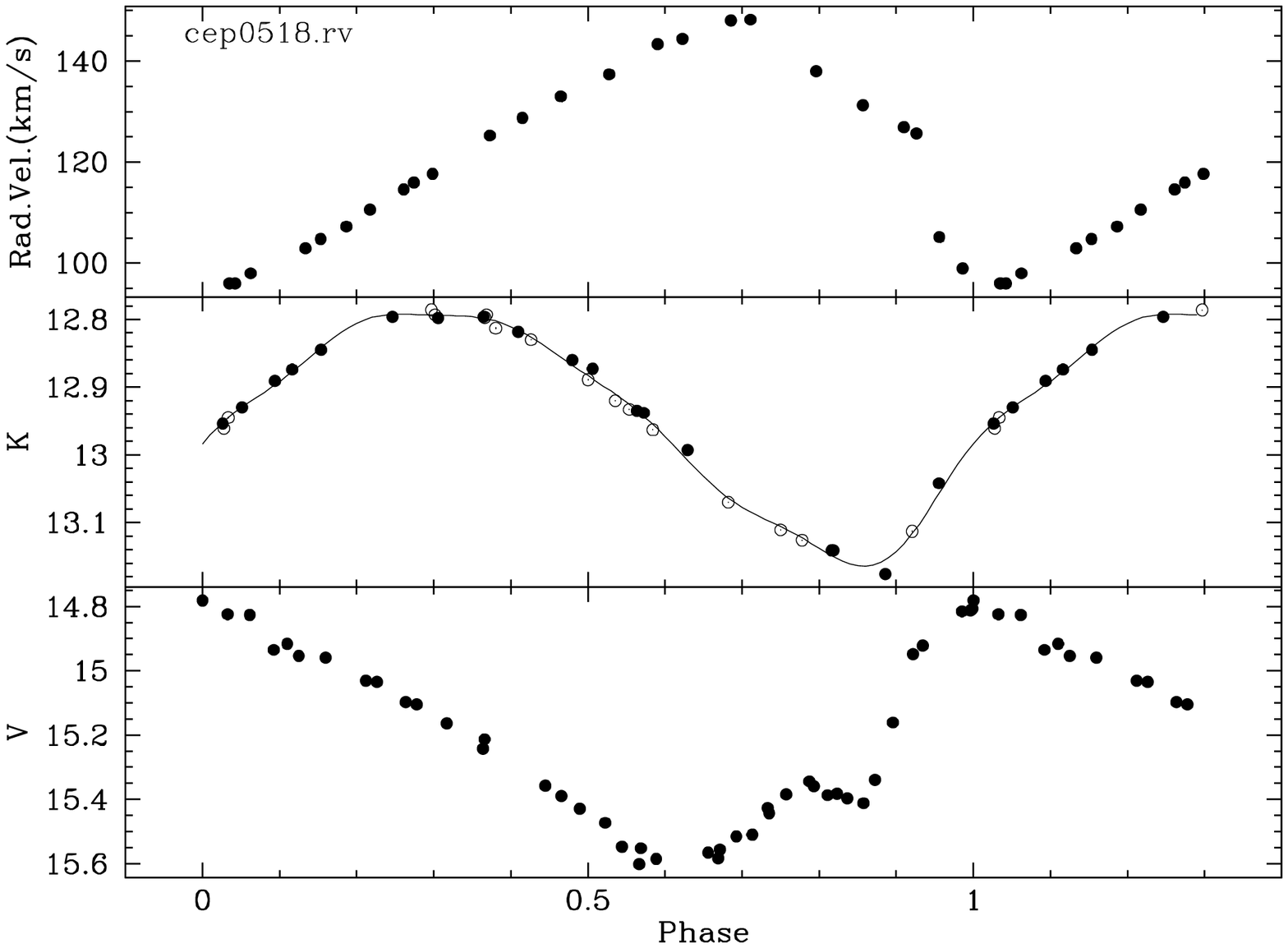}
\caption{The light and radial velocity curves for the star OGLE-SMC-CEP0518.
\label{fig.cep0518-data}}
\end{figure}

\begin{figure}
\centering
\includegraphics[width=9cm]{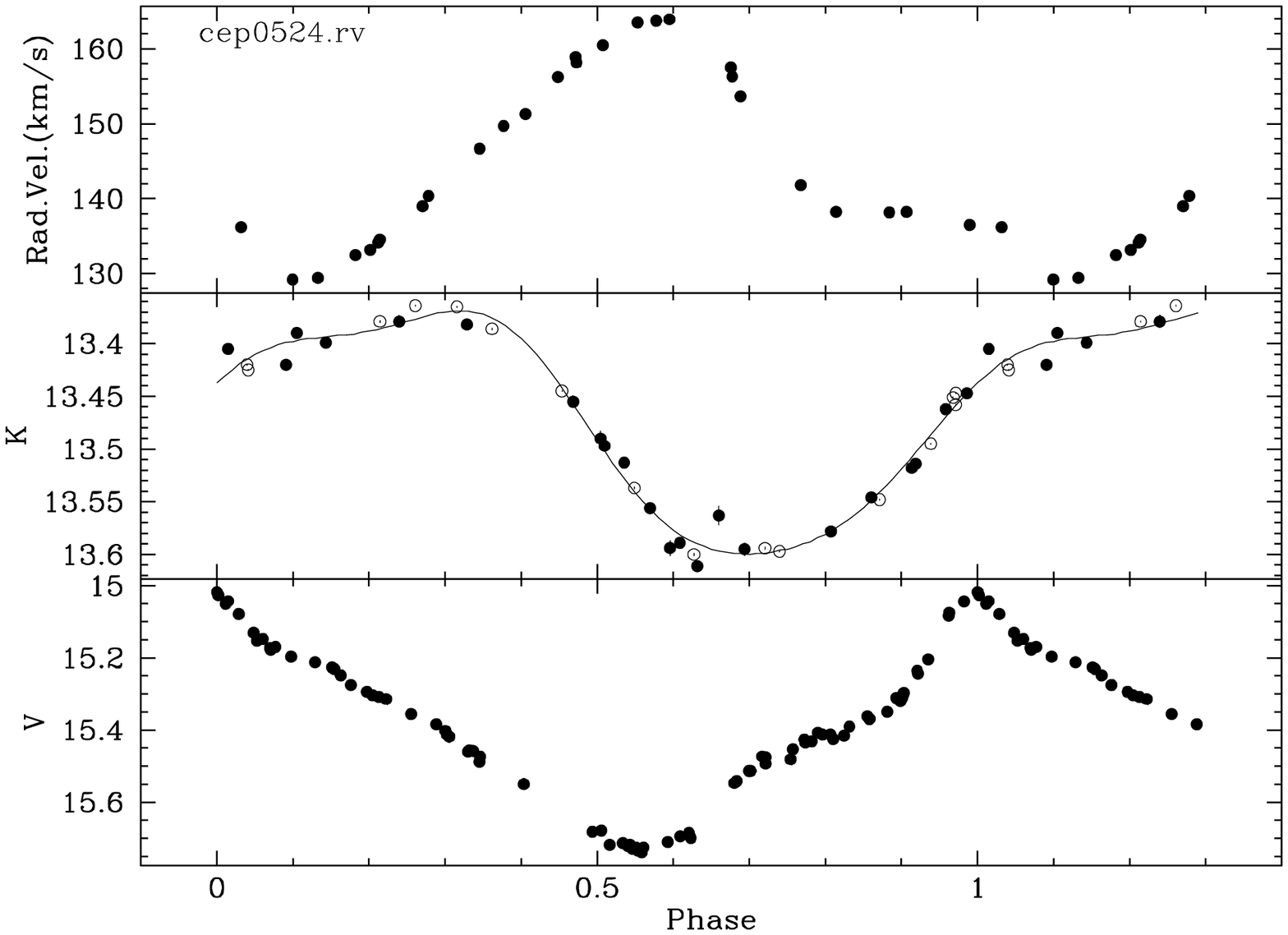}
\caption{The light and radial velocity curves for the star OGLE-SMC-CEP0524.
\label{fig.cep0524-data}}
\end{figure}

\begin{figure}
\centering
\includegraphics[width=9cm]{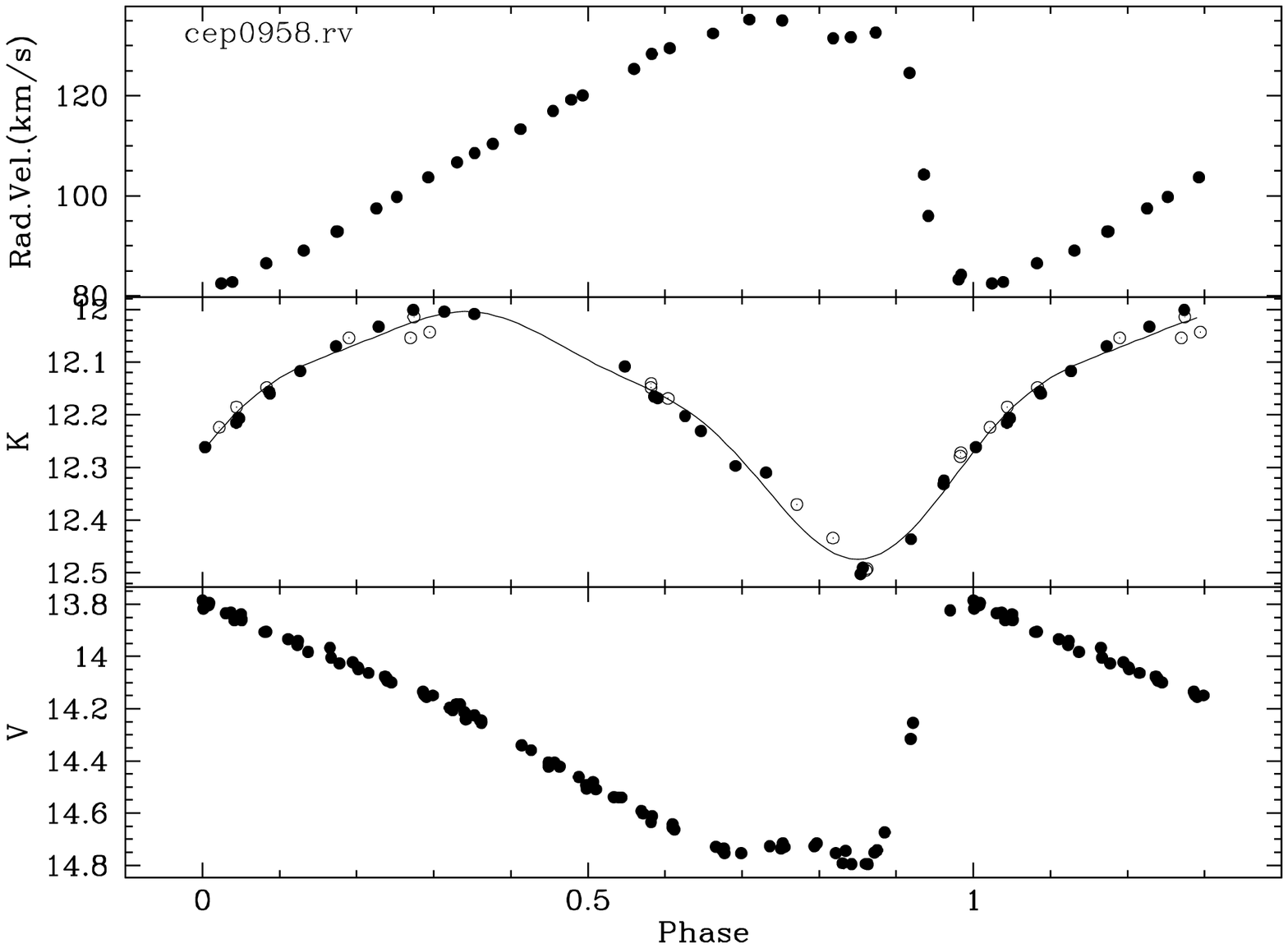}
\caption{The light and radial velocity curves for the star OGLE-SMC-CEP0958.
\label{fig.cep0958-data}}
\end{figure}

\begin{figure}
\centering
\includegraphics[width=9cm]{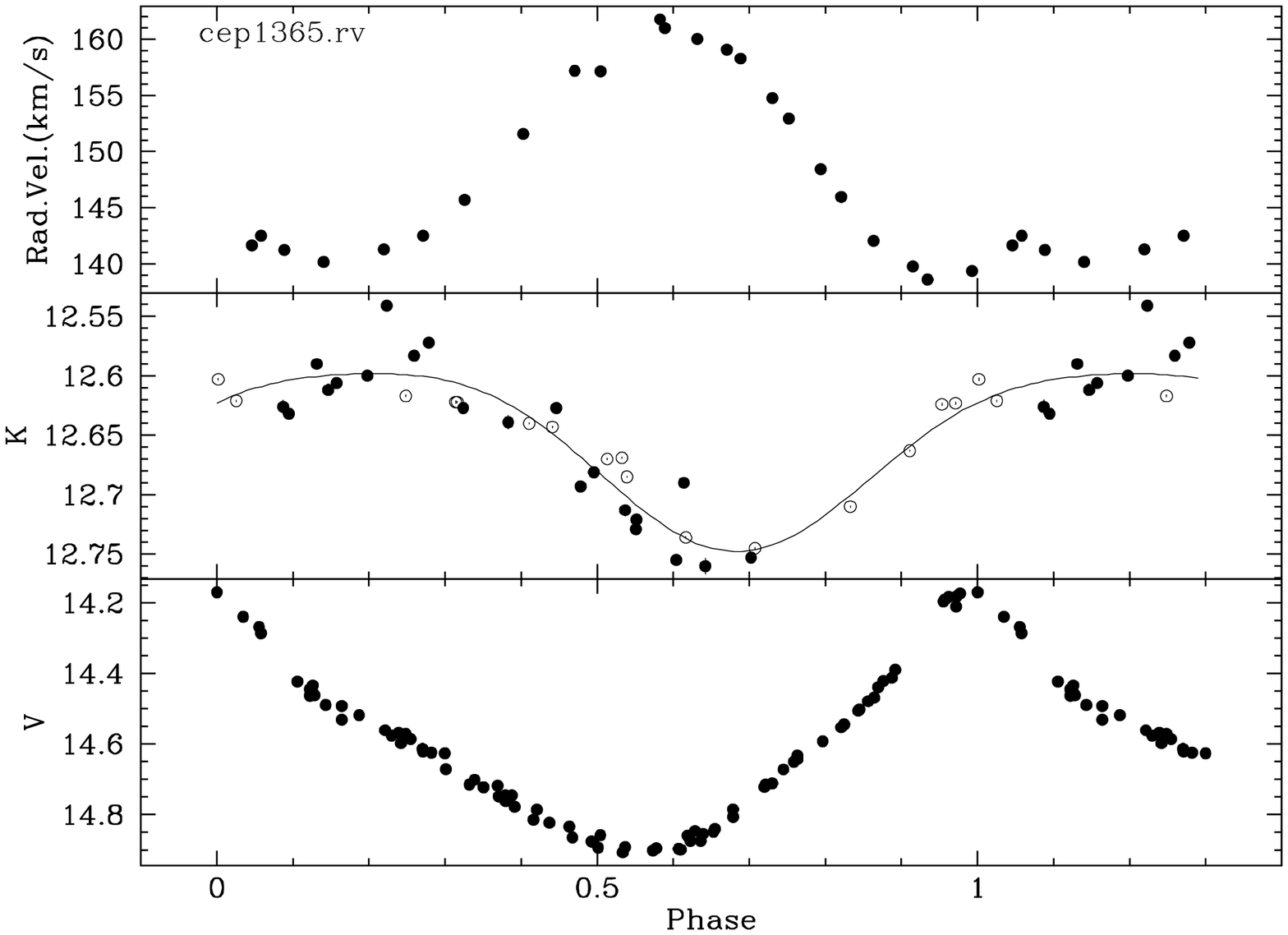}
\caption{The light and radial velocity curves for the star OGLE-SMC-CEP1365.
\label{fig.cep1365-data}}
\end{figure}

\begin{figure}
\centering
\includegraphics[width=9cm]{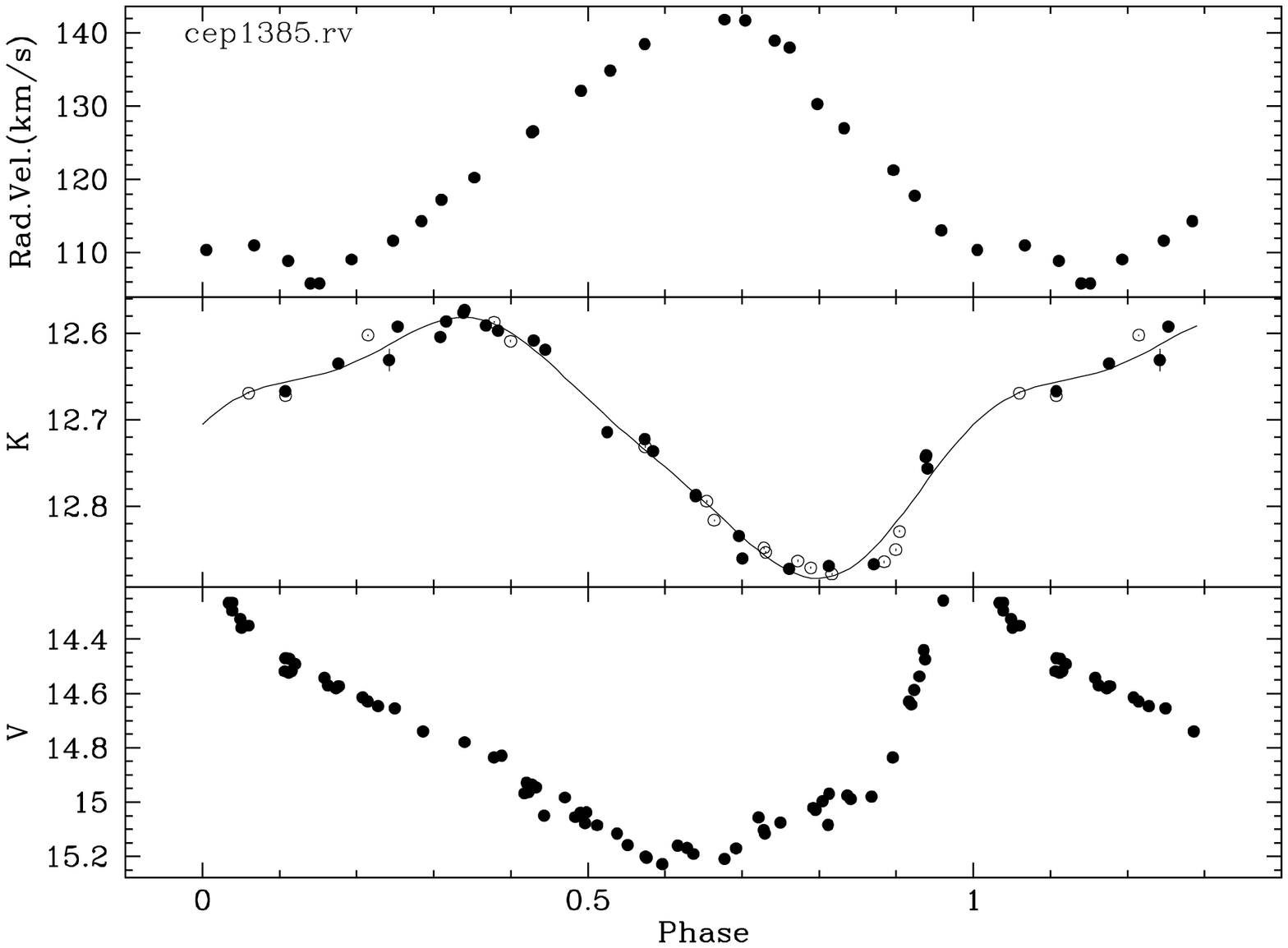}
\caption{The light and radial velocity curves for the star OGLE-SMC-CEP1385.
\label{fig.cep1385-data}}
\end{figure}

\begin{figure}
\centering
\includegraphics[width=9cm]{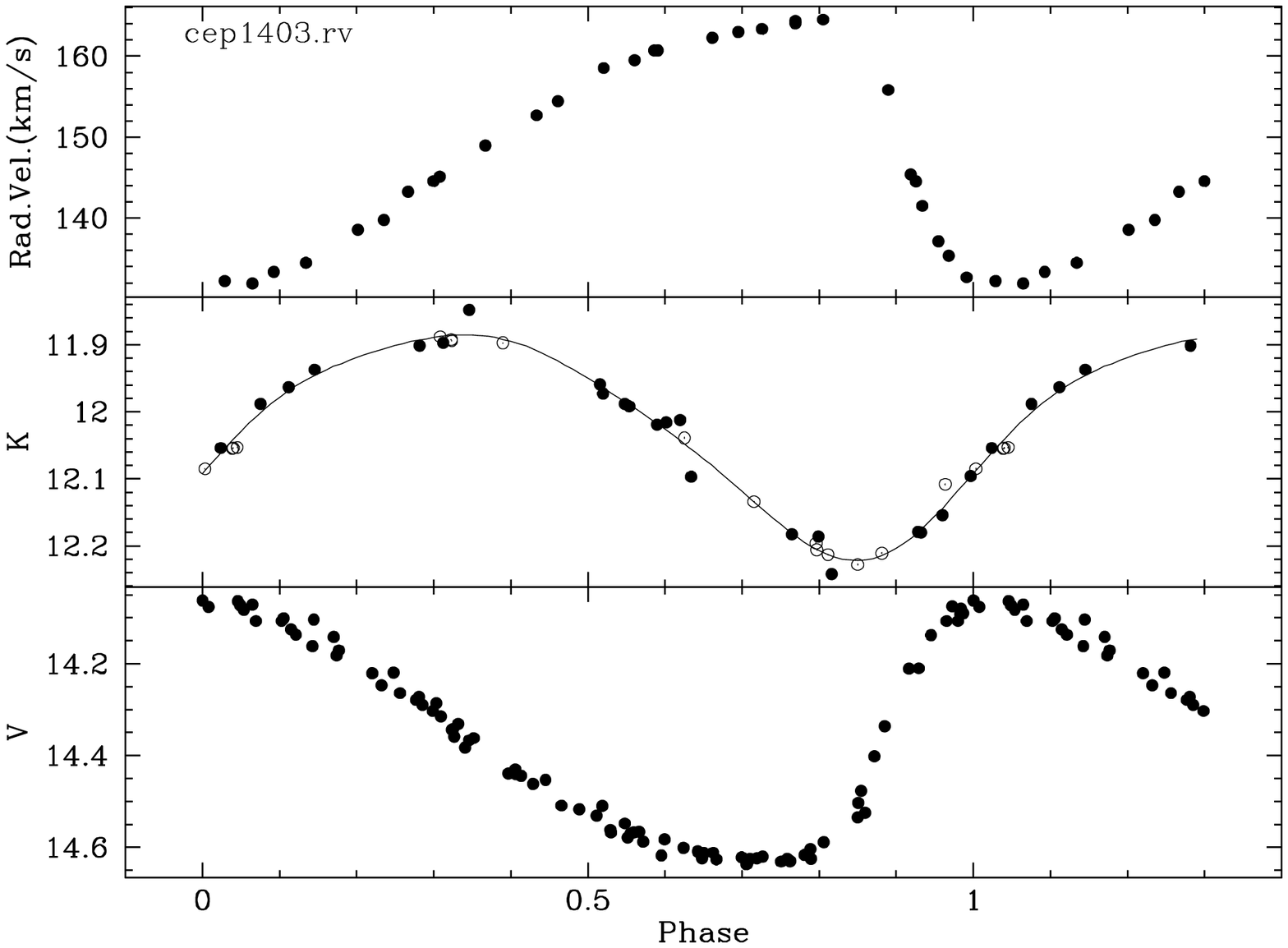}
\caption{The light and radial velocity curves for the star OGLE-SMC-CEP1403.
\label{fig.cep1403-data}}
\end{figure}

\begin{figure}
\centering
\includegraphics[width=9cm]{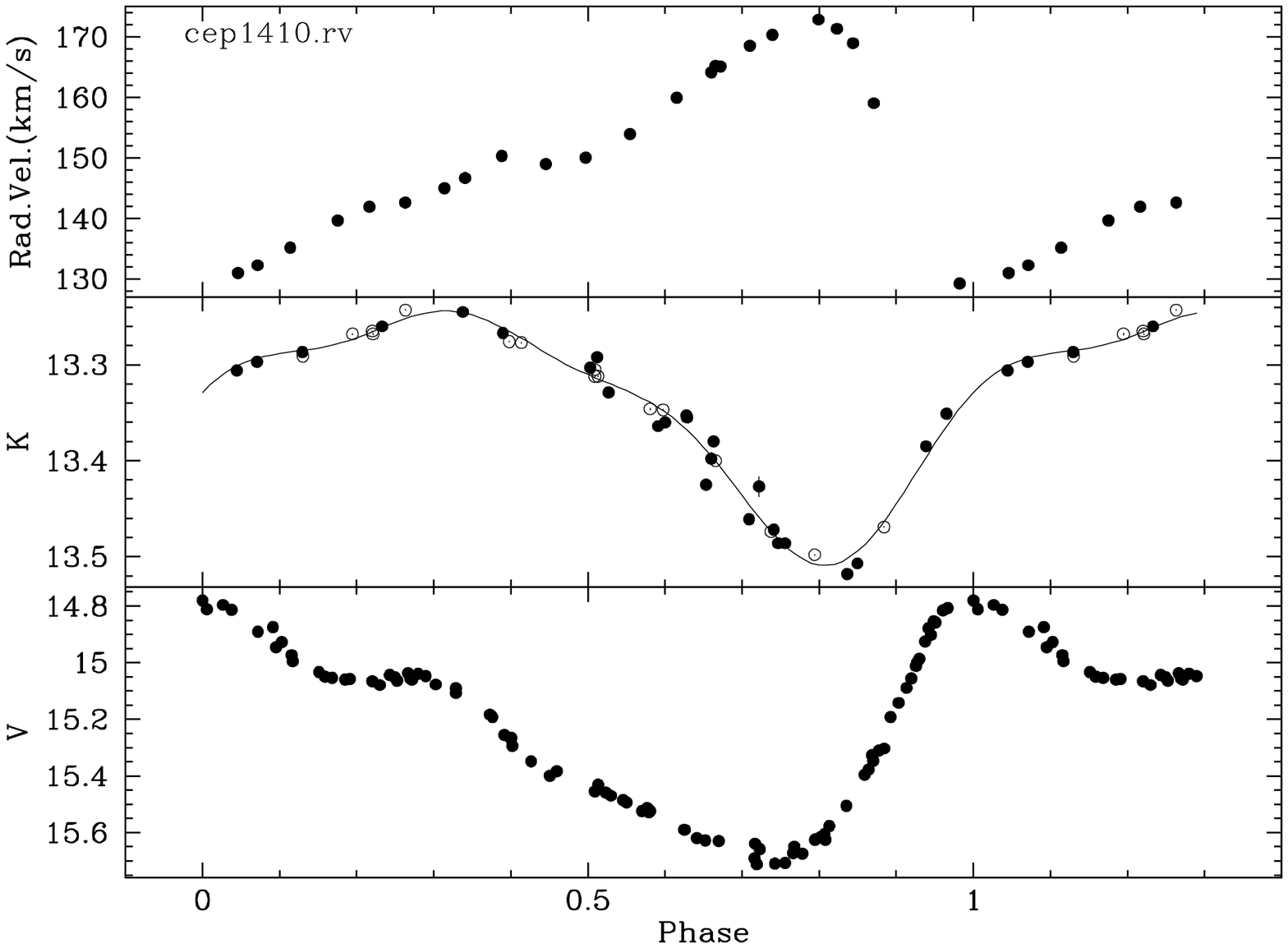}
\caption{The light and radial velocity curves for the star OGLE-SMC-CEP1410.
\label{fig.cep1410-data}}
\end{figure}

\begin{figure}
\centering
\includegraphics[width=9cm]{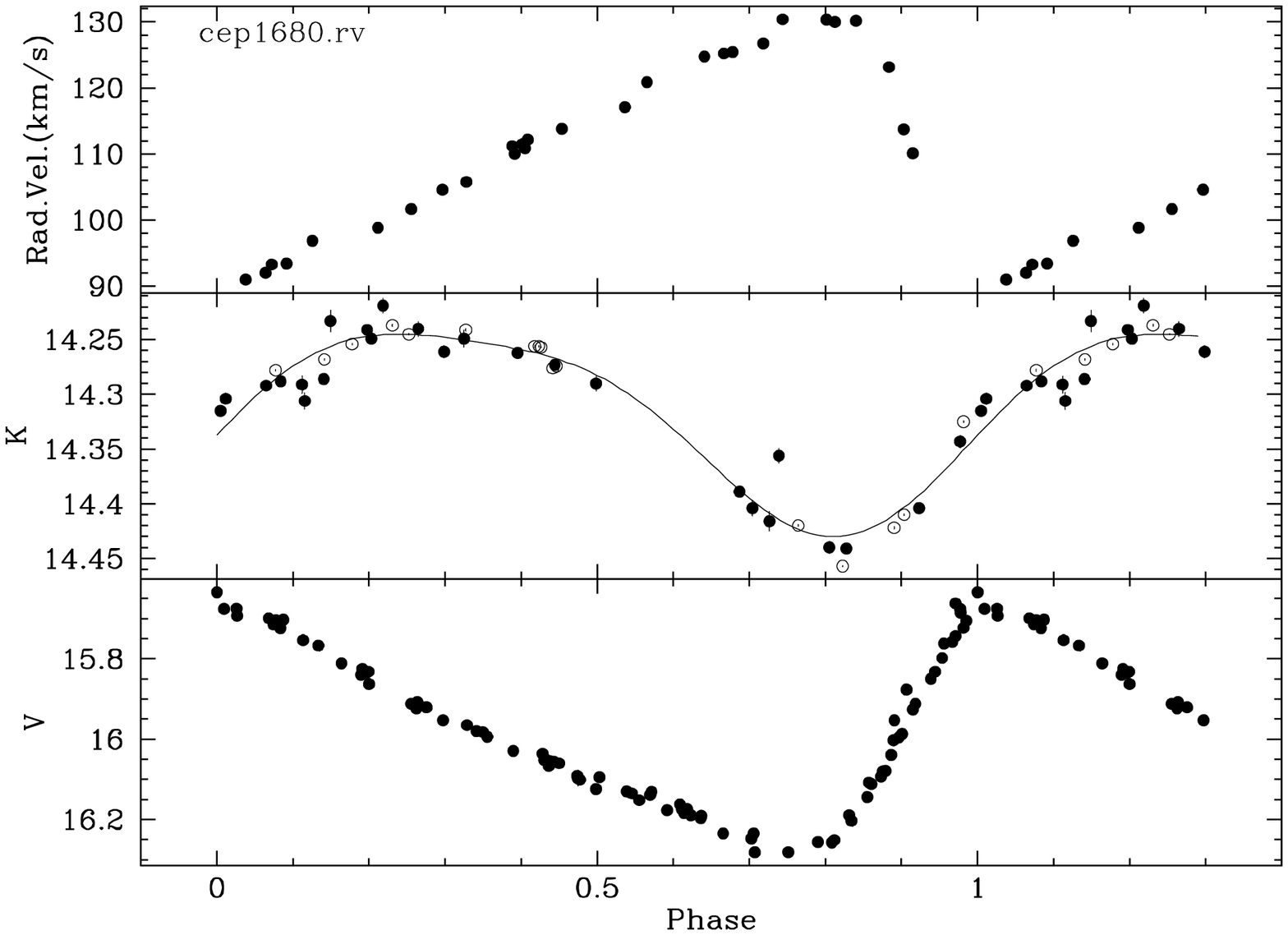}
\caption{The light and radial velocity curves for the star OGLE-SMC-CEP1680.
\label{fig.cep1680-data}}
\end{figure}

\begin{figure}
\centering
\includegraphics[width=9cm]{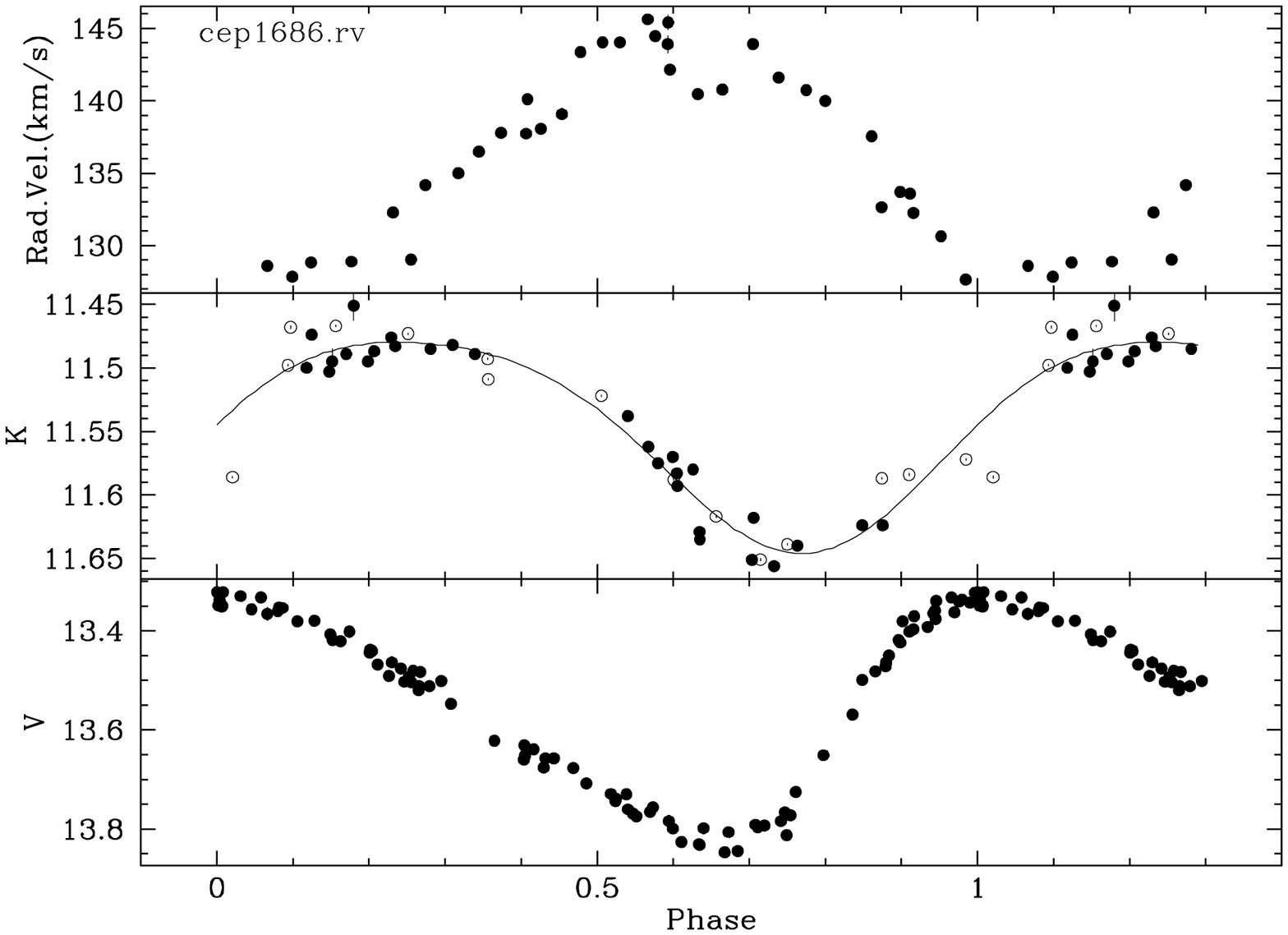}
\caption{The light and radial velocity curves for the star OGLE-SMC-CEP1686.
\label{fig.cep1686-data}}
\end{figure}

\begin{figure}
\centering
\includegraphics[width=9cm]{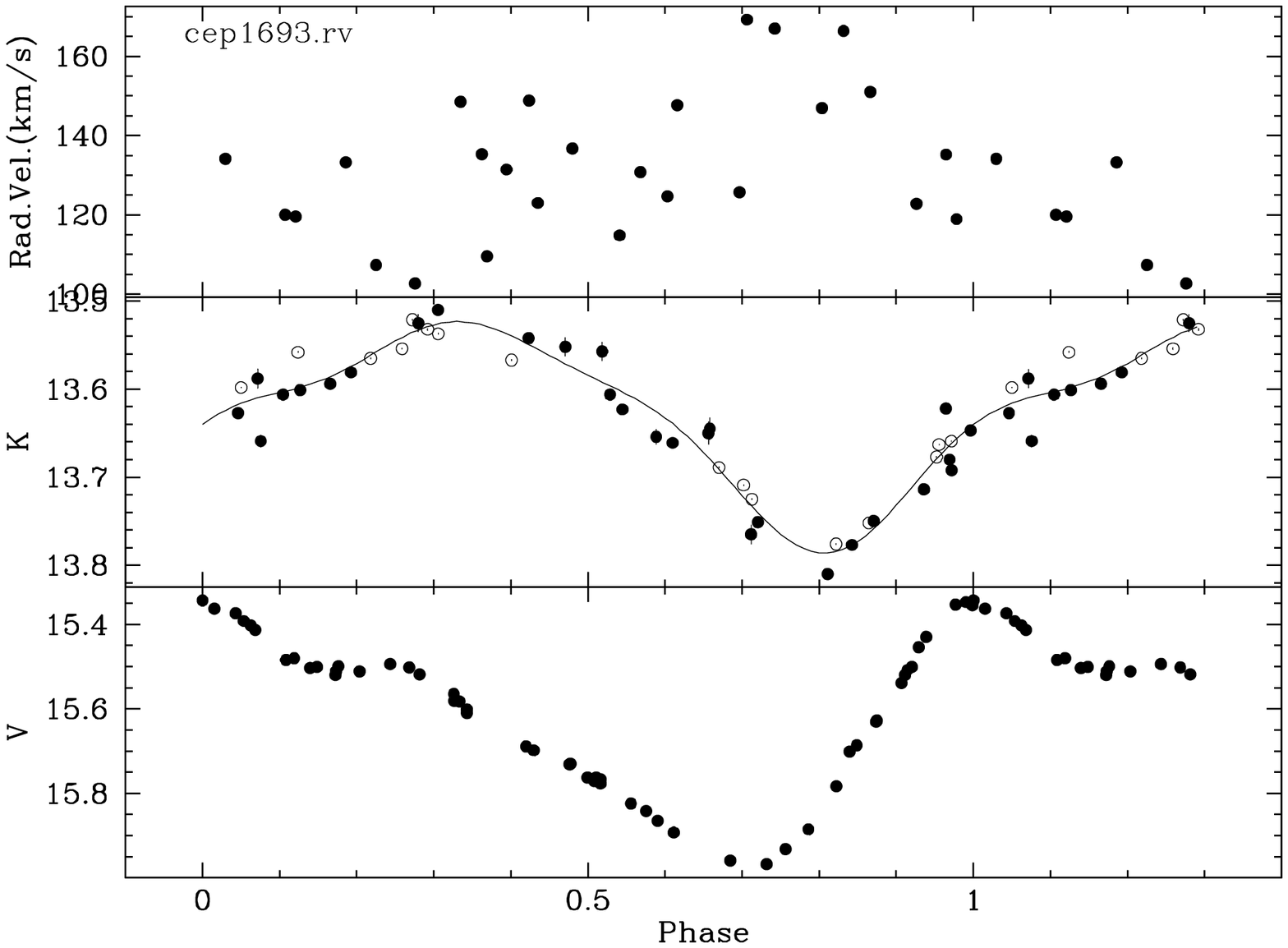}
\caption{The light and radial velocity curves for the star OGLE-SMC-CEP1693.
\label{fig.cep1693-data}}
\end{figure}

\begin{figure}
\centering
\includegraphics[width=9cm]{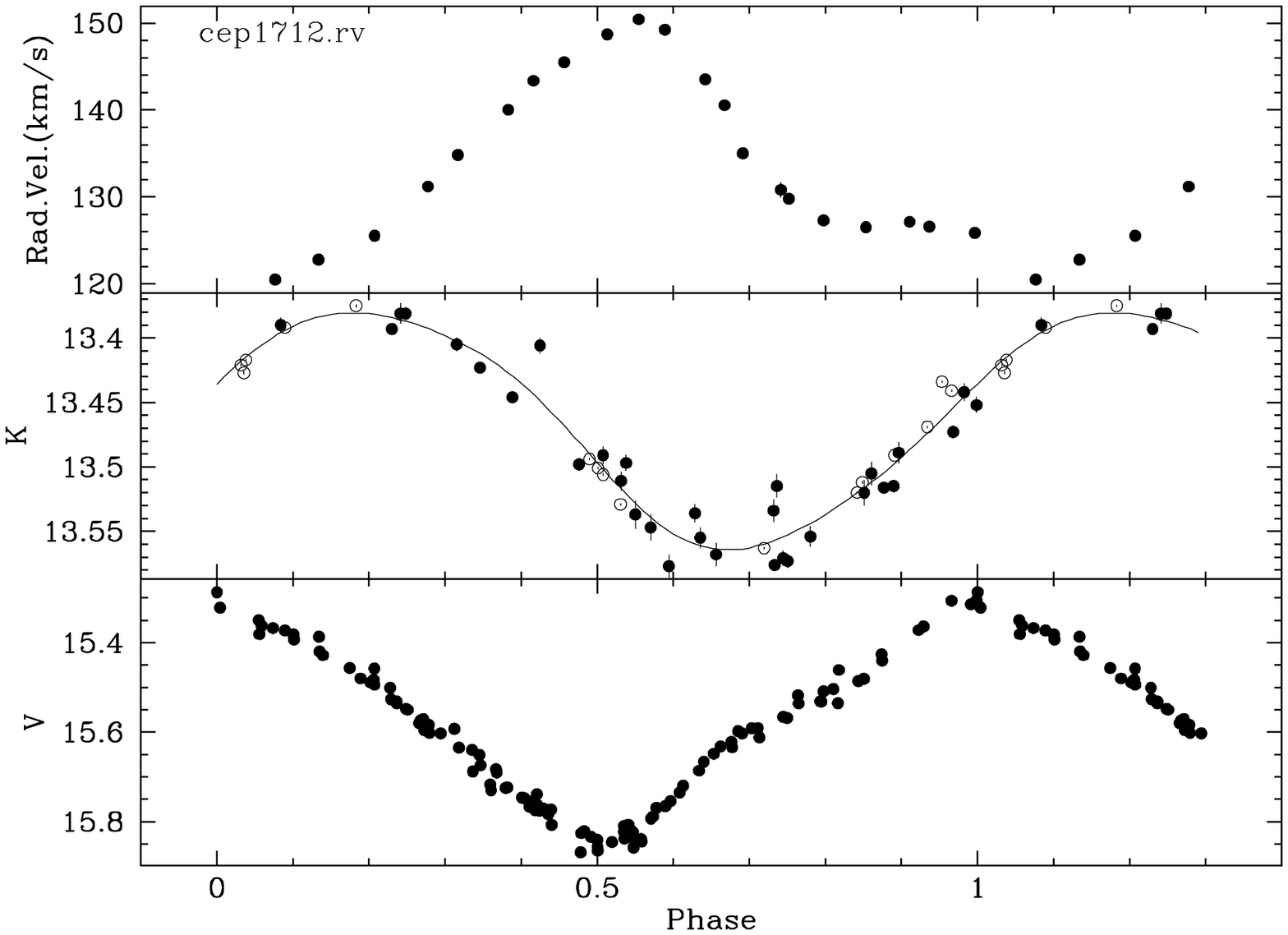}
\caption{The light and radial velocity curves for the star OGLE-SMC-CEP1712.
\label{fig.cep1712-data}}
\end{figure}

\begin{figure}
\centering
\includegraphics[width=9cm]{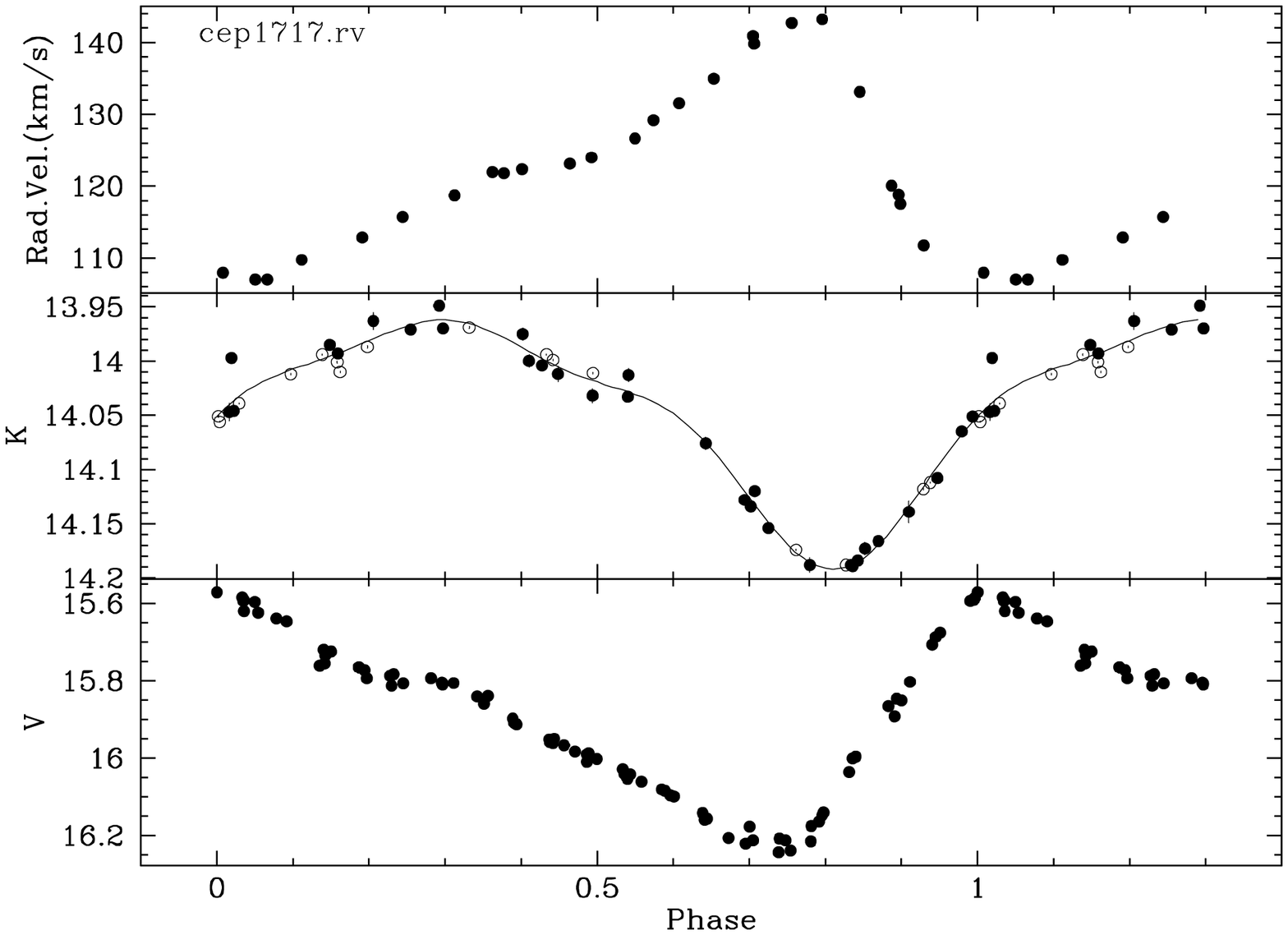}
\caption{The light and radial velocity curves for the star OGLE-SMC-CEP1717.
\label{fig.cep1717-data}}
\end{figure}

\begin{figure}
\centering
\includegraphics[width=9cm]{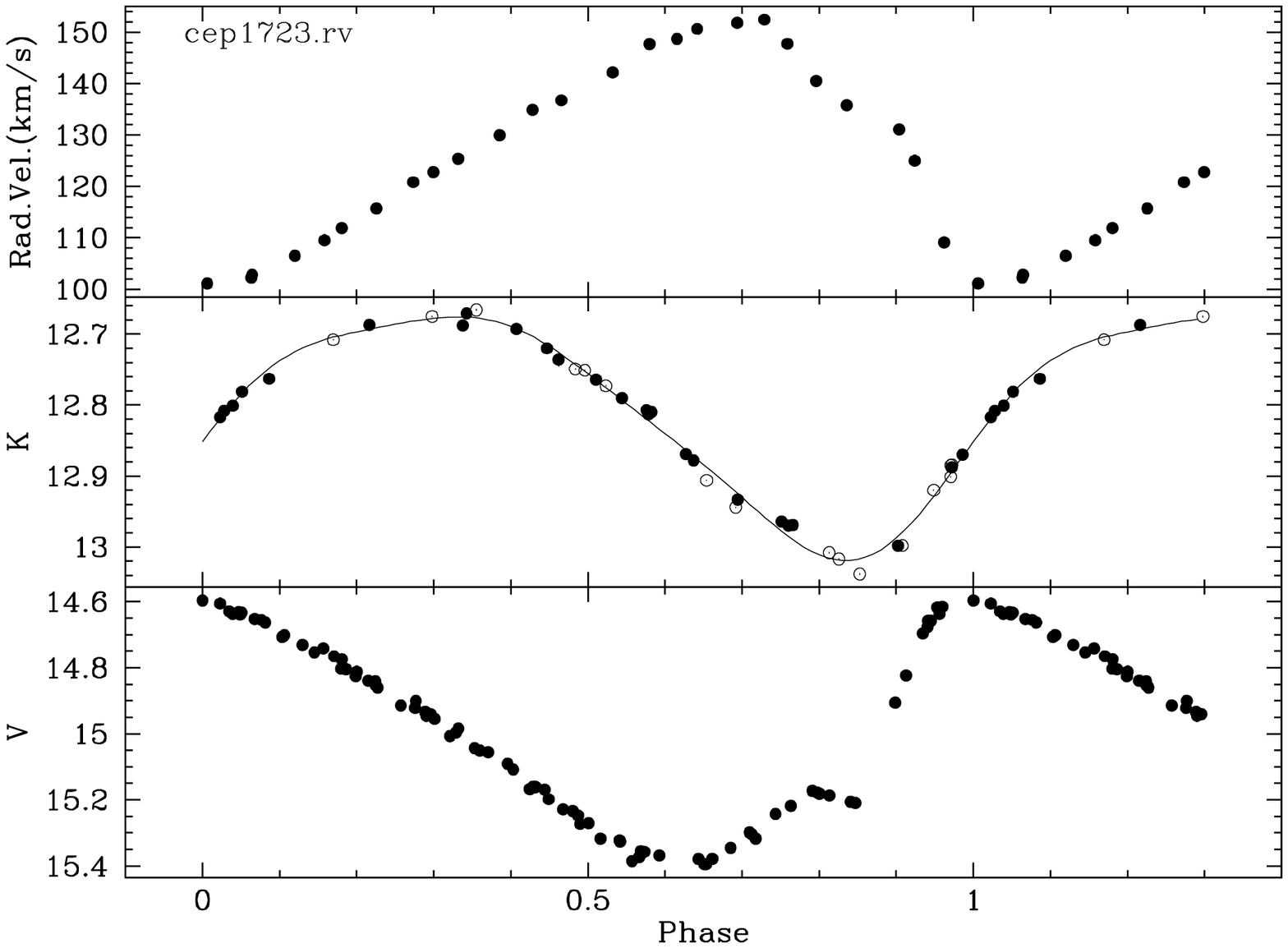}
\caption{The light and radial velocity curves for the star OGLE-SMC-CEP1723.
\label{fig.cep1723-data}}
\end{figure}

\begin{figure}
\centering
\includegraphics[width=9cm]{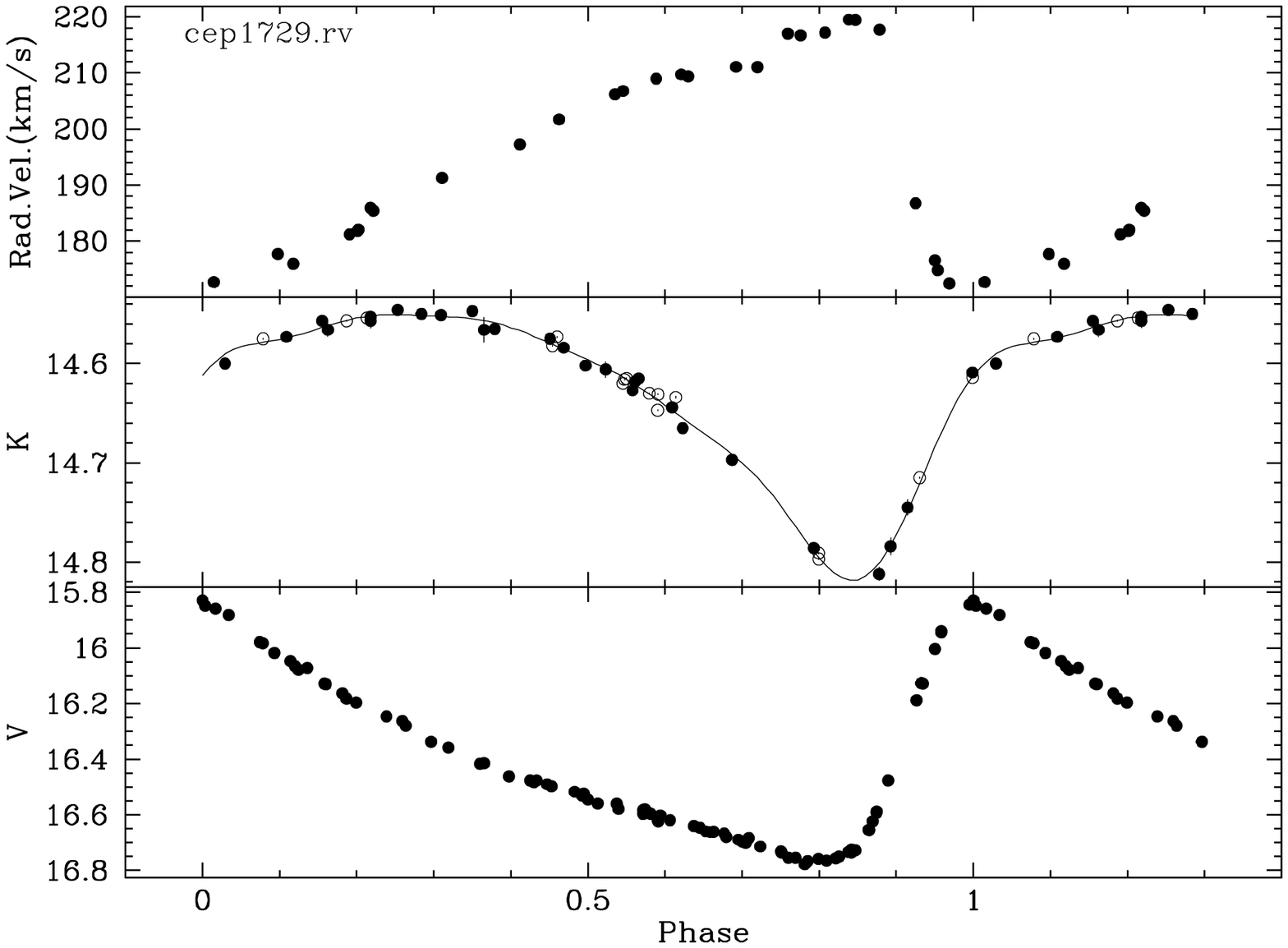}
\caption{The light and radial velocity curves for the star OGLE-SMC-CEP1729.
\label{fig.cep1729-data}}
\end{figure}

\begin{figure}
\centering
\includegraphics[width=9cm]{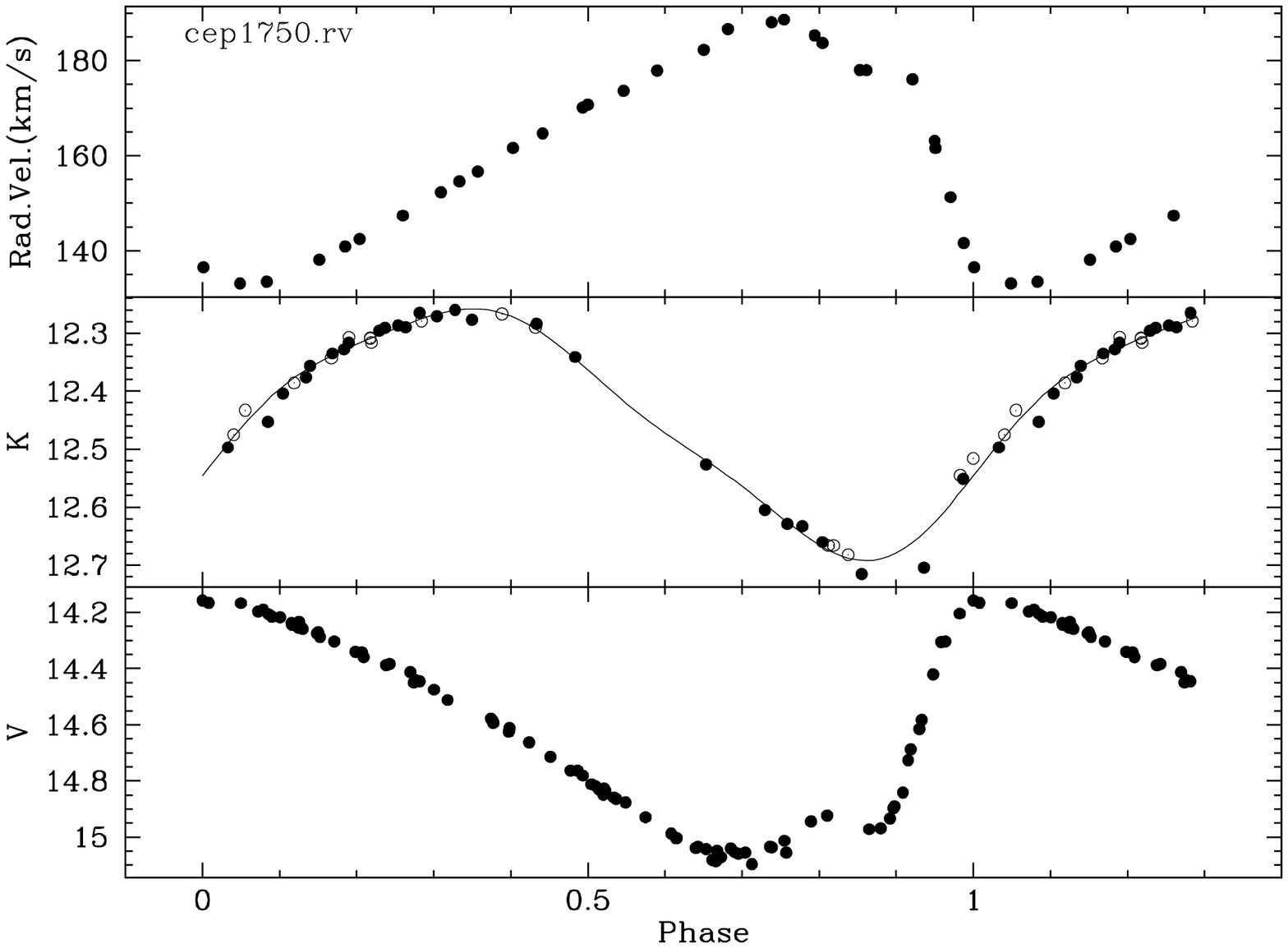}
\caption{The light and radial velocity curves for the star OGLE-SMC-CEP1750.
\label{fig.cep1750-data}}
\end{figure}

\begin{figure}
\centering
\includegraphics[width=9cm]{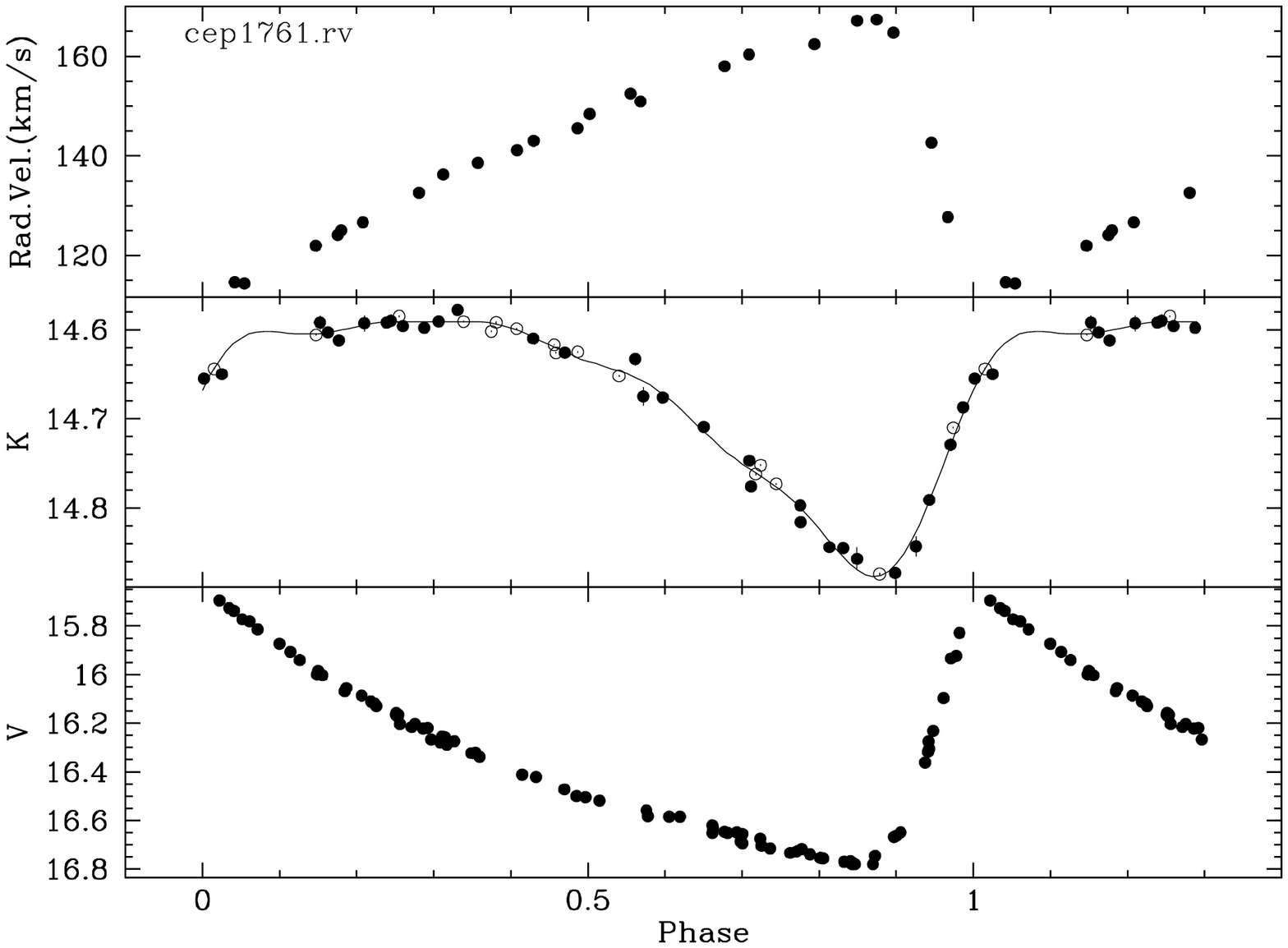}
\caption{The light and radial velocity curves for the star OGLE-SMC-CEP1761.
\label{fig.cep1761-data}}
\end{figure}

\begin{figure}
\centering
\includegraphics[width=9cm]{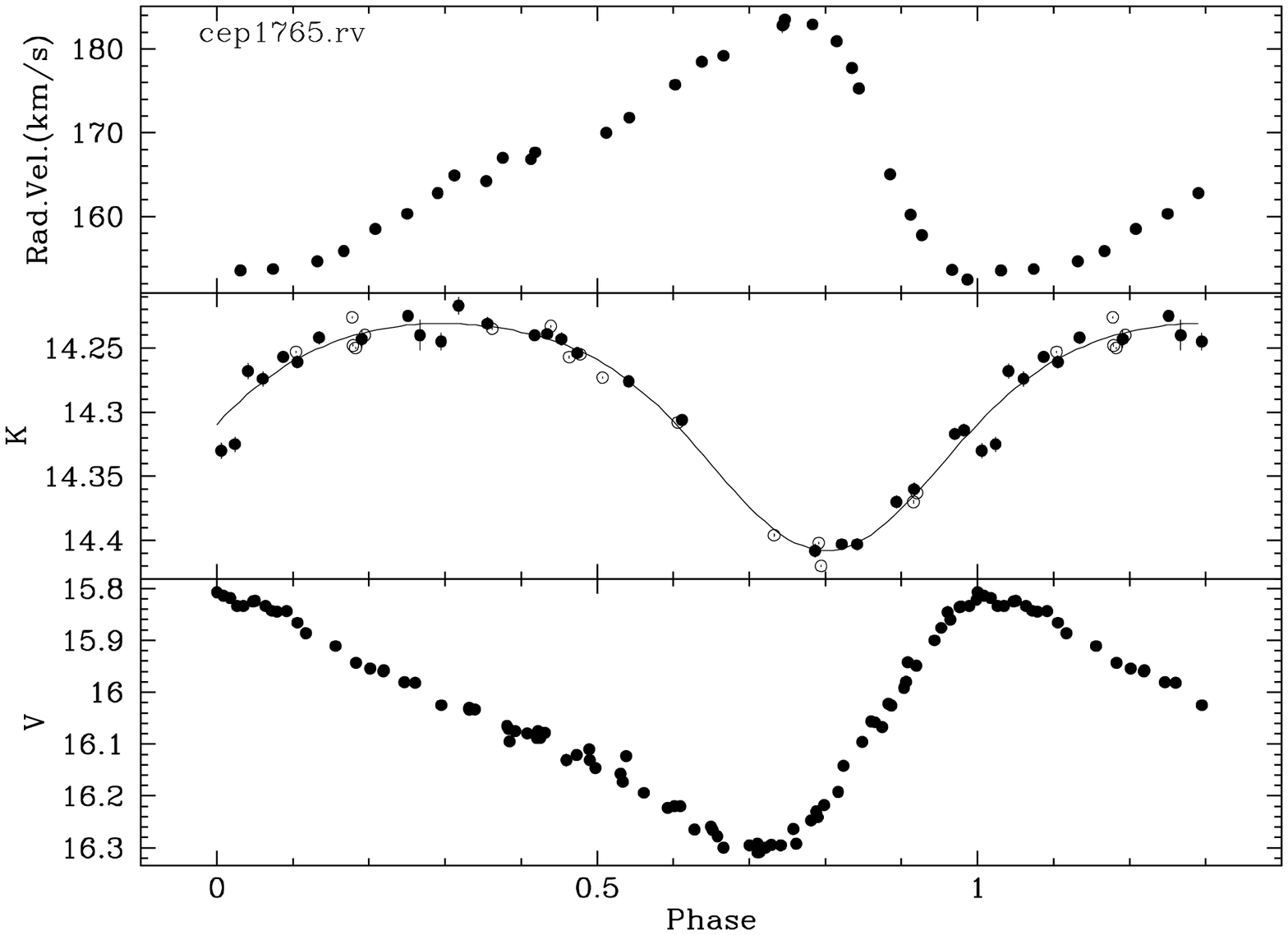}
\caption{The light and radial velocity curves for the star OGLE-SMC-CEP1765.
\label{fig.cep1765-data}}
\end{figure}

\begin{figure}
\centering
\includegraphics[width=9cm]{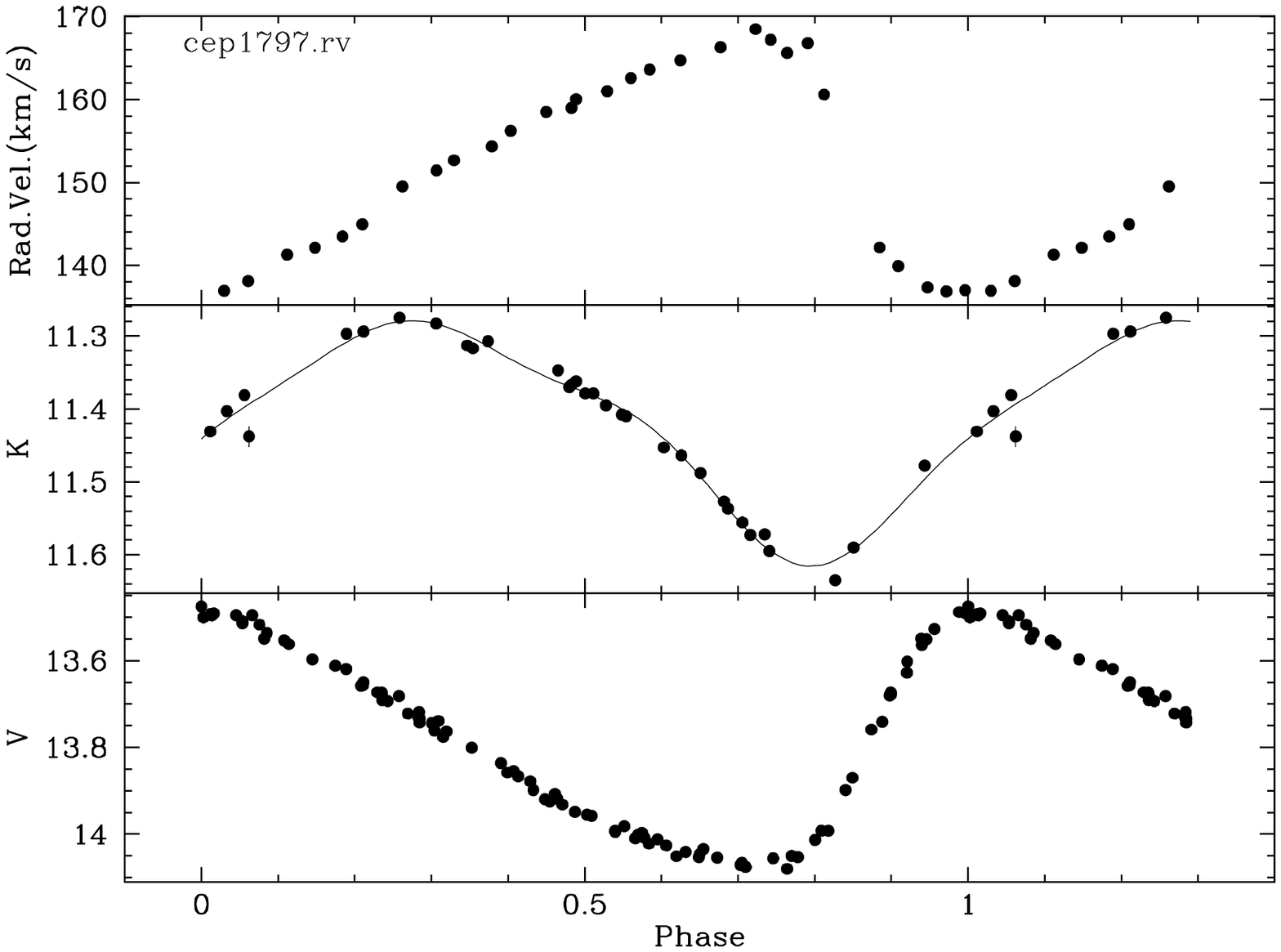}
\caption{The light and radial velocity curves for the star OGLE-SMC-CEP1797.
\label{fig.cep1797-data}}
\end{figure}

\begin{figure}
\centering
\includegraphics[width=9cm]{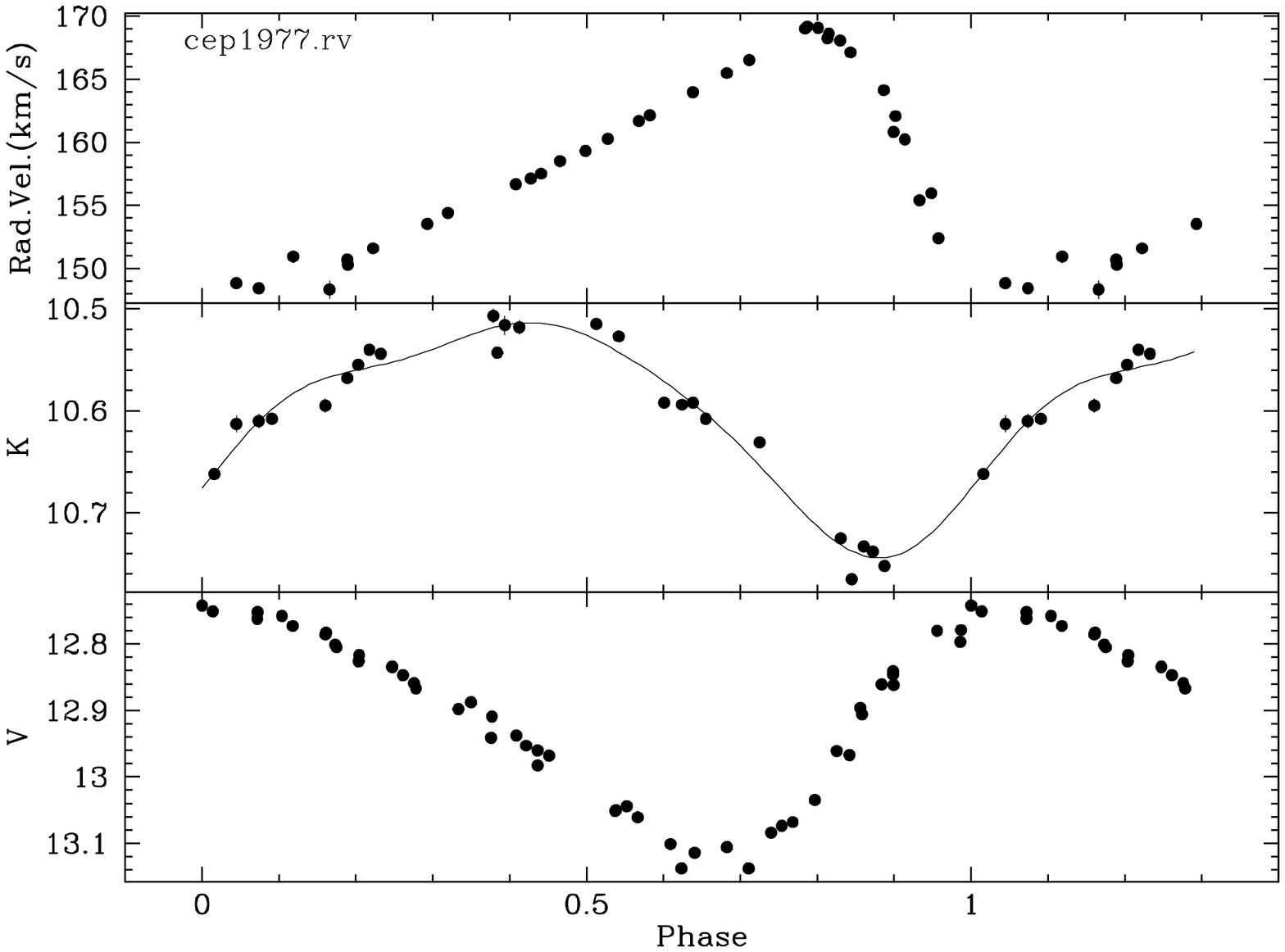}
\caption{The light and radial velocity curves for the star OGLE-SMC-CEP1977.
\label{fig.cep1977-data}}
\end{figure}

\begin{figure}
\centering
\includegraphics[width=9cm]{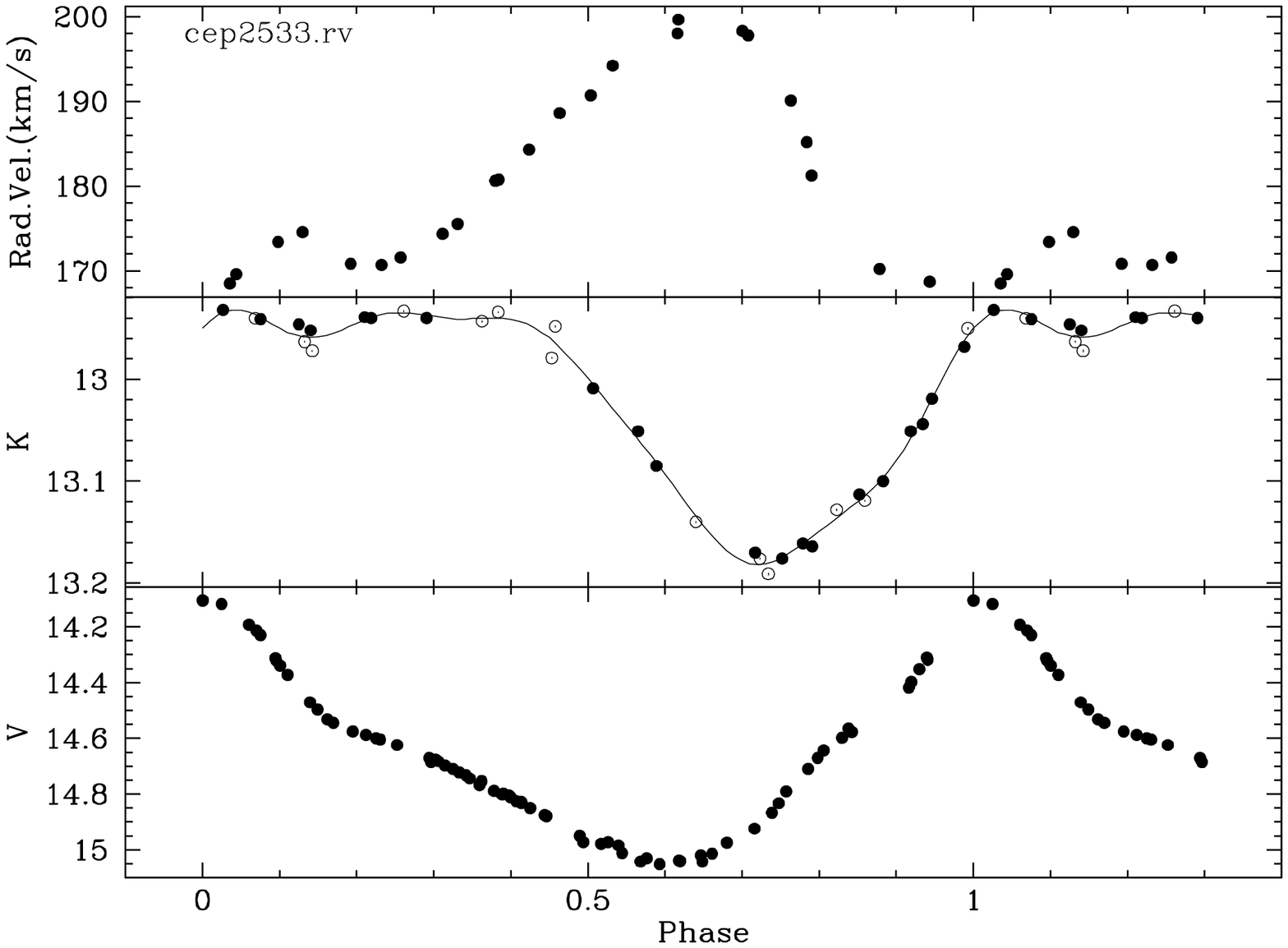}
\caption{The light and radial velocity curves for the star OGLE-SMC-CEP2533.
\label{fig.cep2533-data}}
\end{figure}

\begin{figure}
\centering
\includegraphics[width=9cm]{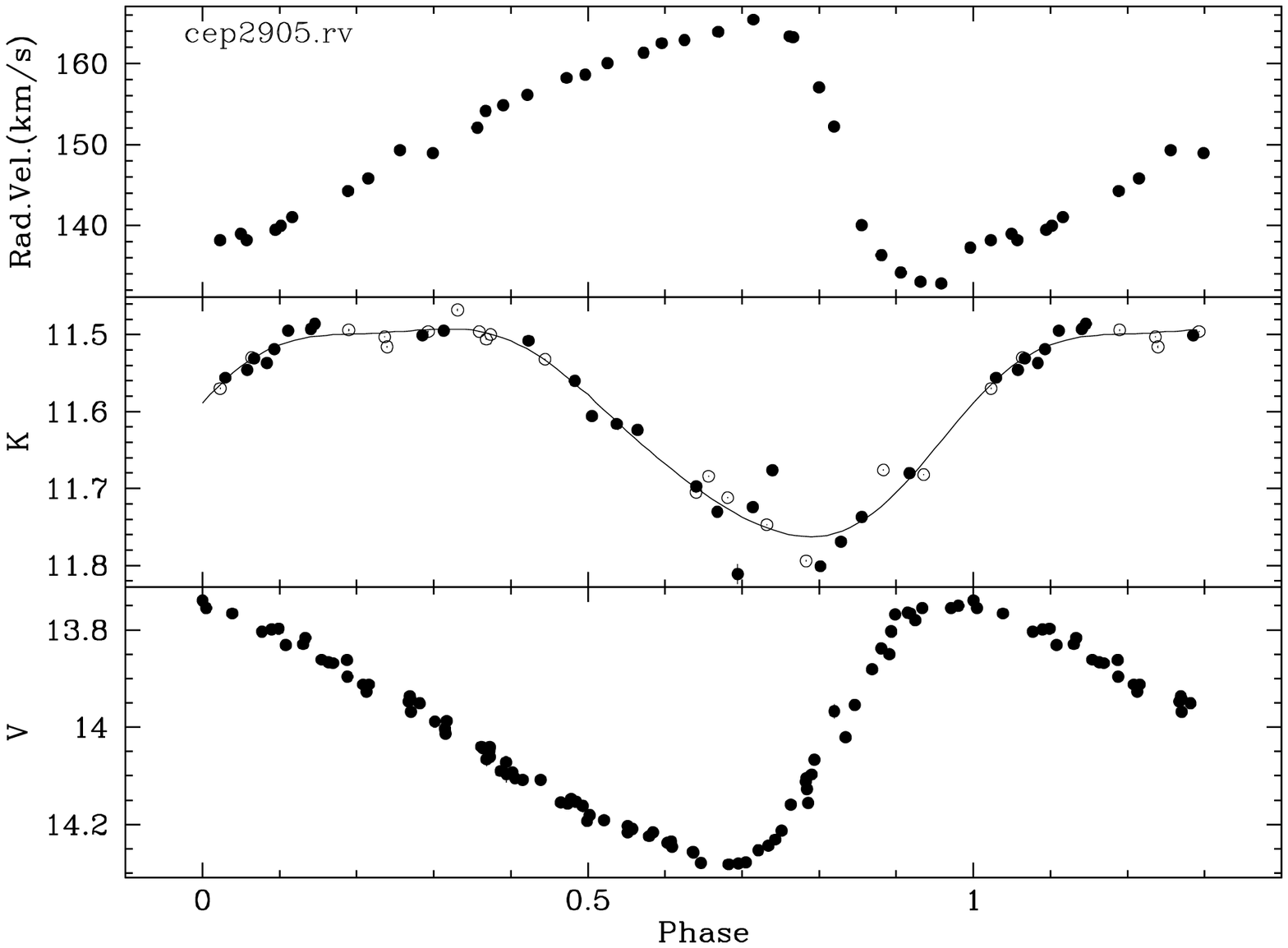}
\caption{The light and radial velocity curves for the star OGLE-SMC-CEP2905.
\label{fig.cep2905-data}}
\end{figure}

\begin{figure}
\centering
\includegraphics[width=9cm]{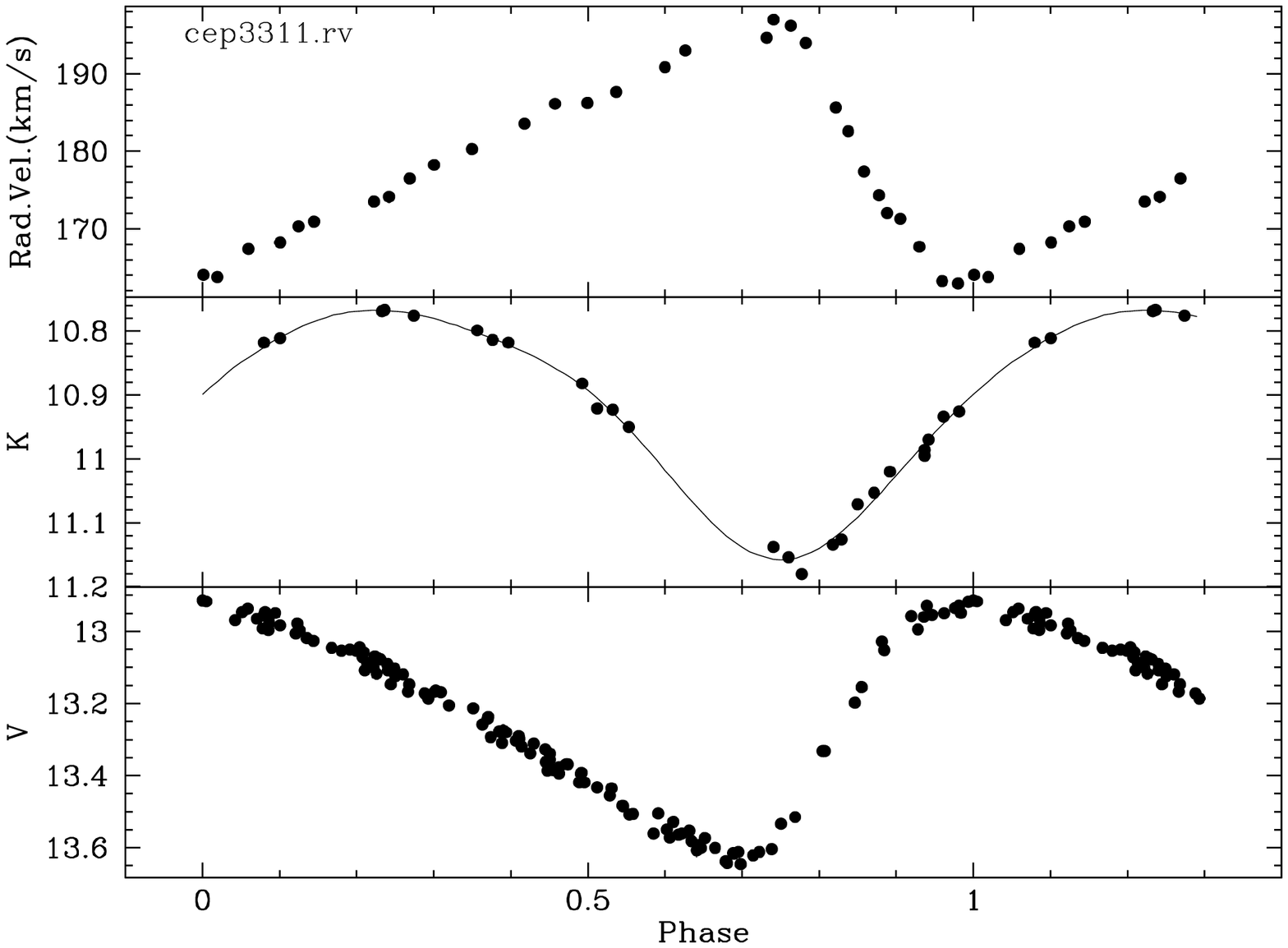}
\caption{The light and radial velocity curves for the star OGLE-SMC-CEP3311.
\label{fig.cep3311-data}}
\end{figure}

\begin{figure}
\centering
\includegraphics[width=9cm]{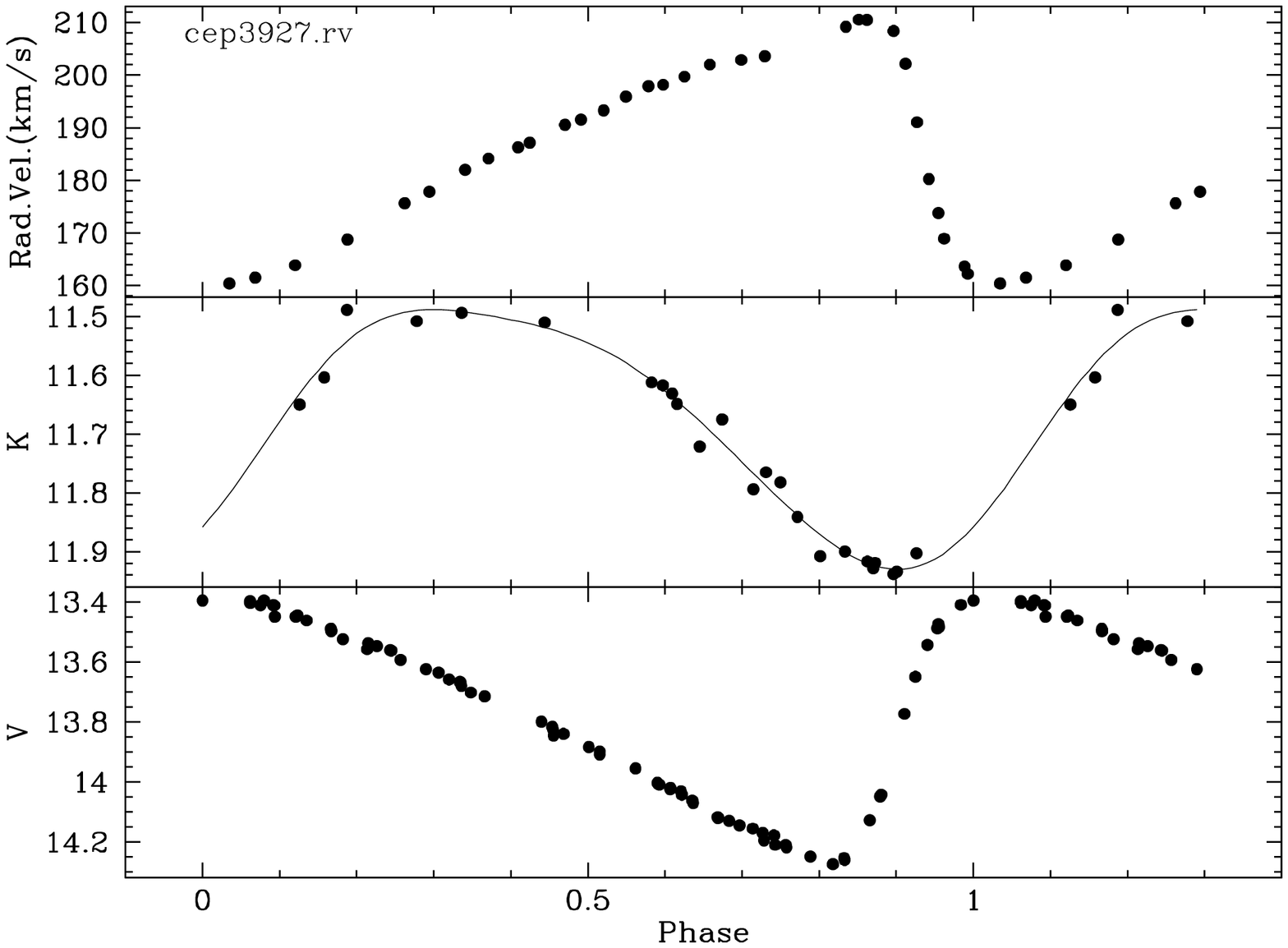}
\caption{The light and radial velocity curves for the star OGLE-SMC-CEP3927.
\label{fig.cep3927-data}}
\end{figure}

\begin{figure}
\centering
\includegraphics[width=9cm]{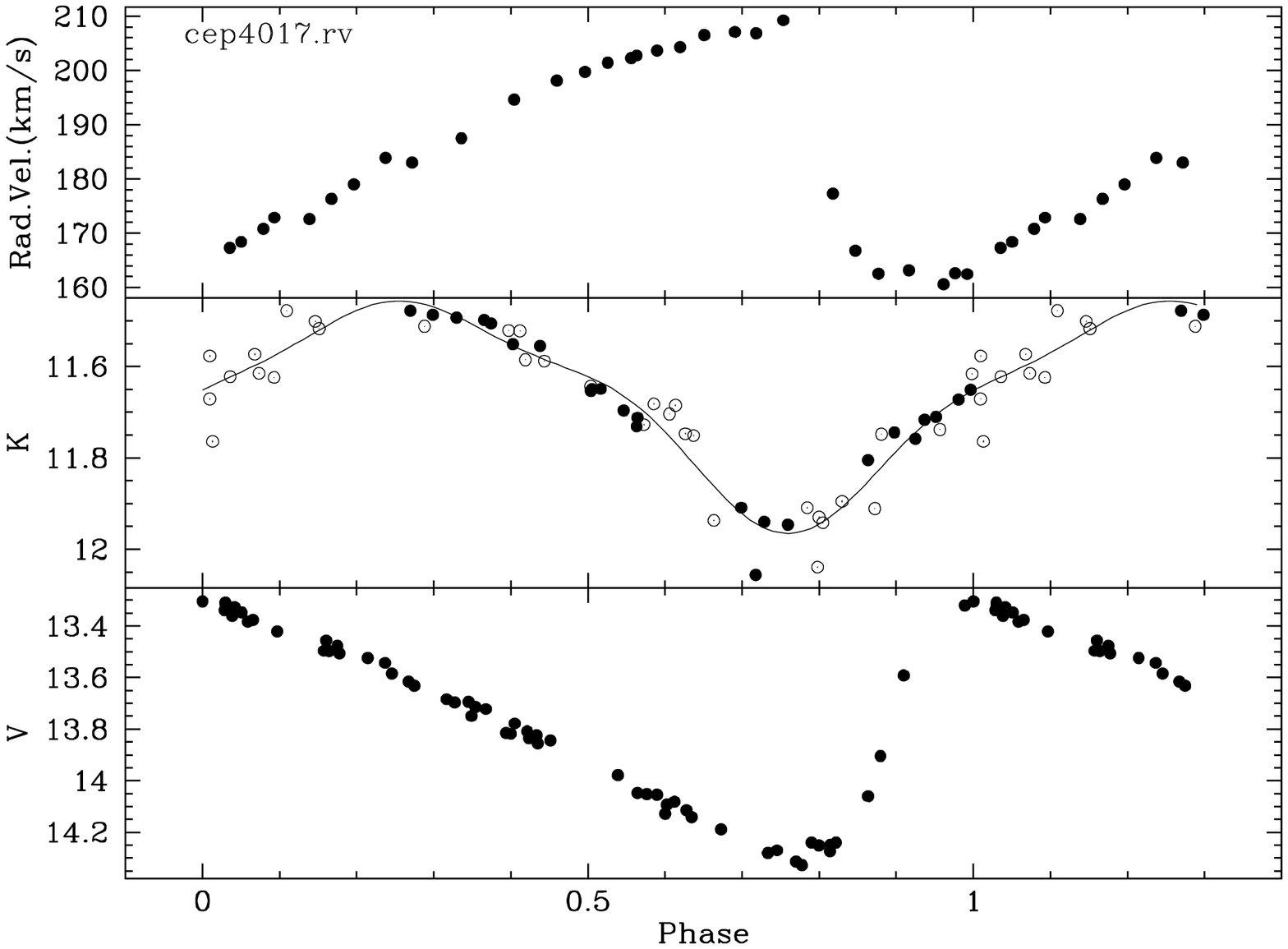}
\caption{The light and radial velocity curves for the star OGLE-SMC-CEP4017.
\label{fig.cep4017-data}}
\end{figure}

\begin{figure}
\centering
\includegraphics[width=9cm]{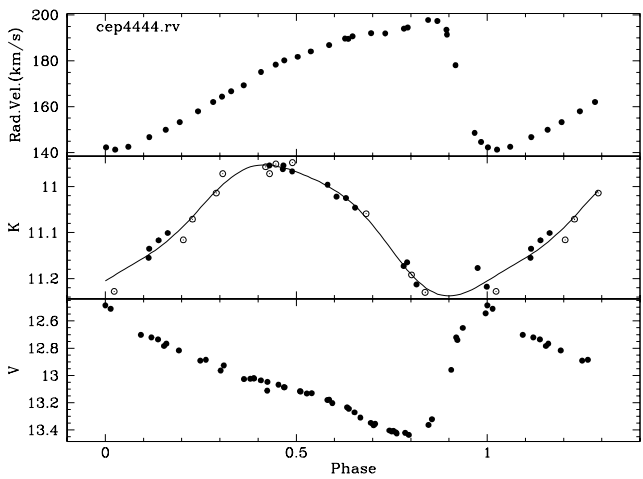}
\caption{The light and radial velocity curves for the star OGLE-SMC-CEP4444.
\label{fig.cep4444-data}}
\end{figure}

In the figures \ref{fig.cep0320-data}-\ref{fig.cep4444-data} we show the
data used for the IRSB analysis as described in Sec.\ref{sec.data} for
each of the stars. We distinguish in the $K$-band light curve between
the new data presented here (filled circles) and the appropriately
shifted VMC data (open circles). The adopted fourier fit to the $K$-band
light curve has also been overplotted.

\clearpage
\section{Removing orbital motion from the radial velocity curves}
\label{app.orbitalrv}
\begin{figure}
\centering
\includegraphics[width=9cm]{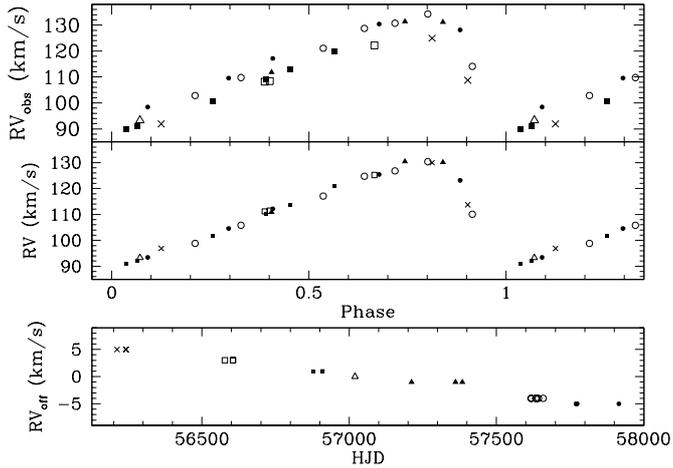}
\caption{The observed radial velocities for the star OGLE-SMC-CEP-1680
are shown in the upper panel. In the middle panel the adopted velocities
are shown after applying radial velocity offsets to the data from a
given epoch. The velocity offsets adopted are shown as function of time
in the lower panel. Different symbols refer to different epochs.
\label{fig.rvoff_cep1680_app}}
\end{figure}

\begin{figure}
\centering
\includegraphics[width=9cm]{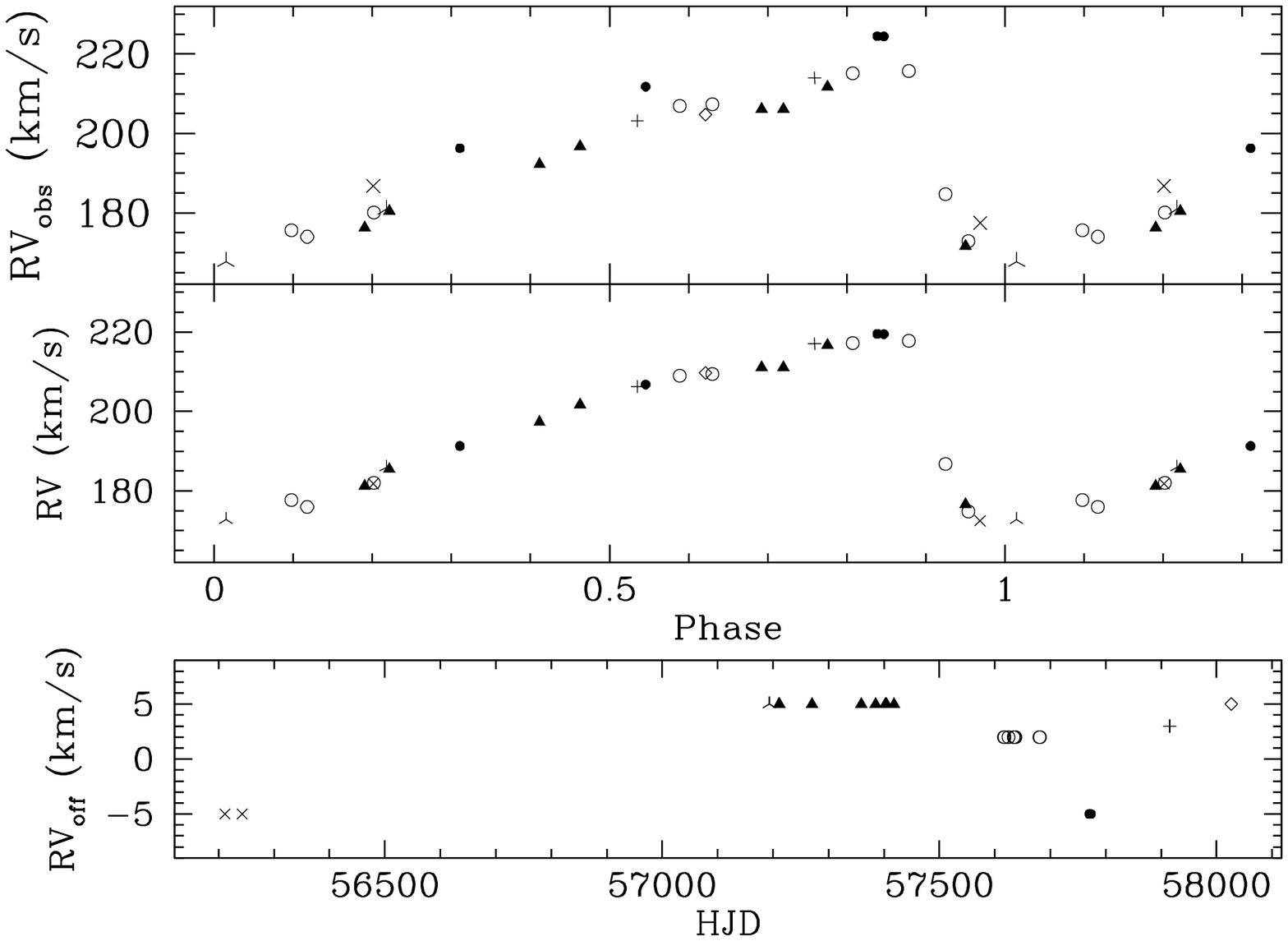}
\caption{The observed radial velocities for the star OGLE-SMC-CEP-1729
are shown in the upper panel. In the middle panel the adopted velocities
are shown after applying radial velocity offsets to the data from a
given epoch. The velocity offsets adopted are shown as function of time
in the lower panel. Different symbols refer to different epochs.
\label{fig.rvoff_cep1729_app}}
\end{figure}

\begin{figure}
\centering
\includegraphics[width=9cm]{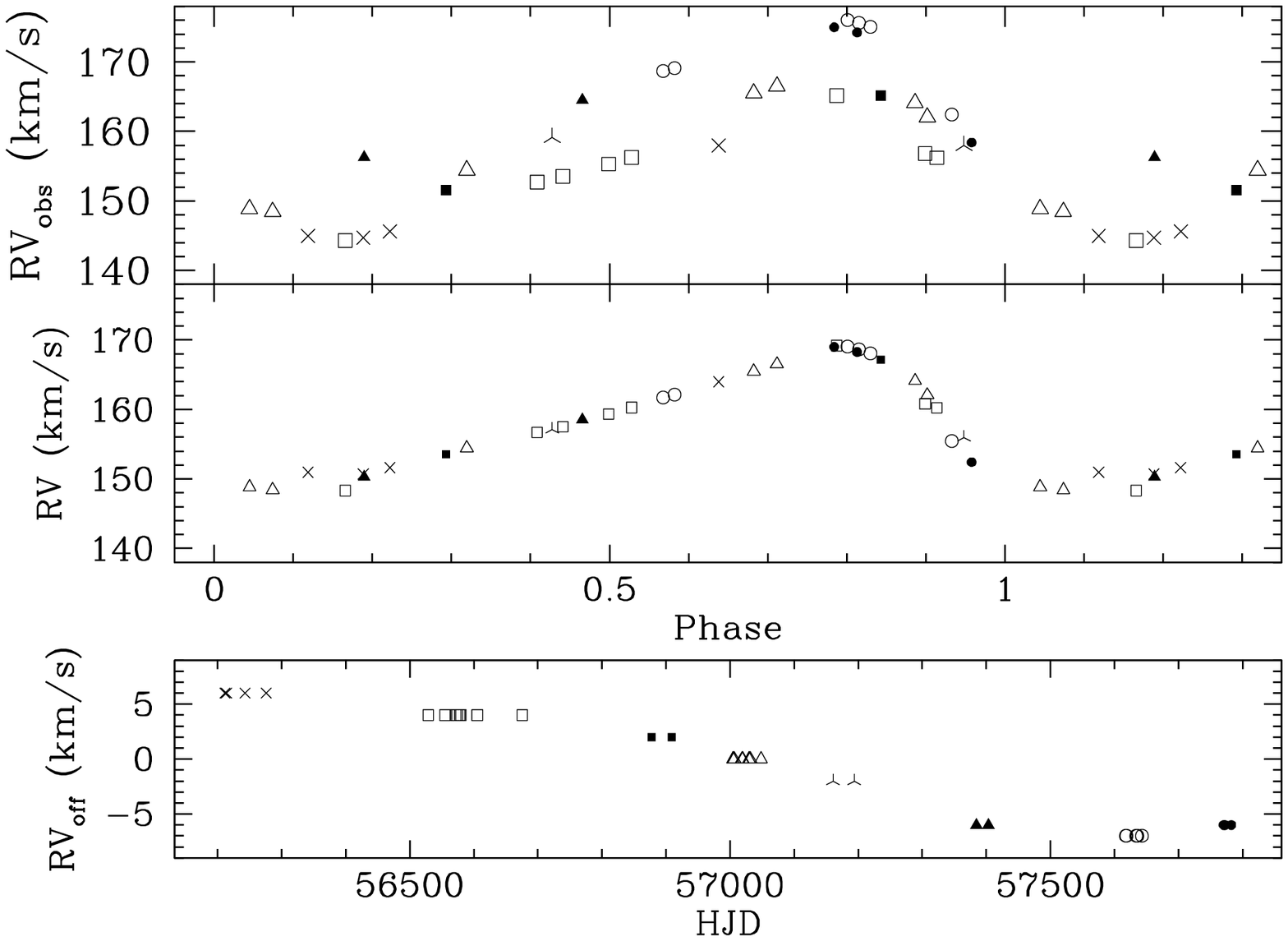}
\caption{The observed radial velocities for the star OGLE-SMC-CEP-1977
are shown in the upper panel. In the middle panel the adopted velocities
are shown after applying radial velocity offsets to the data from a
given epoch. The velocity offsets adopted are shown as function of time
in the lower panel. Different symbols refer to different epochs.
\label{fig.rvoff_cep1977_app}}
\end{figure}

\begin{figure}
\centering
\includegraphics[width=9cm]{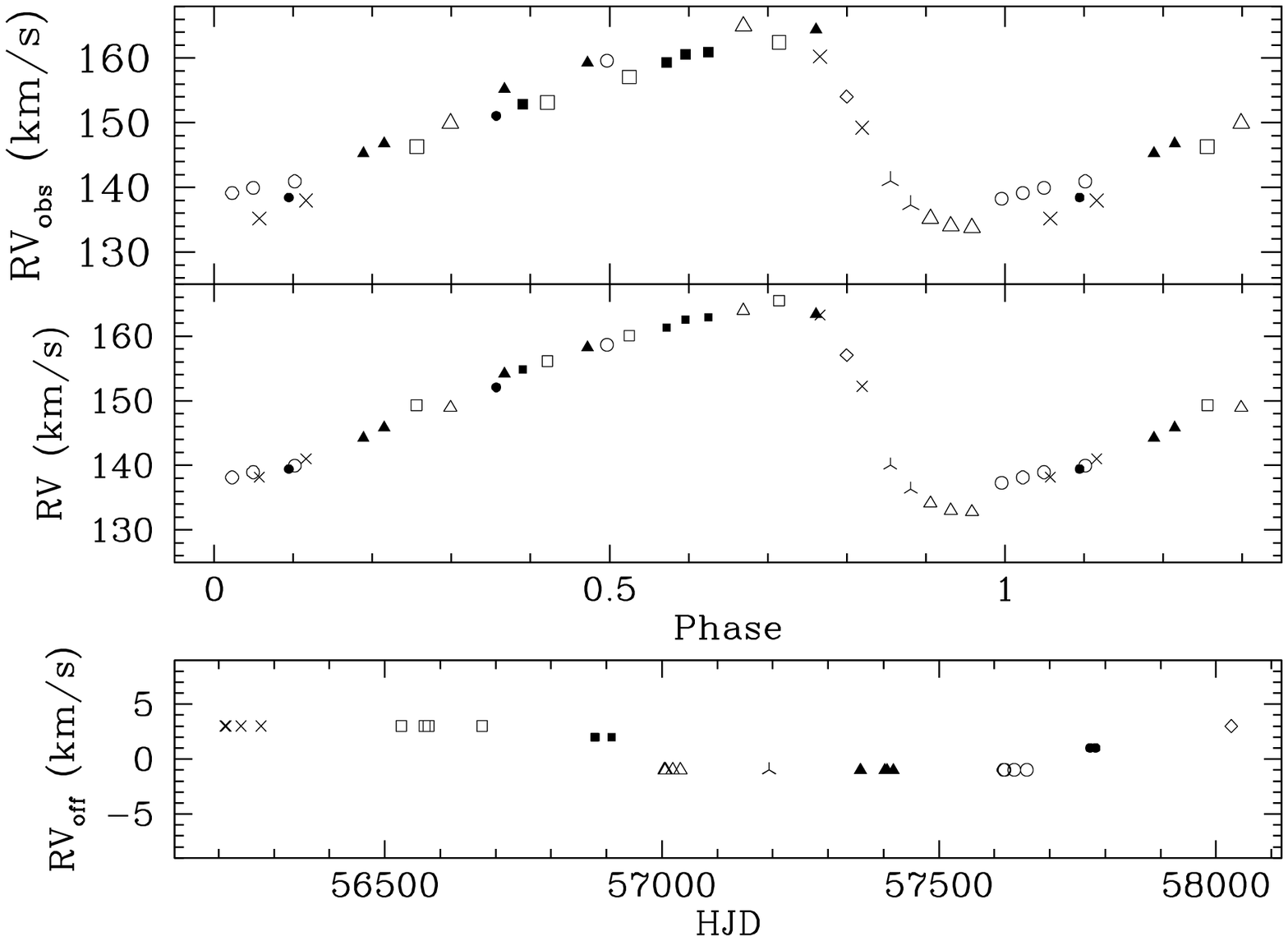}
\caption{The observed radial velocities for the star OGLE-SMC-CEP-2905
are shown in the upper panel. In the middle panel the adopted velocities
are shown after applying radial velocity offsets to the data from a
given epoch. The velocity offsets adopted are shown as function of time
in the lower panel. Different symbols refer to different epochs.
\label{fig.rvoff_cep2905_app}}
\end{figure}

\begin{figure}
\centering
\includegraphics[width=9cm]{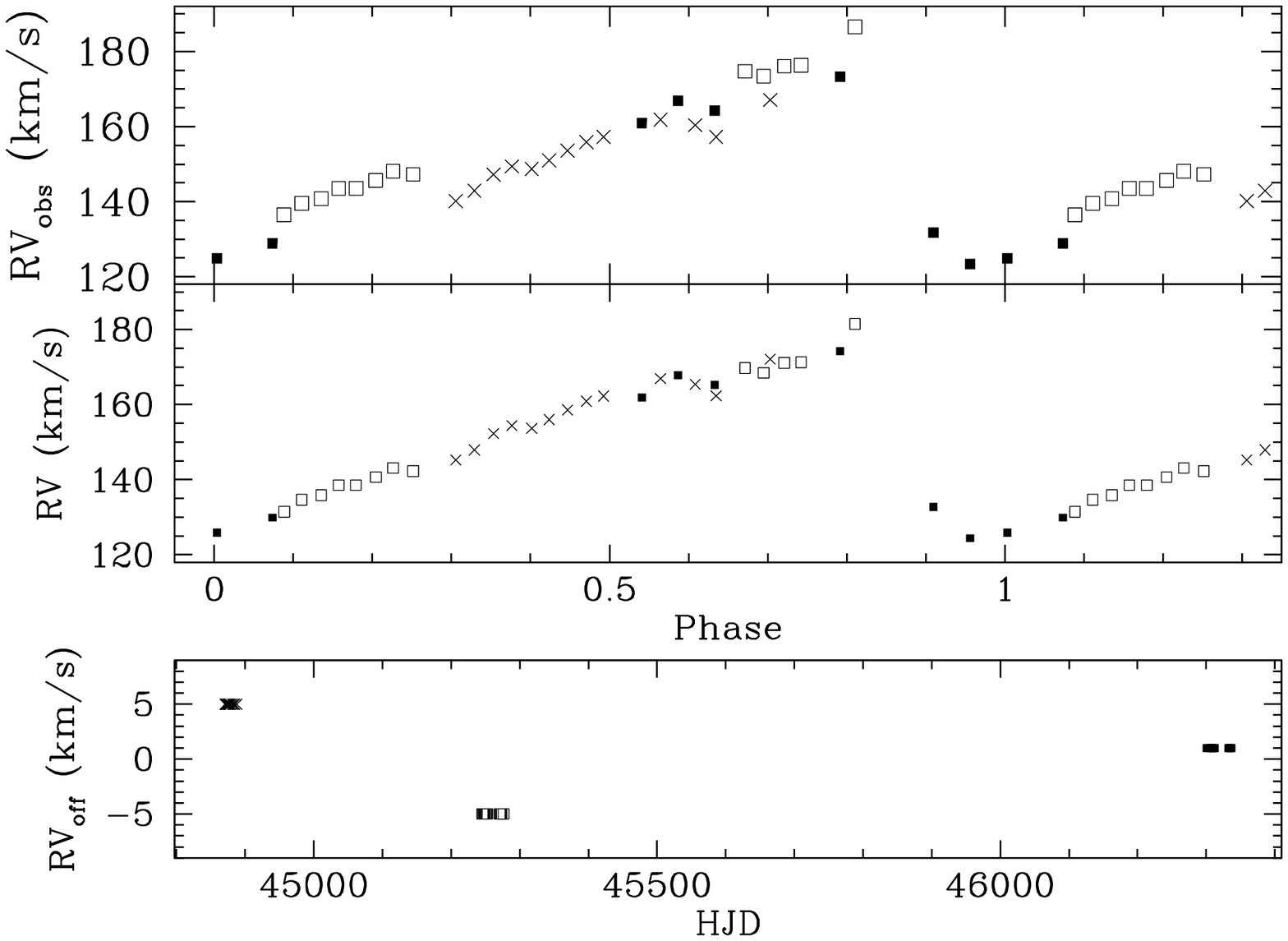}
\caption{The observed radial velocities for the star HV837
are shown in the upper panel. In the middle panel the adopted velocities
are shown after applying radial velocity offsets to the data from a
given epoch. The velocity offsets adopted are shown as function of time
in the lower panel. Different symbols refer to different epochs.
\label{fig.rvoff_HV837_app}}
\end{figure}

\begin{table*}
\caption{\label{tab.RVorb}Radial velocities, $RV$, corrected for the putative
orbital motion. Tabulated is the identfier,
the pulsational phase, the HJD, the actually observed radial velocity, 
$RV_{\mbox{\scriptsize obs}}$,
the estimated statistical uncertainty, $\sigma$, the corrected radial
velocity, $RV$, and the applied velocity offset, $RV_{\mbox{\scriptsize
off}}$.
The full table is available in the electronic version of the paper as 
well as from the CDS.}
\begin{tabular}{c c c r r r r}
\hline\hline
Identifier & phase & HJD & 
\multicolumn{1}{c}{$RV_{\mbox{\scriptsize obs}}$} &
\multicolumn{1}{c}{$\sigma$} & 
\multicolumn{1}{c}{$RV$} & 
\multicolumn{1}{c}{$RV_{\mbox{\scriptsize off}}$}\\
\hline
OGLE-SMC-CEP-1680 & 0.038 & 2456877.71500 & 89.99 & 0.37 & 90.99 & 1.0 \\
OGLE-SMC-CEP-1680 & 0.064 & 2456877.84320 & 91.02 & 0.40 & 92.02 & 1.0 \\
OGLE-SMC-CEP-1680 & 0.072 & 2457019.64830 & 93.30 & 0.08 & 93.30 & 0.0 \\
OGLE-SMC-CEP-1680 & 0.091 & 2457772.57380 & 98.44 & 0.07 & 93.44 & -5.0 \\
OGLE-SMC-CEP-1680 & 0.126 & 2456242.63810 & 91.89 & 0.41 & 96.89 & 5.0 \\
OGLE-SMC-CEP-1680 & 0.212 & 2457616.72960 & 102.84 & 0.50 & 98.84 & -4.0 \\
OGLE-SMC-CEP-1680 & 0.256 & 2456878.77960 & 100.66 & 0.22 & 101.66 & 1.0 \\
OGLE-SMC-CEP-1680 & 0.297 & 2457773.57740 & 109.59 & 0.06 & 104.59 & -5.0 \\
OGLE-SMC-CEP-1680 & 0.328 & 2457636.85270 & 109.78 & 0.34 & 105.78 & -4.0 \\
OGLE-SMC-CEP-1680 & 0.388 & 2456605.67110 & 108.20 & 0.30 & 111.20 & 3.0 \\
...\\
\hline
 & & days & \kms & \kms & \kms & \kms \\
\hline
\end{tabular}
\end{table*}

In Sec.\ref{sec.rvdata} it was shown that some of the stars
exhibit secular  radial velocity changes which are most likely
due to orbital motion. We have attempted to minimize the impact
of these secular drifts by applying velocity offsets to data from
different time intervals by trying to obtain continous readial
velocity curves as a function of phase and also assuming that such
drifts are slow compared to the pulsational period of the stars. In
Fig.\ref{fig.rvoff_cep1680_app}-\ref{fig.rvoff_HV837_app} we have plotted
the observed data (upper panels), the corrected data (middle panels),
and the applied radial velocity offsets (lower panels).  The symbols
correspond to certain time interval and we have estimated the offsets
in steps of 1\kms. It can be seen that the scatter around the radial
velocity curves are strongly reduced and from the lower panels it
can be seen that the offsets follow a slow secular change compared
to the pulsational periods.  For two of the suspected binary stars,
OGLE-SMC-CEP-1686 and -1693 we did not find good shifts. In the first
case the offsets were small so we have simply adopted the data as they
were observed. For the second star however, the offsets seemed to be
very large compared to the pulsational amplitude and the time scale aslo
seemed to be comparable to the pulsational period. We consider this
star to be peculiar and have disregarded it in the further analysis.
The velocity offsets as well as the corrected velocities for the five
stars can be found in Tab.\ref{tab.RVorb}.

\end{appendix}

\end{document}